\documentclass[sigconf]{acmart}
\usepackage{enumitem}
\usepackage{todonotes}
\usepackage{rotating}
\usepackage{multirow}
\usepackage{natbib}
\usepackage{wasysym}
\RequirePackage[skins]{tcolorbox}
\newtcolorbox{custombox}[1]{
	colback=gray!10,
	colframe=gray!70,
	left=1mm,
	right=1mm,
	top=1mm,
	bottom=1mm,
	fonttitle=\bfseries,
	arc=0mm,
	leftrule=1mm,
	rightrule=0mm,
	toprule=0mm,
	bottomrule=0mm,
	notitle,
	before=\par\smallskip\noindent,
	before upper={\textbf{#1: } },
}

\newcommand{\ignore}[1]{}

\usepackage{subcaption}

\AtBeginDocument{%
  \providecommand\BibTeX{{%
    \normalfont B\kern-0.5em{\scshape i\kern-0.25em b}\kern-0.8em\TeX}}}

\copyrightyear{2023}
\acmYear{2023}
\setcopyright{rightsretained}
\acmConference[ITiCSE 2023]{Proceedings of the 2023 Conference on Innovation
and Technology in Computer Science Education V. 2}{July 8--12, 2023}{Turku,
Finland}
\acmBooktitle{Proceedings of the 2023 Conference on Innovation and
Technology in Computer Science Education V. 2 (ITiCSE 2023), July 8--12, 2023,
Turku, Finland}\acmDOI{10.1145/3587103.3594206}
\acmISBN{979-8-4007-0139-9/23/07}

\usepackage{color}

\begin{document}

\newcommand{\LLMS}[0]{{large language models }}

\title[The Robots are Here: The Generative AI Revolution in Computing Education]{The Robots are Here: \\Navigating the Generative AI Revolution in Computing Education}


%
%

\author[Prather]{James Prather}
\authornotemark[1]
\orcid{0000-0003-2807-6042}
\affiliation{
  \institution{Abilene Christian University}
  \city{Abilene, Texas}
  \country{USA}
}
\email{james.prather@acu.edu}

\author[Denny]{Paul Denny}
\authornotemark[1]
\orcid{0000-0002-5150-9806}
\affiliation{
  \institution{University of Auckland}
  \city{Auckland}
  \country{New Zealand}
}
\email{paul@cs.auckland.ac.nz}

\author[Leinonen]{Juho Leinonen}
\authornotemark[1]
\orcid{0000-0001-6829-9449}
\affiliation{
  \institution{University of Auckland}
  \city{Auckland}
  \country{New Zealand}
}
\email{juho.leinonen@auckland.ac.nz}

\author[Becker]{Brett A. Becker}
\authornote{Randomly-ordered Co-leaders}
\orcid{0000-0003-1446-647X}
\affiliation{%
  \institution{University College Dublin}
  \city{Dublin}
  \country{Ireland}
}
\email{brett.becker@ucd.ie}

%
%

\author{Ibrahim Albluwi}
\orcid{0000-0003-1816-3943}
\affiliation{
  \institution{Princess Sumaya University for Technology}
  \streetaddress{P.O. Box 1438}
  \city{Amman}
  \country{Jordan}
}
\email{i.albluwi@psut.edu.jo}


\author{Michelle Craig}
\orcid{0000-0001-8283-0072}
\affiliation{
  \institution{University of Toronto}
  \city{Toronto}
  \country{Canada}
}
\email{mcraig@cs.toronto.edu}

\author{Hieke Keuning}
\orcid{0000-0001-5778-7519}
\affiliation{%
   \institution{Utrecht University}
   \streetaddress{Heidelberglaan 8, 3584 CS Utrecht}
   \city{Utrecht}
   \country{The Netherlands}
}
\email{h.w.keuning@uu.nl}

\author{Natalie Kiesler}
\orcid{0000-0002-6843-2729}
\affiliation{%
   \institution{DIPF Leibniz Institute for Research and Information in
Education}
   \streetaddress{Rostocker Straße 6}
   \city{Frankfurt am Main}
   \country{Germany}
   \postcode{60323}
}
\email{kiesler@dipf.de}

\author{Tobias~Kohn}
\orcid{0000-0002-9251-8944}
\affiliation{%
  \institution{TU~Wien}
  \city{Vienna}
  \country{Austria}
}
\email{tobias.kohn@tuwien.ac.at}

\author{Andrew Luxton-Reilly}
\orcid{0000-0001-8269-2909}
\affiliation{
  \institution{University of Auckland}
  \city{Auckland}
  \country{New Zealand}
}
\email{a.luxton-reilly@auckland.ac.nz}

\author{Stephen MacNeil}
\orcid{0000-0003-2781-6619}
\affiliation{%
  \institution{Temple University} 
  \streetaddress{1801 N Broad St}
  \city{Philadelphia}
  \state{Pennsylvania}
  \country{USA}
  \postcode{19122}
}
\email{stephen.macneil@temple.edu}

\author{Andrew Petersen}
\orcid{0000-0003-1337-7985}
\affiliation{
  \institution{University of Toronto Mississauga}
  \city{Toronto}
  \country{Canada}
}
\email{andrew.petersen@utoronto.ca}

\author{Raymond Pettit}
\orcid{0000-0001-9675-025X}
\affiliation{
  \institution{University of Virginia}
  \city{Charlottesville}
  \state{Virginia}
  \country{USA}
}
\email{rp6zr@virginia.edu}

\author{Brent N. Reeves}
\orcid{0000-0001-5781-1136}
\affiliation{%
  \institution{Abilene Christian University}
  \city{Abilene}
  \state{Texas}
  \country{USA}
}
\email{brent.reeves@acu.edu}

\author{Jaromir Savelka}
\orcid{0000-0002-3674-5456}
\affiliation{%
  \institution{Carnegie Mellon University}
  \city{Pittsburgh}
  \state{Pennsylvania}
  \country{USA}
}
\email{jsavelka@andrew.cmu.edu}

\renewcommand{\shortauthors}{Prather, Denny, Leinonen, Becker, et al.}


\begin{abstract}

Recent advancements in artificial intelligence (AI) are fundamentally reshaping computing, with large language models (LLMs) now effectively being able to generate and interpret source code and natural language instructions. These emergent capabilities have sparked urgent questions in the computing education community around how educators should adapt their pedagogy to address the challenges and to leverage the opportunities presented by this new technology.  In this working group report, we undertake a comprehensive exploration of LLMs in the context of computing education and make five significant contributions.  First, we provide a detailed review of the literature on LLMs in computing education and synthesise findings from 71 primary articles, nearly 80\% of which have been published in the first 8 months of 2023. Second, we report the findings of a survey of computing students and instructors from across 20 countries, capturing prevailing attitudes towards LLMs and their use in computing education contexts. Third, to understand how pedagogy is already changing, we offer insights collected from in-depth interviews with 22 computing educators from five continents who have already adapted their curricula and assessments. Fourth, we use the ACM Code of Ethics to frame a discussion of ethical issues raised by the use of large language models in computing education, and we provide concrete advice for policy makers, educators, and students.  Finally, we benchmark the performance of LLMs on various computing education datasets, and highlight the extent to which the capabilities of current models are rapidly improving.  There is no doubt that LLMs and other forms of generative AI will have a profound impact on computing education over the coming years.  However, just as the technology will continue to improve, so will our collective knowledge about how to leverage these new models and tools in educational settings. We expect many important conversations around this topic will emerge as the community explores how to provide more effective, inclusive, and personalised learning experiences. Our aim is that this report will serve as a focal point for both researchers and practitioners who are exploring, adapting, using, and evaluating LLMs and LLM-based tools in computing classrooms.

\end{abstract}

\ccsdesc[500]{Social and professional topics~Computing education}
\ccsdesc[500]{Computing methodologies~Artificial intelligence}




\keywords{AI; artificial intelligence; code generation; Codex; computer programming; Copilot; CS1; GitHub; GPT; large language models; LLM; novice programming; OpenAI; pedagogical practices}



\settopmatter{printfolios=true}

\maketitle

\section{Introduction}



Many disruptions to computing education -- and education globally -- have occurred in the past few years. During the COVID-19 pandemic, students adapted to learning online in unprecedented ways. It was during this time Generative AI became available to the public with the November 2022 release of ChatGPT being the main catalyst. Suddenly students are not just learning \textit{about} AI in advanced computer science courses, but \textit{using} it. Unlike before, they are not using it just passively where AI powers some aspect of the tools they might use (such as Google Translate where AI \textit{transforms} data), but in an active manner where students are knowingly and intentionally using and interacting with AI as a tool to \textit{generate} new data with natural language prompts. These generative tools have much broader capabilities than what was available just a few years ago and can be used in all disciplines including computing for myriad tasks. 

In computing education, researchers have demonstrated that these models have an increasing capacity to perform source code generation and interpretation through a natural language interface \cite{denny2023computing}. For instance, it is likely that pair programming might evolve in some cases from two students working together to a student and their LLM working together~\cite{ITiCSE_Keynote_2023}. Many of these models are easily available and free for students, and early reports reveal that students are already using them for assistance on their assignments \cite{bird2023taking}. In addition, there is now at least one textbook, published in September 2023, which features the use of Generative AI -- specifically GitHub Copilot and ChatGPT -- from day 1 of introductory programming courses~\cite{porter2023learn}. The profound impacts of LLMs on computing education are still not entirely known but are already being felt by educators~\cite{lau2023from}.     

The evidence gathered over the past few decades about how students learn best supports the commonly adopted approach of having students write many small programs checked by automated assessment tools over the course of their introductory terms~\cite{allen2019msp}. However, this approach may have become obsolete given how easily most LLMs can now solve introductory computing problems with simple prompts~\cite{finnieansley2022robots, finnieansley2023my,reeves2023evaluating,savelka2023thrilled}. Furthermore, generative AI models can provide wrong or biased answers, and students may also become over-reliant on LLM tools or generate code plagiarised from online sources by the model~\cite{becker2023programming}. The models might generate code students do not understand~\cite{kazemitabaar2023studying} or may distract them with large blocks of text they did not write~\cite{prather2023its}. Teachers may look to AI detectors to enforce some semblance of normal, but evidence is mounting that these tools are currently ineffective~\cite{orenstrakh2023detecting}.

However, these models offer computing educators opportunities in addition to the aforementioned challenges. Recent research has shown promising possibilities for providing students with LLM partners in pair programming, given the right context and with the right scaffolding and support \cite{kazemitabaar2023studying, bird2023taking,prather2023its}. LLMs can also provide detailed code explanations to support students working through difficult problems ~\cite{macneil2023experiences,leinonen2023comparing} and can even explain error messages \cite{leinonen2023using} known to have vexed students for decades \cite{Becker2019wgpaper}. Instructors can also benefit as these models can rapidly generate new and personalised teaching materials and programming assignments ~\cite{sarsa2022automatic,denny2022robosourcing}. Most exciting are the opportunities for entirely new types of programming problems utilising LLMs, such as Prompt Problems~\cite{denny2023promptly}.

Large language models will have a profound impact on computing education in the next decade as the technology matures and as teachers and researchers identify opportunities. LLMs will change how, what, and whom we teach not only in computing but in all of education~\cite{ITiCSE_Keynote_2023}. This working group\footnote{\href{https://iticse23-generative-ai.github.io/}{iticse23-generative-ai.github.io/}} report aims to summarise these early movements in computing education to set an agenda for researchers and to collect effective practices for educators.

\section{Contributions}
\label{sec:contributions}
This working group report describes the following efforts that, taken together, aim to describe the current state of LLM issues in computing education and to set out a comprehensive vision of the future of programming education in the LLM era:

\begin{enumerate}[leftmargin=*,align=left]
    \item \textbf{Reviewing Literature (Section~\ref{sec:literature_review}):} We review the existing literature on LLMs in computing education\footnote{Through August 2023.} and present a guide to the opportunities and challenges of LLMs in this domain.
    \item \textbf{Evaluating Current Attitudes (Section~\ref{sec:survey}):} We conducted an international survey of students and instructors to obtain their perspectives of LLMs. From this data, we provide a snapshot of current attitudes toward LLMs and their uses.
    \item \textbf{Identifying New Instructional Approaches (Section~\ref{sec:canda}):} We interviewed instructors in terms of teaching about and/or using LLMs in the classroom. They provide insight into advantages and disadvantages of using LLMs in computing education.
    \item \textbf{Exploring Ethical Implications of LLMs (Sections~\ref{sec:ethics} and~\ref{SEC:ACAD-INTEGRITY}):} We perform an evidence-based ethical analysis on the use of LLMs in computing education by evaluating the AI policies of several leading universities in the context of the ACM Code of Ethics. These examples suggest how universities are responding -- and may in the future further respond -- to the ethical challenges presented by such systems. From this, we furthermore discuss academic integrity issues with LLMs and provide resources for both faculty and students to understand when it may or may not be permissible to utilise LLMs.
    \item \textbf{Encouraging Replication (Section~\ref{sec:benchmarks}):} We replicate prior work using new LLMs, highlighting challenges driven by the speed at which LLMs are improving and with current standards for describing research methods. To encourage comparisons between published work, we identify appropriate, openly available datasets and identify concerns with the quality and type of datasets available.
\end{enumerate}


\section{Review of Literature}\label{sec:literature_review}

The working group aimed to identify prior work that explores how {\LLMS} might impact computing education. We recognise that any such attempt in this nascent and rapidly expanding area of research will quickly become out of date but aim to establish the \textit{status questions} of this new research field and to provide recommendations based on the current scholarly discourse. Furthermore, we used the work we found to inform the other activities of the working group listed in Section \ref{sec:contributions}.

\begin{figure*}[ht!]
\centering
\includegraphics[width=0.7\textwidth]{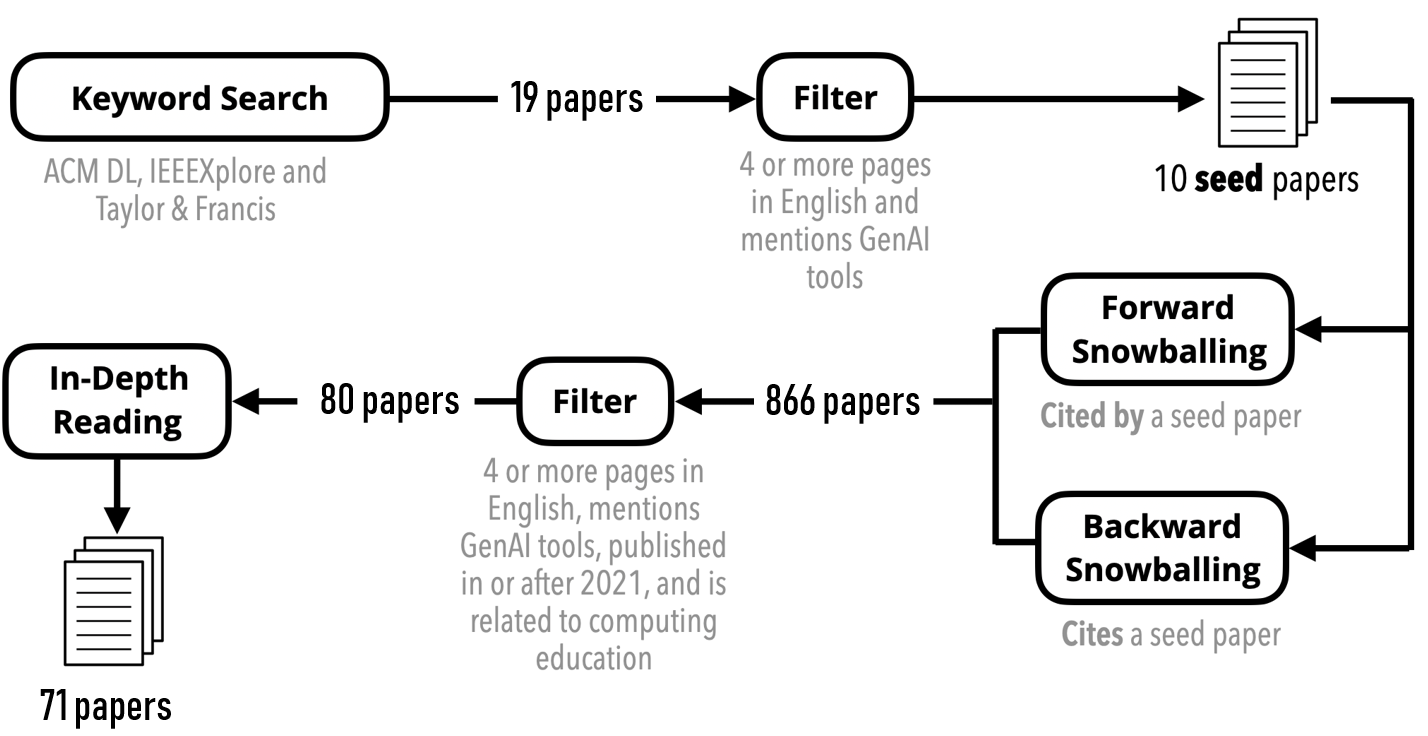}
\caption{Phases of the literature review.
\label{fig:review-process}}
\end{figure*}

\subsection{Method}

We chose to perform a scoping review to rapidly identify gaps and major themes in the literature discussing how \LLMS can support computing education. We explicitly considered but decided not to perform a systematic review, as the research in this area is evolving quickly and relies heavily on dissemination through non-traditional publication channels such as arXiv. We chose to perform one step of forward and backward snowballing~\cite{wohlin2022successful} from a set of reference papers that were identified as being currently significant work in the area of \LLMS in computing education. We decided only one step in the snowballing phase was necessary given the recent advent of \LLMS in computing education. We conducted two separate phases of forward snowballing, one in May 2023 and one in August 2023, with the aim of including as much of recent work as possible.

\subsubsection{Reference papers}

\begin{table*}
\centering
    \caption{Reference papers used to seed the literature review.\label{tab:ref-papers}}
    \begin{tabular}{p{5.15in}|l|c|l}
    \toprule
    \textbf{Title} & \textbf{Venue} & \textbf{Year} & \textbf{Citation} \\\midrule

The Robots Are Coming: Exploring the Implications of OpenAI Codex on Introductory Programming	& ACE & 2022 & \cite{finnieansley2022robots}  \\ \midrule
Automatic Generation of Programming Exercises and Code Explanations Using Large Language Models & ICER	& 2022 & \cite{sarsa2022automatic}  \\ \midrule
Github copilot in the classroom: learning to code with AI assistance & JCSC & 2022 & \cite{puryear2022github} \\ \midrule
Programming Pedagogy and Assessment in the Era of AI/ML: A Position Paper & COMPUTE	& 2022 & \cite{raman2022programming} 	\\ \midrule
My AI Wants to Know If This Will Be on the Exam	& ACE & 2023 & \cite{finnieansley2023my}  \\ \midrule
Using Large Language Models to Enhance Programming Error Messages & SIGCSE TS& 2023 & \cite{leinonen2023using} \\ \midrule
Experiences from Using Code Explanations Generated by Large Language Models in a Web Software Development E-Book	& SIGCSE TS& 2023 & \cite{macneil2023experiences} 	\\ \midrule
Conversing with Copilot: Exploring Prompt Engineering for Solving CS1 Problems Using Natural Language & SIGCSE TS & 2023 & \cite{denny2023conversing}  \\ \midrule
Using GitHub Copilot to Solve Simple Programming Problems & SIGCSE TS& 2023 & \cite{wermelinger2023using} \\  \midrule 
Programming Is Hard - Or at Least It Used to Be: Educational Opportunities and Challenges of AI Code Generation	& SIGCSE TS& 2023 & \cite{becker2023programming}\\

\bottomrule
    \end{tabular}
\end{table*}

We collected a set of reference papers using keyword searches over three databases: (1) ACM Digital Library (Full-Text Collection), (2) Taylor \& Francis Online, and (3) IEEE Xplore.  These choices were guided by the book ``Past, Present and Future of Computing Education Research: A Global Perspective''~\cite{apiola2023past} which includes a chapter on venues that have shaped computing education research (pp 121-150).  This chapter lists 13 conference and magazine venues and two journals dedicated to computing education research literature, and our database searches were scoped to cover these venues: ACM SIGCSE Sponsored (SIGCSE Technical Symposium, ITiCSE, ICER, CompEd); ACM SIGCSE In-Cooperation (ACE, Koli Calling, COMPUTE, WiPSCE, CCSC); ACM Journal (TOCE); Taylor and Francis (CSE); and IEEE (FiE, ToE, TLT).

The keywords used included ``large language models'' and ``generative AI'' as well as three common models.  Queries were refined as appropriate for the different databases, and filters were used as appropriate when scoping the search, such as filtering by ``SIGCSE sponsored'' venues in the ACM Digital Library.  In addition, the searches were conducted using a filter for dates beginning in January 2021.  
The start date of January 2021 was chosen based on the technological timeline of LLMs and their relevance to computing education. By January 2021, LLMs, especially with the advent of models like GPT-3 in mid-2020, started gaining significant traction and recognition in broader research and application areas. Furthermore, the integration of such advanced LLMs into computing education, pedagogically and practically, was still in nascent stages. By setting January 2021 the start date of the literature search, we aimed to capture the most recent and relevant research insights right from the outset of substantial scholarly attention towards the intersection of LLMs and computing education. 
As an example, the final query used when searching the ACM Digital Library was:

\begin{verbatim}
        [All: "large language models"] OR 
        [All: "generative AI"] OR 
        [All: "Codex"] OR 
        [All: "GPT-3"] OR 
        [All: "GPT-4"]
\end{verbatim}

The search was conducted on 26th April 2023 and resulted in 19 papers.  For each paper, the following inclusion criteria was applied:

\begin{enumerate}
  \item \textit{Must mention generative AI, large language models, or a specific tool using those technologies, such as GitHub Copilot.}
  \item \textit{At least 4 pages in length (inclusive).\footnote{This criterion rules out posters and abstracts.}}
  \item \textit{Written in English.}
\end{enumerate}

A total of 3 papers were excluded based on length and 5 for not being aligned with the topic. 
The resulting set of reference papers ("seed papers"), listed in Table~\ref{tab:ref-papers}, includes 10 papers. By necessity due to the age of the research area, these papers are largely published in 2022 and 2023.

\subsubsection{Snowballing (phase 1)}

Each paper that cites or is cited by at least one of the reference papers was evaluated for inclusion by two working group members. The backward snowballing phase, which considered all papers in the reference list for each paper in the reference set, resulted in 381 papers. For forward snowballing, we used the ``cited by'' feature in Google Scholar at the beginning of May 2023, resulting in 132 papers. There were duplicates in this list, but we decided not to identify duplicates until the final review. Each of these 513 snowballed papers were assigned to two members of the group. 

At this stage of the review, the papers were not read in detail; rather, the evaluators searched for evidence that a paper should be given deeper consideration.  The inclusion criteria included the three criteria used to filter the reference papers plus a publication date criterion and a content criterion:

\begin{enumerate}
  \item \textit{Must mention generative AI, large language models, or a specific tool using those technologies, such as Copilot, AND }
  \item \textit{At least 4 pages in length (inclusive), AND} 
  \item \textit{Written in English, AND }
  \item \textit{``Published'' in or after 2021.} For papers published in non-traditional venues such as arXiv, this is the upload date, AND 
\item \textit{Must have direct applicability to computing education.} This criterion was refined to the following and was interpreted generously:
  \begin{enumerate}
      \item \textit{The paper explicitly states a relation to computing education, OR}
      \item \textit{the participants include students working on problems typical of a computing education context, OR}
      \item \textit{the problems or inputs featured are drawn from a computing education context, OR}
      \item \textit{the resource or tool being created is specifically designed for computing education.}
  \end{enumerate}
\end{enumerate}

Each of the five criteria had to be satisfied to include the paper. Each paper was independently evaluated twice; the evaluator could flag the paper for inclusion, exclusion, or discussion. If there was disagreement between the two evaluators or if they flagged the paper for discussion, it was evaluated by a third evaluator who made a final decision. In addition, given the subjectivity of criterion (5), all papers that were marked as not being included because of this criterion were reviewed by a third evaluator. As a result, papers were included if (a) both initial evaluators flagged it for inclusion or (b) if the third evaluator intervened due to a disagreement between the initial reviewers or when they reviewed criterion (5). All evaluators were instructed to interpret the criteria generously, to avoid exclusion of potentially relevant work. After these reviews, 46 (9.0\%) papers were identified for inclusion. 

Finally, each of these papers selected for inclusion were read in depth by a single member of the working team. The goal of this step was to identify potential impact on future research in the area or on computing education practice. In addition, we extracted some details about the work being performed, such as the location of authors, the type of work published, and evidence of research quality. Data extraction was guided by a set of questions implemented as an online form to help standardise the process. The questions on the form are presented in Appendix~\ref{app:paper-extraction}.

During this final step and upon a more thorough reading of the papers, 8 additional papers were flagged as not being relevant. The exclusion of some papers at this stage was expected, as the reviewers had been instructed to identify any \textit{potentially} relevant work. These exclusions left us with 38 relevant papers (including the original ten reference papers).

\subsubsection{Snowballing (phase 2)}

We ran a second phase of forward snowballing using the ``cited by'' feature in Google Scholar at the end of August 2023. After removing the duplicate papers that appeared in the first snowballing phase, and applying the same process described above for paper inclusion and exclusion, we ended up with a total of 71 papers (10 reference papers, 28 papers from the first snowballing phase, and 33 papers from the second). The final list of papers (including the 10 original reference papers and papers from the two snowballing phases) is shown in Table~\ref{tab:lit}. Interestingly, while the second snowballing phase covered only a few months, it resulted in a number of papers that is very close to the number of papers that resulted from the first snowballing phase, which covered a period of around two and a half years. Including a second snowballing phase was motivated by the very fast pace at which the literature is growing in this area, which this finding supports. 

\subsection{Descriptive statistics}

\begin{table*}

\scriptsize

	\begin{center}
 	\caption{The list of papers included in the literature review. 
  }
	\label{tab:lit}
		\begin{tabular}{|p{2cm}|p{12cm}|p{2cm}|l|}
			\hline
			Author & Title & Venue & Year \\
   \hline


\hline
\citeauthor{ahmed2022synshine} &SYNSHINE: improved fixing of Syntax Errors &IEEE Trans. Softw. Eng. &2021\\
\hline
\citeauthor{austin2021program} &Program Synthesis with Large Language Models &arXiv &2021\\
\hline


\hline
\citeauthor{brennan2022exploring} &	 	Exploring the Implications of OpenAI Codex on Education for Industry 4.0 &SOHOMA &2022\\ 
\hline
\citeauthor{dakhel2022github} &	 	GitHub Copilot AI Pair Programmer: Asset or Liability? &J. Syst. Softw. &2022\\
\hline
\citeauthor{denny2022robosourcing} &Robosourcing Educational Resources -- Leveraging Large Language Models for Learnersourcing &arXiv &2022\\
\hline
\citeauthor{ernst2022ai-driven} &	 	AI-Driven Development Is Here: Should You Worry? &{IEEE} Software &2022\\
\hline
\citeauthor{finnieansley2022the} &	 	The Robots Are Coming: Exploring the Implications of OpenAI Codex on Introductory Programming &ACE &2022\\
\hline
\citeauthor{gherciu2022net} &	 	Net Impact of Large Language Models Trained on Code &Student conf. 
&2022\\
\hline
\citeauthor{li2022competition} &	 	Competition-level code generation with AlphaCode &Science &2022\\
\hline
\citeauthor{puryear2022github} &	 	Github copilot in the classroom: learning to code with AI assistance &J. Comput. Sci. Coll. &2022\\
\hline
\citeauthor{raman2022programming} &	 	Programming Pedagogy and Assessment in the Era of AI/ML: A Position Paper &COMPUTE &2022\\
\hline
\citeauthor{sarsa2022automatic} &	 	Automatic Generation of Programming Exercises and Code Explanations Using Large Language Models &ICER  &2022\\
\hline
\citeauthor{vaithilingam2022expectation} &	 	Expectation vs. Experience: Evaluating the Usability of Code Generation Tools Powered by Large Language Models &CHI EA &2022\\
\hline
\citeauthor{zhang2022repairing} &	 	Repairing Bugs in Python Assignments Using Large Language Models &arXiv &2022\\
\hline


\hline
\citeauthor{al2023socratic} &	 	Socratic Questioning of Novice Debuggers: A Benchmark Dataset and Preliminary Evaluations &BEA & 2023\\
\hline
\citeauthor{alves2023centaur} &The centaur programmer - How Kasparov’s Advanced Chess spans over to the software development of the future &arXiv &2023\\
\hline
\citeauthor{babe2023studenteval} &	 	StudentEval: A Benchmark of Student-Written Prompts for Large Language Models of Code &arXiv &2023\\
\hline
\citeauthor{balse2023investigating} & Investigating the Potential of GPT-3 in Providing Feedback for Programming Assessments & ITiCSE & 2023\\
\hline
\citeauthor{barke2023grounded} &Grounded Copilot: How Programmers Interact with Code-Generating Models &OOPSLA &2023\\
\hline
\citeauthor{becker2023programming} &Programming Is Hard - Or at Least It Used to Be: Educational Opportunities and Challenges of AI Code Generation &SIGCSE TS &2023\\
\hline
\citeauthor{bellettini2023davinci} & Davinci Goes to Bebras: A Study on the Problem Solving Ability of GPT-3 &CSEDU &2023\\
\hline
\citeauthor{brusilovsky2023the} &	 	The Future of Computing Education Materials & (in draft) &2023\\
\hline
\citeauthor{bull2023generative} & Generative AI Assistants in Software Development Education: A vision for integrating Generative AI into educational practice, not instinctively defending against it. & IEEE Software & 2023\\
\hline
\citeauthor{cipriano2023gpt} & GPT-3 vs Object Oriented Programming Assignments: An Experience Report & ITiCSE & 2023\\
\hline
\citeauthor{denny2023can} & Can We Trust AI-Generated Educational Content? Comparative Analysis of Human and AI-Generated Learning Resources & arXiv & 2023\\
\hline
\citeauthor{denny2023computing} & Computing Education in the Era of Generative AI & arXiv & 2023\\
\hline
\citeauthor{denny2023conversing} &Conversing with Copilot: Exploring Prompt Engineering for Solving CS1 
Problems Using Natural Language& SIGCSE TS &2023\\
\hline
\citeauthor{denny2023promptly} & Promptly: Using Prompt Problems to Teach Learners How to Effectively Utilize AI Code Generators & arXiv & 2023\\
\hline
\citeauthor{dobslaw2023experiences} & Experiences with Remote Examination Formats in Light of GPT-4 & arXiv & 2023\\
\hline
\citeauthor{druga2023scratch} &	 	Scratch Copilot Evaluation: Assessing AI-Assisted Creative Coding for Families &arXiv &2023\\
\hline
\citeauthor{finnieansley2023my} &	 	My AI Wants to Know If This Will Be on the Exam&ACE &2023\\
\hline
\citeauthor{french2023creative} & Creative Use of OpenAI in Education: Case Studies from Game Development & Multi-modal Tech. \& Interaction & 2023\\
\hline
\citeauthor{hellas2023exploring} & Exploring the Responses of Large Language Models to Beginner Programmers' Help Requests & arXiv & 2023\\
\hline
\citeauthor{idialuwhodunnit} &	 	Whodunnit: Human or AI?& - &2023\\
\hline
\citeauthor{jaipersaud2023decomposed} & Decomposed Prompting to Answer Questions on a Course Discussion Board & AI in Education & 2023\\
\hline
\citeauthor{jalil2023chatgpt} &	 	ChatGPT and Software Testing Education: Promises \& Perils& IEEE ICSTW & 2023\\
\hline
\citeauthor{kazemitabaar2023studying} &	 	Studying the effect of AI Code Generators on Supporting Novice Learners in Introductory Programming&CHI &2023\\
\hline
\citeauthor{kendon2023ai} & AI-Generated Code Not Considered Harmful & WCCCE & 2023\\
\hline
\citeauthor{kiesler2023large} & Large Language Models in Introductory Programming Education: ChatGPT's Performance and Implications for Assessments & arXiv & 2023\\
\hline
\citeauthor{lau2023from} & From "Ban It Till We Understand It" to "Resistance is Futile": How University Programming Instructors Plan to Adapt as More Students Use AI Code Generation & ICER & 2023\\
\hline
\citeauthor{leinonen2023comparing} &	 	Comparing Code Explanations Created by Students and Large Language Models &arXiv &2023\\
\hline
\citeauthor{leinonen2023using} &		Using Large Language Models to Enhance Programming Error Messages &SIGCSE TS &2023\\
\hline
\citeauthor{liffiton2023codehelp} & CodeHelp: Using Large Language Models with Guardrails for Scalable Support in Programming Classes & arXiv & 2023\\
\hline
\citeauthor{ma2023ai} & Is AI the better programming partner? Human-Human Pair Programming vs. Human-AI pAIr Programming & arXiv & 2023\\
\hline
\citeauthor{macneil2023experiences} &	 	Experiences from Using Code Explanations Generated by Large Language Models in a Web Software Development E-Book&SIGCSE TS &2023\\
\hline
\citeauthor{matelsky2023large} &	 	A large language model-assisted education tool to provide feedback on open-ended responses &arXiv &2023\\
\hline
\citeauthor{nam2023in} & In-IDE Generation-based Information Support with a Large Language Model & arXiv & 2023\\
\hline
\citeauthor{orenstrakh2023detecting} &	 	Detecting LLM-Generated Text in Computing Education: A Comparative Study for ChatGPT Cases &arXiv &2023\\
\hline
\citeauthor{padurean2023neural} & Neural Task Synthesis for Visual Programming & arXiv & 2023\\
\hline
\citeauthor{pankiewicz2023large} & Large Language Models (GPT) for automating feedback on programming assignments & arXiv & 2023\\
\hline
\citeauthor{philbin2023exploring} &	 	Exploring the Potential of Artificial Intelligence Program Generators in Computer Programming Education for Students &Inroads & 2023\\
\hline
\citeauthor{phung2023generating} &	 	Generating High-Precision Feedback for Programming Syntax Errors using Large Language Models&arXiv &2023\\
\hline
\citeauthor{phung2023generative} &	 	Generative AI for Programming Education: Benchmarking ChatGPT, GPT-4, and Human Tutors &International J. of Management &2023\\
\hline
\citeauthor{piccolo2023many} &	 	Many bioinformatics programming tasks can be automated with ChatGPT &arXiv &2023\\
\hline
\citeauthor{poldrack2023aiassisted} &	 	AI-assisted coding: Experiments with GPT-4&arXiv &2023\\
\hline
\citeauthor{prather2023its} &	 	``It's Weird That it Knows What I Want'': Usability and Interactions with Copilot for Novice Programmers&TOCHI &2023\\
\hline
\citeauthor{rajabi2023exploring} & Exploring ChatGPT's impact on post-secondary education: A qualitative study & WCCCE & 2023\\
\hline
\citeauthor{reeves2023evaluating} &	 	Evaluating the Performance of Code Generation Models for Solving Parsons Problems With Small Prompt Variations &ITiCSE & 2023\\
\hline
\citeauthor{ross2023a} &	 	A Case Study in Engineering a Conversational Programming Assistant's Persona&ACM IUI&2023\\
\hline
\citeauthor{sandoval2023lost} &	 	Lost at C: A User Study on the Security Implications of Large Language Model Code Assistants &USENIX  &2023\\
\hline
\citeauthor{savelka2023can} &	 	Can Generative Pre-trained Transformers (GPT) Pass Assessments in Higher Education Programming Courses?&arXiv &2023\\
\hline
\citeauthor{savelka2023large} &	 	Large Language Models (GPT) Struggle to Answer Multiple-Choice Questions about Code&arXiv &2023\\
\hline
\citeauthor{savelka2023thrilled} & Thrilled by Your Progress! Large Language Models (GPT-4) No Longer Struggle to Pass Assessments in Higher Education Programming Courses & arXiv & 2023\\
\hline
\citeauthor{singla2023evaluating} & Evaluating ChatGPT and GPT-4 for Visual Programming & arXiv & 2023\\
\hline
\citeauthor{sridhar2023harnessing} & Harnessing llms in curricular design: Using gpt-4 to support authoring of learning objectives & arXiv & 2023\\
\hline
\citeauthor{wang2023exploring} & Exploring the Role of AI Assistants in Computer Science Education: Methods, Implications, and Instructor Perspectives & VL/HCC & 2023\\
\hline
\citeauthor{wermelinger2023using} &	 	Using GitHub Copilot to Solve Simple Programming Problems&SIGCSE TS &2023\\
\hline
\citeauthor{widjojo2023addressing} &	 	Addressing Compiler Errors: Stack Overflow or Large Language Models? &arXiv& 2023\\
\hline
\citeauthor{yan2023practical} &	 	Practical and Ethical Challenges of Large Language Models in Education: A Systematic Literature Review&arXiv &2023\\
\hline
\citeauthor{zan2023large} & Large language models meet NL2Code: A survey & Annual Meeting of the ACL & 2023\\
\hline
\citeauthor{zastudil2023generative} & Generative AI in Computing Education: Perspectives of Students and Instructors & arXiv & 2023\\
\hline

		\end{tabular}
	\end{center}
\end{table*}

\begin{table}
	\begin{center}
 	\caption{Venues presenting the work included in our literature review.}
	\label{tab:venues}
		\begin{tabular}{l|c}
			\toprule
			Venue & Count\\\midrule
   ACM & 22\\
   ArXiV & 32\\
   IEEE & 5\\\midrule
   Other Publishers &  9\\
   Grey Literature & 3\\\bottomrule
		\end{tabular}
	\end{center}
\end{table}

Statistics about the papers included in our analysis are presented in Table~\ref{tab:venues} and Table~\ref{tab:LLMs}. The work has been presented in a range of venues, including traditional conferences and journals. However, due to rapid changes in this field, a large number of papers were published only on arXiv. Some of this work was later published in a conference or journal, but some only remains visible -- and is cited from -- that site.

\ignore{
===== Article type =====
Evaluation paper (evaluating an existing tool / LLM), 20
Supervised study, 9
Position / discussion paper, 7
New tool paper (presenting a new tool / LLM), 3
Unsupervised study, 3
systematic literature review, 1
} 

\ignore{
===== Affiliation =====
Academic, 33
Industry, 6

===== Country =====
USA, 25
New Zealand, 11
Finland, 5
Canada, 5
Ireland, 5
United Kingdom, 3
India, 2
Australia, 1
Argentina, 1
Portugal, 1
Switzerland, 1
Unspecified but probably USA, 1
Germany, 1
Austria, 1
Italy, 1
Belgium, 1
} 

\ignore{
===== Human Participants Location =====
None, 22
USA, 6
unclear, 3
New Zealand, 3
Not possible to determine, 2
Canada, 1
Finland, 1

===== Human Participants Level =====
Not applicable, 22
Tertiary, 11
Not possible to determine, 2
Secondary, 1
Professional developers, 1
Uncontextualized, 1

===== Human Participants Count =====
Not applicable, 22
Unclear, 2
58, 2
32, 1
20 in person + 5 developers' public video streams, 1
71, 1
264, 1
963, 1
69, 1
24, 1
5, 1
9, 1
19, 1
>50000, 1
} 


\begin{table}
	\begin{center}
 	\caption{LLMs and languages featured in the reviewed literature. Some papers reported on more than one (or no specific) LLM or language, so the counts do not match the number of papers reviewed.}
	\label{tab:LLMs}
		\begin{tabular}{l|cp{1cm}l|c}
			\toprule
LLM         & Count & & Language    & Count\\\midrule
Codex       & 20    & & Python      & 37\\
Copilot     & 12    & & Java        & 6\\
GPT-3/3.5   & 28    & & C/C++       & 6\\
Other       & 7     & & Javascript  & 2\\
GPT-4       & 11    & & C\#         & 2\\\bottomrule
		\end{tabular}
	\end{center}
\end{table}


\ignore{
\begin{table}
	\begin{center}
 	\caption{\colorbox{red!60}{To be updated} LLMs and languages featured in the reviewed literature. Some papers reported on more than one (or no specific) LLM or language, so the counts do not match the number of papers reviewed.}
	\label{tab:LLMs}
		\begin{tabular}{l|cp{1cm}l|c}
			\toprule
			LLM & Count & & Language & Count\\\midrule
Codex & 13 & & Python & 21\\
Copilot & 10 & & Java & 4\\
GPT-3/3.5 & 8 & & C/C++ & 4\\
Other & 2 & & Javascript & 2\\
GPT-4 & 1 & & & \\\bottomrule
		\end{tabular}
	\end{center}
\end{table}
}

\ignore{
===== LLM Used =====
Codex, 13
Copilot, 10
GPT-3/3.5, 7
GPT-4, 1
Synfix, 1
AlphaCode, 1
ChatGPT, 1
N/A, 3

===== Programming Languages =====
Python, 21
Java, 4
C/C++, 4
Javascript, 2
Multiple (unspecified), 2
Rust, 1
Haskell, 1
Typescript, 1
HTML/CSS, 1
N/A, 7
} 

Despite the relative recency of this area of research, the papers we reviewed also used a wide range of LLMs, which is described in Table~\ref{tab:LLMs}. The rapid pace of the field is a potential threat, however, to the results being published. For example, the most commonly used LLM considering papers from the first snowballing phase only (i.e. up to May 2023) was Codex, which is now no longer available, and the most recent version of GPT (GPT-4) had only a single piece of research using it. Table~\ref{tab:LLMs} shows the results considering all the papers we analysed (i.e. up to August 2023), where the most commonly used LLM has become GPT-3/3.5, and the most recent version of GPT has 11 papers using it.

Table~\ref{tab:LLMs} also describes the languages being investigated. The majority of the research focuses on Python, with some work being done on Java and C. The table omits languages only explored by one paper in our set; most of these come from a single paper that investigates multiple languages. Python being the most popular language used is not too surprising, however, as popular LLMs such as Codex have been reported to be most proficient in Python~\cite{chen2021evaluating}.

\ignore{
===== LLM Used =====
Codex, 13
Copilot, 10
GPT-3/3.5, 7
GPT-4, 1
Synfix, 1
AlphaCode, 1
ChatGPT, 1
N/A, 3

===== Programming Languages =====
Python, 21
Java, 4
C/C++, 4
Javascript, 2
Multiple (unspecified), 2
Rust, 1
Haskell, 1
Typescript, 1
HTML/CSS, 1
N/A, 7
} 


\begin{table*}
	\begin{center}
  	\caption{Assessment of quality metrics adapted from \citet{hellas2018predicting}}
	\label{tab:qual-metrics}
		\begin{tabular}{l|l}
\toprule
\multirow{3}{*}{Is there a clearly defined research question/hypothesis?}   & Yes: 44\\
                                                                            & No: 18\\
                                                                            & Vague / Unclear: 9 \\\midrule
\multirow{3}{*}{Is the research process clearly described?} & Yes: 55\\
                                                            & No: 10\\
                                                            & Vague / Unclear: 6\\\midrule

\multirow{3}{*}{Are the results presented with sufficient detail?}  & Yes: 57\\
                                                                    & No: 6\\
                                                                    & Vague / Unclear: 8\\\midrule

\multirow{3}{*}{Are threats to validity / limitations addressed in an explicit (sub)section?} 
                                                            & Yes, in a separate (sub)section: 38\\
                                                            & Yes, but not in a separate (sub)section: 15\\
                                                            & No: 18\\

\bottomrule
        \end{tabular}
    \end{center}
\end{table*}

\ignore{

\begin{table*}
	\begin{center}
  	\caption{\colorbox{red!60}{To be updated} Assessment of quality metrics adapted from \citet{hellas2018predicting}}
	\label{tab:qual-metrics}
		\begin{tabular}{l|l}
\toprule
\multirow{3}{*}{Is there a clearly defined research question/hypothesis?} & Yes: 23\\
& No: 10\\
& Vague / Unclear: 4 \\\midrule
\multirow{3}{*}{Is the research process clearly described?} & Yes: 29\\
& No: 5\\
& Vague / Unclear: 3\\\midrule
\multirow{3}{*}{Are the results presented with sufficient detail?} & Yes: 29\\
& No: 3\\
& Vague / Unclear: 5\\\midrule
\multirow{3}{*}{Are threats to validity / limitations addressed in an explicit (sub)section?} & Yes, in a separate (sub)section: 18\\
& Yes, but not in a separate (sub)section: 7\\
& No: 12\\

\bottomrule
        \end{tabular}
    \end{center}
\end{table*}

===== Quality metrics =====
--> Research question
Yes, 23
No, 10
Vague / Unclear, 4
--> Description of process
Yes, 29
No, 5
Vague / Unclear, 3
--> Description of results
Yes, 29
Vague / Unclear, 5
No, 3
--> Explicit listing of threats
Yes, 18
No, 12
Vague / Unclear, 7
} 

Table~\ref{tab:qual-metrics} contains our evaluation of four quality metrics reported in \citet{hellas2018predicting}. They reported these metrics as part of a review of performance prediction research, so several of their questions are focused on work from that domain. For example, they ask, ``is the value being predicted clearly defined?'' We selected the most generally applicable questions, and we updated their question about threats to validity to focus specifically on whether they were discussed in an explicit subsection. Compared to their results, we find that the work in this area is reported more clearly in all four aspects measured. In particular, threats to validity are explicitly discussed in the majority, rather than minority, of cases, and slightly more of the work we examined present an explicit research question.


\subsection{Classification of literature}
\label{sec:litcategories}


The papers we reviewed broadly fall into five categories, with respect to the role that the LLM plays in the study: (i) assessing the performance, capabilities, and limitations of LLMs, (ii) using LLMs to generate teaching materials, (iii) using LLMs to analyse student work (e.g. identifying errors and repairing bugs), (iv) studying the interactions between programmers and LLMs, 
and (v) position papers and surveys/interviews. Category (i) is by far the largest group, indicating a strong desire to assess the current capabilities and limitations of LLMs in computing education contexts. We acknowledge that some papers would fit into more than one category; in these cases, we classified the paper into the most fitting category.

We now briefly summarize the main contributions of the papers included in our review, organised into these five categories.




\subsubsection{Assessing the performance, capabilities and limitations of LLMs (35): }
More than thirty papers looked into assessing the performance or capabilities of large language models. Most of these looked into the performance of LLMs in generating code, often for programming exercises~\cite{cipriano2023gpt,denny2023conversing,finnieansley2022robots,finnieansley2023my,li2022competition,padurean2023neural,piccolo2023many,poldrack2023aiassisted,puryear2022github,reeves2023evaluating,savelka2023can,singla2023evaluating,wermelinger2023using,austin2021program,brennan2022exploring,dakhel2023github}. 
Some looked into other types of exercises such as multiple-choice questions~\cite{savelka2023can,savelka2023large,savelka2023thrilled,wang2023exploring}, textbook questions~\cite{jalil2023chatgpt}, exam questions~\cite{dobslaw2023experiences}, computational thinking tasks~\cite{bellettini2023davinci}, and textual reports~\cite{brennan2022exploring}. In general, LLMs seem to perform at a level that is equivalent to or better than that of average students, at least for code generation tasks. For other tasks, such as answering MCQs~\cite{savelka2023large}, Parsons problems~\cite{reeves2023evaluating} and computational thinking tasks~\cite{bellettini2023davinci}, the performance of LLMs is currently not as great. When used to generate questions or hints to support students in solving problems, the research is inconclusive~\cite{al2023socratic,balse2023investigating,druga2023scratch,phung2023generative,widjojo2023addressing}, although LLMs are able to provide better explanations of code than students~\cite{leinonen2023comparing}, and are able to explain their answers to textbook questions in about half of the cases~\cite{jalil2023chatgpt}. A study found that code generated by ChatGPT had enough differences from student code that it could be detected very accurately (between 96-98\% accuracy)~\cite{idialuwhodunnit}. However, tools that assess whether a given text was generated by an LLM show a large number of false positives and should not be trusted blindly~\cite{orenstrakh2023detecting}. LLMs can also be prompted to act as a programming assistant, even ones originally trained for code generation~\cite{ross2023a}, but struggle with questions that require semantic understanding of code, such as predicting the output of a program~\cite{austin2021program} or analysis/reasoning about code~\cite{savelka2023large}.

\subsubsection{Position papers and surveys/interviews (17): }
Twelve papers were surveys or position papers, summarising or discussing research conducted by others~\cite{alves2023centaur,becker2023programming,brusilovsky2023the,denny2023computing,kendon2023ai,ma2023ai,philbin2023exploring,zan2023large,ernst2022ai-driven,gherciu2022net,raman2022programming,yan2023practical}. Some of the papers focused solely on education~\cite{becker2023programming,brusilovsky2023the,raman2022programming,yan2023practical}, while others included discussion on both the professional and the educational contexts~\cite{alves2023centaur,ernst2022ai-driven,gherciu2022net}.  The education-focused survey papers discuss both positive impacts---such as potentially increased productivity for both students and instructors~\cite{becker2023programming,alves2023centaur}---and negative impacts---such as over-reliance~\cite{becker2023programming,gherciu2022net}---of large language models. All papers suggest that LLMs will have substantial impact on computing education and programming more generally.

Five papers in the literature review used interviews to understand user perceptions and attitudes towards LLMs.  The authors of one paper interviewed professional software developers on how they use code generation tools~\cite{bull2023generative}. They found that the interviewees thought that generative AI has many use cases in software development. They note that while the tools do not require training to use, developers will need to understand the generated code for quality assurance, and to avoid over-reliance as the quality of code produced by these tools can vary. Three other papers report on interviews with students and instructors about their experiences with, and attitudes towards LLMs~\cite{lau2023from,rajabi2023exploring,zastudil2023generative}, finding that there is no consensus about the use of LLMs in higher education, its benefits or risks, but there is general awareness of the problem of academic integrity in the light of LLMs used by students. One study reports on the experiences of five students using generative AI for assignments~\cite{french2023creative}, highlighting both aspects of where it offered effective support, but also many limitations.

\subsubsection{Studying the interactions between programmers and LLMs (9): }
A total of nine papers looked into interactions between programmers and LLMs. Some focused on finding interaction patterns~\cite{prather2023its,vaithilingam2022expectation,barke2023grounded} while others focused more on how productivity is impacted by the use of models~\cite{kazemitabaar2023studying,nam2023in}, whether code produced when using AI code generators is less secure than when not~\cite{sandoval2023lost}, and how students use code explanations generated by LLMs~\cite{macneil2023experiences,pankiewicz2023large}. 

Based on the findings of this research, students engage in different interaction modes when using AI code generators. These include exploration~\cite{barke2023grounded}, acceleration~\cite{barke2023grounded}, shepherding~\cite{prather2023its}, and drifting~\cite{prather2023its}. In exploration, the programmer is unsure of what to do next, using the code generator for exploring potential approaches to tackle the problem. In acceleration, the programmer knows what they are doing and uses the LLM for producing the desired code faster. In shepherding, the programmer spends the majority of their time on guiding the LLM to produce the desired code. In drifting, the programmer drifts from one incorrect code suggestion to the next, indicating struggles in understanding the generated code.

Kazemitabaar et al. studied how novice programmer productivity and learning is affected by the use of AI code generators~\cite{kazemitabaar2023studying}. They found that students who used AI code generators performed significantly better (1.15x progress, 0.59x errors, 1.8x higher correctness, 0.57x time spent) without negative effects on learning. Sandoval et al. found that using code generator tools did not seem to introduce new security risks~\cite{sandoval2023lost}. MacNeil et al. found that students generally found code explanations generated by LLMs useful for learning, but the perceived utility of the explanations and students' engagement with them varied by explanation type~\cite{macneil2023experiences}.

One study looked at how students write prompts for LLMs~\cite{denny2023promptly}. To this effect, the problems had to be stated in a more visually oriented way to prevent them from copying the problem statement directly, but rather write prompts on their own. Most students found the prompt writing to be beneficial, while a few voiced concerns. Another study (which we classified in group (i)) used student written prompts in order to evaluate LLMs and found it to be an effective benchmark~\cite{babe2023studenteval}.

\subsubsection{Using LLMs to analyse student work (5): }
Five papers used LLMs to analyse student work, for example, by looking into using LLMs to fix bugs or errors in student work~\cite{leinonen2023using,phung2023generating,zhang2022repairing,ahmed2022synshine}. Two papers looked at repairing programming errors~\cite{ahmed2022synshine,zhang2022repairing}, one at enhancing programming error messages~\cite{leinonen2023using}, and one at providing feedback to students based on a buggy student program~\cite{phung2023generating}. The studies that examined the performance in bug repair both reported that their results surpassed previous state-of-the-art automated program repair results. Zhang et al. reported an overall repair rate of up to 96.5\% using Codex with few-shot examples and iterative prompting ~\cite{zhang2022repairing} and Ahmed et al. achieved a repair rate of 89.4\%~\cite{ahmed2022synshine}. Leinonen et al. found that Codex could enhance programming error messages -- which are notoriously hard for students to understand -- about 54\% of the time on average, noting that this performance is not good enough for using the model directly with students~\cite{leinonen2023using}. Phung et al. propose a method where instructors could balance the `coverage' of feedback, i.e. whether a student receives feedback at all, and the `precision' of feedback, i.e. whether the feedback is of good quality. They found that in the best case, their proposed method can achieve a precision of 72.4\% with a coverage of 64.2\% for one of the datasets they used and a precision of 76\% and a coverage of 31.2\% for the other dataset included in the study.

One paper presents a tool to help out students with issues without revealing the full solution~\cite{liffiton2023codehelp} and reports positive feedback from both students and instructors. Other studies have also looked into mitigating risks of LLMs such as wrong answers~\cite{jaipersaud2023decomposed} or guiding the students with hints rather than full solutions~\cite{al2023socratic} (we categorised these papers as assessing LLMs).

\subsubsection{Using LLMs to generate teaching materials (5): }
Five papers looked into using LLMs to generate teaching materials~\cite{denny2023can,denny2022robosourcing,sarsa2022automatic,sridhar2023harnessing,matelsky2023large}. In two cases, the teaching materials being generated were programming exercises~\cite{denny2022robosourcing,sarsa2022automatic}. One of the main findings in both papers was that LLMs can be coaxed into generating exercises with prescribed themes (such as basketball or cooking) and programming concepts (such as loops or conditionals). In addition, the exercises generated by LLMs were novel and sensible, although the authors cautioned that the quality might not be good enough to provide the LLM-generated exercises directly to students. Another study that compared LLM-generated content with student-generated content concluded that the quality is comparable, but still recommends further research~\cite{denny2023can}.
Similarly, LLMs were found to be able to generate reasonable learning objectives~\cite{sridhar2023harnessing}. One paper presents a new tool, but does not offer an actual evaluation of it~\cite{matelsky2023large}. All papers suggest that using LLMs could help instructors save time in generating teaching materials.


\subsection{Educational opportunities and risks}

The broader literature on LLMs and their potential effects is often organized around the dichotomy of opportunities and risks \cite{kasneci2023chatgpt, bommasani2022opportunities, sok2023chatgpt, qadir2023engineering}.  For example, Bommasani et al. produced an extensive report documenting the opportunities and risks of foundation models across a broad variety of domains, including education \cite{bommasani2022opportunities}.  
Kasneci et al. documented similar opportunities, risks and mitigation strategies, specifically focusing on the use of ChatGPT in education \cite{kasneci2023chatgpt}.

Among the papers in our dataset, there was broad agreement that LLMs would have a major impact on teaching and learning in computing courses.
Authors identified various opportunities and risks for both students and teachers and we present these in the following sections.

\subsection{Opportunities}

The papers we reviewed identified a number of potential opportunities that could positively impact computing education.  One of the prominent opportunities that emerged was related to reducing instructor workload, for example by generating large banks of diverse learning resources and support materials \cite{sarsa2022automatic, macneil2023experiences,denny2023can}, automating various aspects of the grading process \cite{yan2023practical}, and providing personalised help to students who are struggling and who would otherwise consume considerable instructor effort \cite{brusilovsky2023future,al2023socratic,pankiewicz2023large}.  Related to this theme, Bull and Kharrufa argue that the type of scaffolding that AI tools can provide are able to ``support the student in their learning and ... offload some of that ... burden from the educator'' \cite{bull2023generative}.

Improving the learning experience for students was another common opportunity that emerged.  Several papers described the idea of using an LLM as an assistant or pair programmer, which represents a significant change from current pedagogical practice \cite{ross2023a, piccolo2023many, leinonen2023comparing}.  As a concrete example of the kind of assistance that could be provided while students are programming, Leinonen et al. suggest that LLMs could help students understand terse error messages \cite{leinonen2023using} which have traditionally been a source of difficulty for novice learners \cite{Becker2019wgpaper}.
Several groups also identified opportunities for creating new tools around LLMs, e.g., to support repair of syntactically incorrect code~\cite{ahmed2022synshine}, to help answer questions~\cite{kazemitabaar2023studying,hellas2023exploring} or provide hints~\cite{pankiewicz2023large}, or even to support the crowdsourcing of new questions~\cite{denny2022robosourcing}. Indeed, some recent papers found in the second phase snowballing focused on introducing tools around LLMs, such as CodeHelp~\cite{liffiton2023codehelp} and Promptly~\cite{denny2023promptly}. 
Despite the promise of using AI tools for learning support, Dakhel et al. and Prather et al. caution that although they can be a great asset for professional developers, they may be less helpful for novices if the tools generate non-optimal or erroneous outputs which could cause confusion \cite{dakhel2023github, prather2023its}.  As the quality and performance of the models improve, this may be less of an issue as time goes by. 

Another recurring opportunity mentioned in the papers we reviewed was the potential for a renewed focus on problem solving.  For example, Vaithilingam et al. explored the usability of code generation tools and suggest they can be used to rapidly provide a good starting point for a solution, thus allowing programmers to focus on the problem solving process and reducing the need for a focus on lower-level details \cite{vaithilingam2022expectation}.  In a similar vein, Denny et al. \cite{denny2023conversing} and Prather et al. \cite{prather2023its} explore the use of Copilot in two different contexts, suggesting it can be used to teach students how to express problem solutions in natural language, and to focus on guiding students through problem-solving strategies, respectively.

Finally, LLMs present clear opportunities for instructors to rethink assessment practices and reconsider what assessment means in computing courses~\cite{denny2023conversing,bellettini2023davinci,poldrack2023aiassisted,savelka2023large}.  Raman et al. suggest assessments could focus more on code understanding, such as tracing and verification, and less on syntax and code writing \cite{raman2022programming}.  LLMs can also be used to generate a variety of flawed solutions, providing plentiful opportunities for incorporating code review tasks~\cite{finnieansley2022the}.  The systematic literature review of Yan et al. explored practical and ethical challenges of LLMs in education, categorising some of this prior work around the `Assessment and grading' category and argue that grading student assessments is a promising application of LLMs \cite{yan2023practical}. The impressive documented performance of code generation models like Codex on typical CS1 and CS2 problems suggests that some rethinking of assessments is essential \cite{finnieansley2023my}. 

In summary, the papers we reviewed suggest that LLMs present a wide array of opportunities for computing education, improving instructor productivity by reducing workload, enhancing student learning experiences, enabling a greater emphasis on problem-solving, and suggesting new assessment practices.

\subsection{Risks}
Several risks were identified by the papers we reviewed.  Authors were concerned that generative AI could be used in ways that limit student learning or make the work of educators more difficult~\cite{gherciu2022net}.

\subsubsection{Risks for students}
The learning resources produced by generative AI pose significant risks to student success. 
\citet{wermelinger2023using} and \citet{sarsa2022automatic} observe that explanations of code can be a useful learning resource, but if the explanations contain mistakes then learning could be negatively impacted.  Since AI generated content is presented authoritatively (and is frequently correct), students are unlikely to question the content and may learn incorrect information~\cite{bull2023generative,jalil2023chatgpt}. Automatically generated tests may be partially complete, leading students to inadequately test their code~\cite{wermelinger2023using}. The resources created by LLMs might also have less variation than those created by humans~\cite{denny2023can} and thus limit the variety of examples to which students are exposed. 
AI generated content is not curated for specific courses, so learning material generated could potentially include syntax, programming constructs, or other content that is inappropriate for students in a given course~\cite{becker2023programming,hellas2023exploring}.  Example programs used by students for learning could have poor implementation or poor style, which may result in students acquiring undesirable programming habits~\cite{finnieansley2022robots}.

The use of generative AI may also result in students spending time in unproductive ways.  \citet{wermelinger2023using} speculated that students may spend excessive time on prompt engineering in the hope of hitting on a successful solution rather than making process towards the solution, which was indeed observed in the study by \citet{prather2023its} in the `shepherding' interaction pattern. \citet{bull2023generative} suggested that it may take longer to figure out an effective prompt than to write a solution. \citet{hellas2023exploring} found that LLMs tended to hallucinate issues in student code which could cause students to focus on these non-existent issues instead of the actual issues in their code.

\citet{wermelinger2023using} observed that explanations generated by Copilot focused on a line-by-line description of \textit{what} the code did rather than how it achieved the goal desired. This is supported by the findings of \citet{sarsa2022automatic} who noted that Codex seems most proficient at crafting line-by-line code explanations, as opposed to e.g. higher level summaries of the code. 
This is akin to a multi-structural explanation~\cite{sheard2008going}, which may focus student attention at lower levels of the SOLO taxonomy, rather than thinking about the overall purpose of code. However, non-code models such as GPT-3 have been found to be more apt at creating higher level code explanations~\cite{macneil2023experiences}.

The most common concern expressed by authors about student learning was the potential for students to become over-reliant on generative AI tools to solve problems~\cite{becker2023programming,finnieansley2022the,ross2023a,leinonen2023comparing,li2022competition} and assist in debugging code~\cite{leinonen2023using,phung2023generating,zhang2022repairing}.  Students who rely on generative AI may be misled into believing they are making progress, and this illusion of capability may reduce their self-understanding about their level of mastery of the subject matter~\cite{prather2023its, prather2018metacognitive}.

As introductory students realise that generative AI can outperform them on most tasks, they may lose motivation to learn the material, and become demoralised about the future of computing~\cite{leinonen2023comparing}. Further, 
novice learners of programming may become overwhelmed and confused by generated code, which could add to the high levels of frustration that are common in introductory programming courses~\cite{vaithilingam2022expectation,denny2023promptly}.

\subsubsection{Risks for teachers}
Several authors raise concerns about the impact of generative AI on teachers and teaching practice. Unsurprisingly, issues of academic integrity were the most common concern raised.

Generative AI is reported to perform very well in assessments that are commonly used in introductory courses, raising concerns that students will submit solutions that they have not created themselves~\cite{ross2023a,finnieansley2022the,savelka2023can,finnieansley2023my,becker2023programming,denny2023conversing,piccolo2023many,savelka2023large,lau2023from}. The solutions generated by AI cannot be easily identified as plagiarism~\cite{puryear2022github}, which requires teachers to adapt and develop new teaching strategies to ensure academic integrity is maintained~\cite{wermelinger2023using}.

\citet{wermelinger2023using} recommended that educators stop ‘re-dressing’ toy problems because generative AI will provide good solutions which will restrict the learning opportunities for coding, debugging and algorithmic thinking, compared to problems with interesting ‘wrinkles’.  Teachers who shift away from using many small problems to create larger and more authentic problems in an attempt to reduce reliance of generative AI will lose access to the quick and easy assessment methods such as automated grading associated with many introductory programming courses~\cite{finnieansley2023my}.  Educators will need to develop new resources to explicitly address LLMs and guide students instead of leaving them alone with the tool~\cite{vaithilingam2022expectation}. 

Teachers who use generative AI to assist in the creation of learning resources may unintentionally produce exercises that are under-specified or that contain incorrect reference solutions or inadequate/incorrect test cases~\cite{sarsa2022automatic}.
Teachers who are concerned about the impact of generative AI may intentionally modify course delivery in ways that reduce the effectiveness of their teaching practice (e.g., by increasing academic integrity at the cost of scaffolded programming exercises), or adjust the curriculum to shift focus away from code writing, leaving students poorly prepared for subsequent programming courses~\cite{raman2022programming}.

Although there is a growing need to teach students how to use generative AI appropriately, it is unclear how we should do so.  \citet{barke2023grounded} discuss the need to balance the introduction of generative AI too early in the curriculum where over-reliance is a possibility, against introducing generative AI too late and fail to provide an authentic experience relevant to industry practice. \citet{bull2023generative} note that it is challenging for novices to understand generative AI capability and use prompts effectively, suggesting a need to formally teach students to effectively use the tools they have at their disposal.

\subsubsection{Risks for the community}
The rising use of generative AI raises concerns for the broader community.  As more programming code is likely to be generated automatically, there is potential for biases to be unintentionally introduced due to the algorithm used to generate content or due to the source material used in training~\cite{becker2023programming}, More concerningly, generated code may contain security vulnerabilities and bugs~\cite{becker2023programming,bull2023generative,sandoval2023lost}.

Finally,~\citet{kazemitabaar2023studying} found that students with more knowledge benefited more from code generation than students with less knowledge. Similarly,~\citet{nam2023in} found that professionals benefited more from AI code generation than students.  These findings suggest that AI code generators may widen the gap between over- and under-achievers, exacerbating teaching challenges arising from classes with heterogeneous ability levels. In response to this, Prather et al. discuss design considerations for generative AI tools that could eschew these risks and lead to more direct benefit for novice programmers \cite{prather2023its}.


\subsection{Limitations and threats to validity}

Our literature review was conducted from April to August 2023, so work published after this point or with low visibility will have been missed. Due to the relatively recent emergence of powerful large language models, especially their use in the field of computing education, and fast pace of the field, only literature published in less than a three year period (from 2021 to 2023) was considered. Due to this, we also conducted only a single step of snowballing (i.e., we did not do further snowballing on the papers found in the snowballing). This may have omitted some work that failed to reference the most visible early work (our reference papers), but we do not believe that this will include significant numbers of papers or change the general trends identified in our analysis.

The inclusion of results from arXiv and grey literature sources is driven by pragmatism. The machine learning community makes wide use of arXiv due to the fast-paced nature of the field, and if we omitted it, we would miss the most recent results (up to a year of papers) in an already narrow window of time. However, the inclusion of these sources admits work that has not yet undergone peer-review. During our deep review of the included papers, several reviewers raised concerns about the quality of a paper they were reading. We retained these papers as they met the inclusion criteria, and note that, on the whole, the papers appear to be ready for review and demonstrate many of the criteria for quality proposed by~\citet{hellas2018predicting}.
\section{Survey of Student and Instructor Perceptions about GenAI}
\label{sec:survey}

Students and instructors may have quite different views of the use of generative AI tools in computing classrooms.  For example, one of the well-documented concerns regarding generative AI in educational contexts is that students may become over-reliant on them for the generation of answers~\cite{denny2023computing, zastudil2023generative}.  In this case, students who rely on the tools may initially take a more positive view of them when compared with instructors, however, their views may change over time.  Given the speed with which generative AI tools are being developed and adopted, documenting student and instructor perceptions at the current time provides a useful snapshot of current practice and facilitates future explorations of how views may change as these tools become more embedded in the educational sequence.  In this section, we report the findings from two surveys, one with computing instructors and the other with students, that we conducted from July to August 2023 with responses spanning 20 countries.  We first review similar explorations in computing and other disciplines, and then after describing our methods we organize our findings around insights derived from analysis of both quantitative and qualitative data. 

\subsection{Prior explorations of student and instructor perceptions}
\label{sec:survey-prior}

Several recent studies have explored the perceptions of students and teachers toward the potential impact of generative AI in broad educational settings.   Chan and Lee acknowledge a generation gap in how generative AI is perceived \cite{chan2023ai}.   Using an online survey involving 399 students and 184 teachers, predominantly from Hong Kong although across a diverse range of academic disciplines, they examine distinctions in perceptions, experience, and knowledge of generative AI between educators and students across different generations, classified as Gen Z (students) or Gen X and Y (teachers).  They observe that while students are generally optimistic about the use of these new technologies, teachers exhibit more concerns regarding over-dependence and ethical issues, and were also more sceptical about the abilities of generative AI tools.  They emphasise the urgent need for clear policies and guidelines to ensure that academic integrity is maintained and to promote equitable learning experiences.  In follow-up work, Chan addresses this need by proposing an AI policy framework specifically for higher education \cite{chan2023comprehensive}. This encompasses three dimensions: pedagogical, which uses AI to enhance teaching and learning outcomes; governance, which addresses privacy, security and accountability issues; and operational, which pertains to infrastructure and training.  To inform the policy, they conduct an online survey of 457 students and 180 teachers and academic staff from Hong Kong universities.  They argue that the student voice plays an essential role in the drafting and implementation of policy.  In general, both students and teachers reported limited experience with AI tools, suggesting potential for growth in adoption and the need for training on the effective use of AI technologies.

In a related strand of work, Chan and Tsi focused specifically on the capacity of generative AI for replacing human teachers \cite{chan2023airevolution}.  Their rationale for this direct question was to assist educators in preparing for the inevitable integration of AI into educational settings.  The authors review existing literature on the role of AI in the classroom and present a synthesis of its limitations, classifying these into eight categories covering 26 aspects.  For example, the category `Emotional and Interpersonal Skills' highlights the social-emotional competencies of human teachers and covers aspects such as human connection, cultural sensitivity and building trust and rapport.  An online survey consisting of 11 closed items and several open-response questions was distributed to universities in Hong Kong and received responses from 144 teachers and 384 students.  Students generally indicated an appreciation for the unique emotional qualities of human teachers, whereas they expressed concern about student misuse.  Despite some variation in responses, both students and teachers generally agreed that AI is not likely to entirely replace human teachers and in particular the social-emotional competencies.

A recent study by Amani used a survey to measure student and instructor perceptions of generative AI in academia with the goal of capturing perceptions, misconceptions, concerns, and current usage trends \cite{amani2023generative}.  The authors argue that it is essential to report instructor and student perceptions now given the rapid changes and improvements in the tools that are underway.  Two online surveys were created, with the student-oriented survey focusing on current usage and perceptions and the instructor-oriented survey focusing on how it is affecting their current courses and how they think students should use it.  Data was collected from 243 staff and 813 students at Texas A\&M university, revealing a clear perception that resisting these new technologies is not feasible, and that teaching practices must adapt in response.  Students value the high availability of the tools, but recognise the potential for their misuse.  Forman conducted a similar online survey exploring student perceptions of ChatGPT \cite{forman2023chatgpt}.  Analysis of 71 responses to the 7-question survey revealed that students generally had a positive long-term view of the role that such technologies would play in their lives, and that they currently relied on ChatGPT to save time when working on assignments and projects. 

Raman et al. investigate the factors that influence the adoption by university students of ChatGPT \cite{raman2023university}.  Their work, which utilises Rogers' Diffusion of Innovation theory as a conceptual framework, proposes that five attributes of the technology influence its adoption, namely relative advantage, compatibility, ease of use, observability, and trialability.  Their empirical analysis, which is based on a survey of 288 students delivered via Google Forms, supports their hypotheses and indicates gender-based differences in how students prioritise the attributes. 

Although online surveys have been a popular instrument in work exploring student and instructor perspectives, a few recent interview studies~\cite{lau2023from, zastudil2023generative, wang2023towards, rajabi2023exploring} have investigated the impacts of generative AI on computing education research and practice. Notably, Lau and Guo recently conducted in-depth interviews with instructors to understand how they planned to adapt to the emergence of tools like ChatGPT and Copilot \cite{lau2023from}. They conducted Zoom interviews with 20 instructors from nine countries. The interviews were framed around a hypothetical question, where participants were asked to imagine a future where students had access to an AI tool that could both write perfect code for any programming problem and that was undetectable to plagiarism detection methods. Instructors were asked to describe how they would adapt their pedagogical approaches over the short-term and long-term.  In the short-term, the primary concerns centred around cheating and plagiarism, to which instructors have responded by relying more heavily on invigilated exams and educating students about current model limitations.  Longer term perspectives varied, with one school of thought aiming to resist AI tools and teaching in conventional ways, and the other aiming to integrate AI tools into the curriculum to better prepare students for the changing requirements of industry.  Specific examples from this latter category included using AI to provide more personalised help to students, using assignments that focus more on code reading and critique and more open-ended design, and using AI to evaluate new kinds of assessment tasks.  These instructors also viewed AI as being potentially useful for broadening participation and accessibility in computing due to their capacity for providing personalised assistance.  Significantly, whether they tended towards resisting or embracing AI tools, instructors generally agreed that the objectives of computing education will likely need to change to adapt to the growing influence of AI. 

Zastudil et al.~\cite{zastudil2023generative} conducted Zoom interviews with six CS instructors and 12 CS students. The analysis compared and contrasted their experiences, hopes, and concerns about the emergence of generative AI on computing education. Students and instructors aligned on key concerns such as over-reliance, model trustworthiness, and plagiarism; however, they diverged regarding how each group preferred those issues to be addressed. Students stressed the importance of crafting engaging and culturally relevant assignments as well as reducing busy work to address plagiarism, whereas instructors proposed increasing the weight of proctored exams. Students were concerned about the quality of the model's responses and instructors were concerned that students would be unable to identify wrong or misleading responses. Instructors and students were both excited about the potential for GenAI tools to shift course topics toward higher-levels of abstraction, such as design patterns. 

Wang et al.~\cite{wang2023towards} conducted a three-part study which culminated in Zoom interviews with 11 instructors. The authors found that instructors are concerned that students will misuse or over-rely on GenAI tools, but instructors did not have plans to adapt their courses due to a current lack of effective strategies. Instructors believed these problems would be harder to address in the introductory courses. Similar to the findings of Zastudil et al.~\cite{zastudil2023generative}, instructors were concerned about how incorrect model responses might lead students to develop faulty mental models. 

Rajabi et al.~\cite{rajabi2023exploring} interviewed 36 instructors in-person and 4 instructors virtually. Their interviews uncovered four primary themes that related to adapting pedagogy, plagiarism, assessment, and job preparedness. Instructors raised concerns about the trustworthiness of GenAI tools and their capacity to mislead students. However, instructors also argued that GenAI tools should not be banned because students will continue to find ways to use them. Instructors advocated for doing in-class assignments to avoid plagiarism concerns, but acknowledged that this could increase students anxiety about exams and grade weight---a concern that has been previously raised in computing education~\cite{latulipe2015structuring, macneil2016exploring, khan2018active}.   

Based on these prior interview studies, the goal of our survey was to provide a large-scale, systematic, and international overview of the experiences students and instructors have had with generative AI in computing education contexts and to uncover their preferences for how these models should be used in computing classrooms.

\subsection{Methods for data collection and analysis}\label{sec:methods_survey}

To better understand the perceptions and experiences of students and instructors in computing courses as they relate to Generative AI tools, we developed two surveys---one for students and a second for instructors. We designed the survey to have questions that were asked to both groups to facilitate comparisons between these two crucial stakeholders. This method also draws inspiration from previous studies that directly compare the responses from students and instructors to the same questions~\cite{chan2023comprehensive, chan2023ai, chan2023airevolution}. 
We also draw inspiration from previous large-scale surveys in computing education research.  The use of online surveys and recruitment of participants via bulk email such as the SIGCSE mailing list is a common method, and has been used in work by Denny et al. ~\cite{denny2019research}, Schulte and Bennedsen \cite{schulte2006what}, Elarde and Fatt-Fei \cite{elarde2011introductory}, and Dale \cite{dale2006most}.
The following sections describe how participants were recruited and how the survey was constructed. 

\subsubsection{Recruitment and participants}

We recruited 57 instructors and 171 students to complete the corresponding online surveys. 

To recruit instructors ($n=57$), we sent emails to the mailing lists of computing education professional groups including \textit{sigcse-members, sigcse-australasia-members and uki-sigcse-members}. The goal for targeting these mailing lists was to draw a broad sample of computing education practitioners and researchers. However, we recognise that the resulting sample likely results in a selection bias of instructors who are particularly invested in computing education compared with their peers.  This is a well-known challenge, as noted by Schulte and Bennedsen \cite{schulte2006what}. In an attempt to address this, we included a request for them to share the recruitment materials with colleagues in their department. This snowball sampling technique was also used by ITiCSE working group members to share the recruitment materials in their personal networks. 

To recruit students ($n=171$), we also used a snowball sampling method where instructors were requested to share a recruitment announcement with students through their courses and department mailing lists. In this case, it is possible that we may experience a response bias with high-achieving students being more likely to respond to the survey. To address this potential bias we included the phrase ``if you have struggled with your computing courses, your voice is especially appreciated to ensure better experiences for students like you in the future''. 

\subsubsection{Questionnaire design}

We developed the questionnaire to focus on critical topics that have recently emerged across birds-of-a-feather discussions~\cite{macneil2023the}, workshops~\cite{macneil2022automatically}, and position papers~\cite{denny2023computing, becker2023programming}. These topics include calls for curricular and pedagogical changes, consideration of ethics, and a need for replications and benchmarking. We also included questions inspired from work on plagiarism in programming assessments \cite{albluwi2019plagiarism}, student help-seeking behaviour \cite{doebling2021patterins}, and from the related work discussed in Section \ref{sec:survey-prior}.

This resulted in thirty-five survey questions for the student survey (counting all open- and closed-response questions) and forty-two questions for the instructor survey.  The overlap between the student and instructor surveys included twenty-seven questions that were either identical or minor rewordings to improve the readability between groups (e.g. ``... using GenAI tools in ways that your instructors would not approve of?'' on the student survey was reworded as ``... using GenAI tools in ways that you would not approve of?'' for the instructors). 

The questions used in the student survey are listed in Appendix \ref{app:student-survey}, and the instructor survey questions appear in Appendix \ref{app:instructor-survey}.

\subsubsection{Thematic analysis}
To analyse responses from students and instructors for the open-response questions we followed an approach for thematic analysis similar to the reflexive process described by Braun and Clarke \cite{clarke2021thematic}.  The reflexive thematic analysis process is not prescriptive, but provides guidance for the phases needed to robustly explore, interpret and report patterns in qualitative data.  Given the data set size was relatively small (i.e. there were a total of nine open-response questions in common on the instructor and student surveys, and a total of 228 responses across both groups) the questions were divided amongst two researchers who analysed all responses to the questions they were assigned.


Each researcher began by reading the responses to familiarise themselves with the data, and then defining succinct labels that were assigned to each response that captured important features of the data.  Practically, this process used a spreadsheet in which responses were listed on the rows and the labels that were defined for coding the data appeared on the columns.  The final steps of the analysis involved grouping the labels into broader themes suitable for reporting.

\begin{figure*}[ht!]
\centering
\includegraphics[width=1\textwidth]{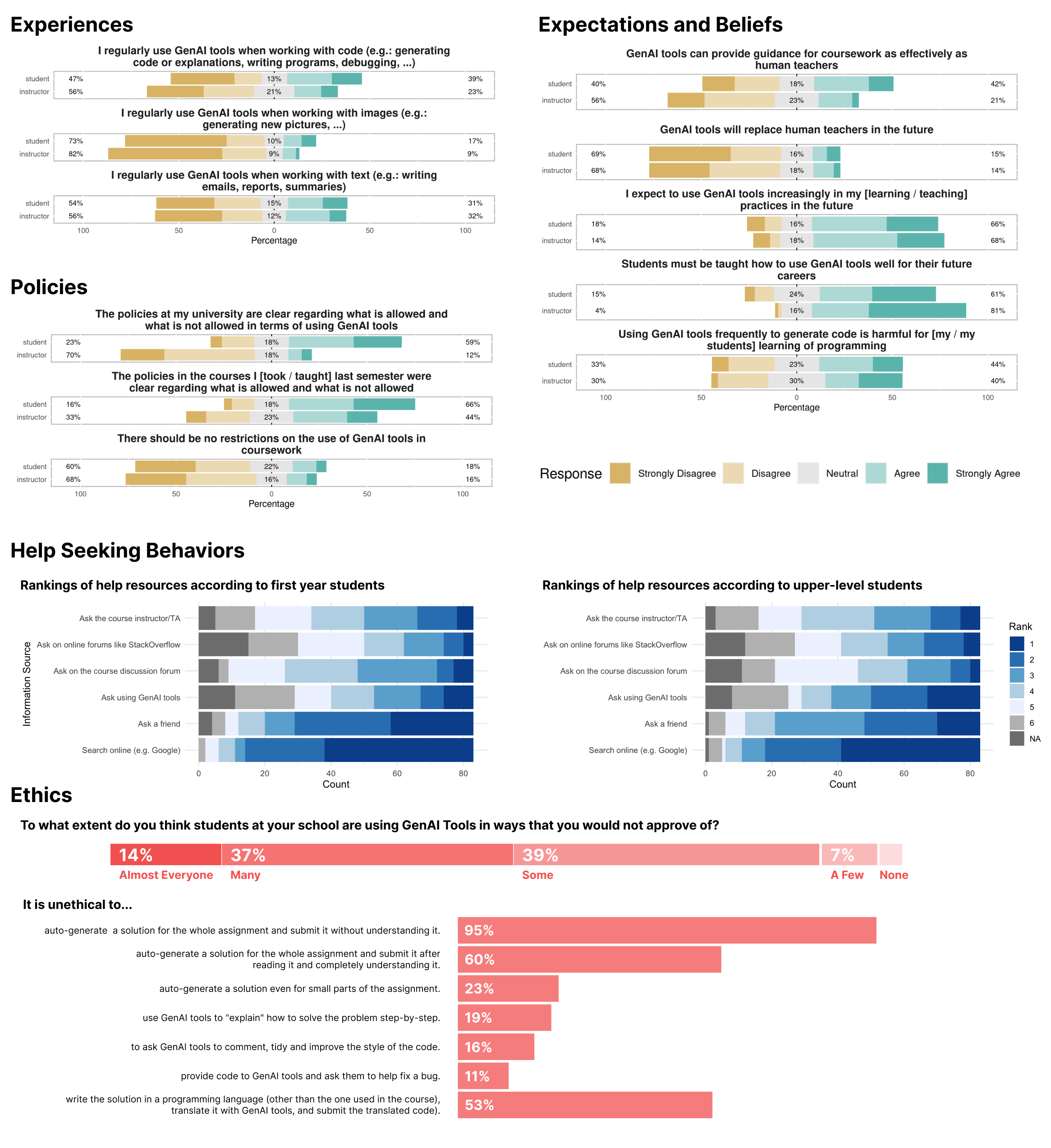}
\caption{Summaries of the survey responses from 171 students and 57 instructors: 1) Students' and instructors' perspectives were compared along likert scale responses, 2) students ranked their help seeking preferences from 1 to 6, and 3) instructors shared their beliefs about the ethical use of Generative AI Tools.
\label{fig:student_responses}}
\end{figure*}

\subsection{Quantitative insights}

\subsubsection{Demographics}

We recruited 57 instructors from 12 countries with an average of 18.2 years of teaching experience. Participants lived primarily in the USA (45.6\%) with others coming from the United Kingdom (17.5\%), Canada (8.8\%), Jordan (5.3\%), and Pakistan (5.3\%). The most common class sizes that instructors reported teaching were in the 11-30 (36.8\%), 31-50 (26.3\%), or 100-250 (24.6\%) ranges. The majority of instructors self-identified as men (77.2\%) with far fewer instructors identifying as women (19.3\%) or non-binary (3.5\%).

Through our snowball sampling method, we additionally recruited 171 student participants across 17 countries. The top five countries included New Zealand (35.7\%), Jordan (17.5\%), USA (14.0\%), Indonesia (8.8\%), and Australia (7\%). About half of the students self-reported being in their first year (48.5\%). Students in their second, third, and fourth years accounted for 21.6\%, 19.9\%, and 7\% of the respondents, respectively. The average number of courses taken was 4.6 with 38\% of students only having taken one course. 90\% of students had taken 10 or fewer courses. Most participants selected computer science as their major (42.7\%). Additional majors included undeclared engineering (15.8\%), software engineering (12.3\%), computer engineering (7.6\%), data science (5.3\%), and information technology (4.1\%). There were five participants who majored in either chemistry, supply chain, math, economics, or psychology; and there were two students who majored in physics.

\subsubsection{Comparison of student and instructor perceptions}

Figure~\ref{fig:student_responses} summarises students' and instructors' responses to the Likert scale questions on the survey. Responses from students and instructors to questions related to experiences and expectations were largely aligned. However, some important differences emerged for questions focusing on course policies. In this subsection, we review the results and briefly discuss the implications.


\textit{Experience and usage.} Students and instructors shared similar experiences with GenAI tools, using them primarily for writing code and working with text. However, fewer individuals in both groups used GenAI tools for tasks involving images. Students had slightly more experience using GenAI for writing code than instructors. While the difference was currently minor, instructors should keep in mind that students may rapidly become more expert at using GenAI tools. In light of this possibility, instructors should proactively stay informed about these tools' capabilities, even if they do not intend to incorporate them into their courses. This proactive approach is crucial to ensure instructors can continue providing meaningful educational experiences to students and remain well-informed about evolving technological advancements.

\textit{Course and institutional policies.} Students and instructors were aligned in their belief that some restrictions should be placed on the use of GenAI tools in their coursework. However, there may be some misalignment around what those restrictions should be. 
While students had mixed opinions about whether the university policies were clear, instructors largely disagreed with the statement that the policies were clear. This misalignment could lead to challenges where students are following implied policies rather than explicit policy guidelines. Given the shared belief that use should be limited, it is important that students, instructors, and institutions be aligned on course and institutional policies. However, it should be noted that students and instructors agreed slightly more strongly that course policies were clearly defined.

\textit{Expectations and beliefs.} Based on the responses from students and instructors, there was a close alignment in expectations and beliefs regarding GenAI tools. Both groups strongly agreed that GenAI tools cannot replace human instructors and that human teachers provide more effective guidance than GenAI tools. However, both students and instructors also expected GenAI tools to play an increasing role in the future of their teaching and learning, as well as in students' future careers. This suggests that while GenAI tools do not currently replace the value provided by instructors, it is important for instructors to reflect on and clearly define their value to students. It may be that students rely less on instructors for explanations and help, but rely on them more for curating the learning environment and ensuring that learning objectives are being achieved. 


\subsubsection{Help-seeking behaviors}

We surveyed students about their help-seeking behaviours to understand how prominently GenAI tools are being used by students when they require assistance. The results indicate that students predominantly continue to favour web searches as their primary resource for help. Nevertheless, GenAI tools are progressively establishing themselves as a dependable resource, surpassing online discussion forums as a preferred source of help. Interestingly, the extent of students' reliance on GenAI tools appears to be influenced by their academic program stage. Upper-level students exhibit a greater tendency to use generative AI tools over other resources, including peers, instructors, and teaching assistants. In contrast, first-year students still exhibit a preference for seeking assistance from their peers when facing challenges. This may reflect differences in the kinds of help students are seeking at different stages in their academic program or differences in their willingness to adopt new technologies, such as GenAI tools. 

\subsubsection{Ethical use of GenAI tools}

To better understand the ethical uses of GenAI tools, we surveyed instructors about the scenarios of use that they considered unethical. The findings reveal a consensus among instructors that auto-generating an entire assignment solution is considered unethical when students lack comprehension of the generated code. However, instructors held differing opinions on the ethics of generating solutions for an entire assignment when students possess a full understanding of the generated code or when students write code in a different programming language and then translate it into the language used in the course. In these cases, approximately half of the instructors deemed such practices ethical, while the other half considered them unethical. This suggests that instructors may be supportive of students using these tools as long as students demonstrate a clear understanding of the task and achieve the intended learning outcomes of the course.  Along this line of reasoning, instructors generally concurred that it is acceptable for students to employ these tools to generate solutions for specific portions of assignments, facilitate code debugging, elucidate concepts, or enhance the style and readability of their code. These are situations where the tool could save time 
without negatively affecting learning outcomes. Finally, when asked about the extent to which instructors believed that their students were using GenAI tools unethically, 
around half (50.8\%) believed that many or almost all of their students were using the tools unethically.  

\subsection{Qualitative insights}








\subsubsection{Instructor use of GenAI}

We asked instructors to describe the ways that they currently make use of GenAI tools, seeking separate responses for text generation and code generation (Questions 16 and 17 in Appendix \ref{app:instructor-survey}).  Overwhelmingly, for both types of content, the most common response from instructors was that they were not currently using GenAI tools.  Half of instructors reported they had not used GenAI tools for text generation and 40\% said they had not used such tools for code generation.  A small number of these instructors (two and four for text and code generation, respectively) indicated they planned to use GenAI tools in the near future.  For example, one instructor planned to use text generation tools to aid in preparing drafts for problems (i.e. \emph{``None currently, but plan to use in the near future for generating ungraded practice problems or first drafts of graded problems''}), and another instructor planned to start using code generating tools in the upcoming semester (i.e. \emph{``So far I have not, but I will next spring to help write code.''}).  This points to an emergent interest in GenAI from instructors in our survey and recognition of the use of GenAI for teaching.

The primary theme to emerge from responses around the current usage of text generation tools was for the creation of a wide variety of learning resources.  Of these resources, equally popular was the production of assessment questions (i.e. \emph{``Occasionally will work with ChatGPT to ideate exam questions''}) and for aiding in various kinds of writing tasks such as report writing and turning brief notes into longer form prose (i.e. \emph{``Generate readable sentences of my brief notes''}).  Other types of text-based artefacts that instructors reported creating were course materials, examples for students, explanations of complex algorithms, and scenarios for highlighting ethical issues in software engineering.

Several other interesting uses of text generation tools were mentioned.  Several instructors described using such tools to support other tasks, such as for performing background research, overcoming writer's block, and for paraphrasing papers when constructing references (i.e. \emph{``creating a reference for paper and paraphrase''}).  Several instructors highlighted the summarisation capacity of GenAI tools since they can effectively condense long-form text content.  One instructor used this feature to extract insights from written student feedback (i.e. \emph{``paste in student feedback about course and ask GenAI to summarise for me''}).  Finally, one instructor reported integrating text generation capabilities into other tools designed to support students (i.e. \emph{``We are actively building a tool to help respond to common questions for students in forums''}).

\vspace*{2mm}
\begin{custombox}{Low uptake of GenAI tools}
The survey revealed that most instructors are not currently using GenAI tools for text or code generation, but some have expressed plans to integrate them in the future. The tools, when utilised, are primarily used for creating diverse educational materials, however, satisfaction regarding the quality of outputs vary.
\end{custombox}

When reporting their use of code generating tools, instructors describe a variety of tasks that involve creating code in varying levels of detail.  Many responses to this question described fairly generic use of such tools (e.g. \emph{``Code writing''}, \emph{``generating part or some of function''}), whereas some were much more specific.  For example, several instructors described generating code examples that they would then give to their students to modify or analyse.  One instructor described generating code as a way to help them understand the suitability of certain topics and common coding patterns (i.e. \emph{``I use GenAI tools to write initial code on topics I am looking into including in coursework or learning more about for course purposes in order to understand common forms of code''}).

Another reported use was for generating programming exercises that could be given to students for practice.  This included some novel ideas, such as asking students to compare their own code with code generated by the AI tool, and asking students to use ChatGPT to generate code and then critique the output that it produces, including highlighting necessary changes.   One instructor noted that attempts to generate exercises suitable for their course were largely unsuccessful due to lack of context regarding the course structure (i.e. \emph{``But because it lacks context about the ordering of course concepts and the goals of the exercises, it has not been much help''}).   Another instructor also mentioned that such tools were not particularly helpful to them for coding, noting that they often found it quicker to write the code themselves, but that they did find value in using it to generate data (i.e. \emph{``I used it fairly heavily in a database course to generate sample data''}).

Overall, most instructors who participated in our survey were not currently using GenAI tools, although several were explicit in their plans to do so in the near future.  Those that were using them were doing so to generate a broad array of educational content, including assessment questions, practice exercises and examples, although not all appeared satisfied in the outcomes.

\subsubsection{Instructor observations of student use of GenAI}

We asked instructors to describe their observations regarding how students are currently using GenAI tools (Question 26 in Appendix \ref{app:instructor-survey}).  The most common response to this question, mirroring the earlier results regarding their own use of the tools, was that they had not observed students using GenAI tools.  However, this was relatively less common (reported by fewer than one-third of participants), suggesting that they have observed their students using GenAI tools more than they use them themselves.

The next most common theme that emerged, appearing in 20\% of responses, was around academic misconduct.  Instructor responses for this theme indicated that students frequently use generative AI tools to cheat on their assignments, in-class exercises, projects and on exams.  They noted students using AI for generating complete solutions, including ``blindly copying and pasting solutions'', and submitting these as their own work even when they sometimes contained advanced elements that were not taught in the course.  One instructor responded to the question about how their students are using GenAI tools with: \emph{``Comprehensively. They are feeding my assignments into ChatGPT and directly copying results and handing them in''}.  This misuse of the tools was a clear concern for instructors, and highlighted problems around over-reliance (i.e. \emph{``they don't check and don't understand the solution generated most of the time''}, and \emph{``they don't realise that it generates something very different from what was asked''}).   

More positive uses of the tools were also reported.  The next most common theme was around using GenAI tools to debug and understand code.  Instructors reported observing students use AI for debugging purposes (i.e. \emph{``they have used it to help fix errors and better understand compiler messages''}), to generate test data and code (i.e. \emph{``writing test cases or code to generate test data''}), and for explaining code that they do not understand.  A similar number of responses also focused on generic code writing help, such as \emph{``to complete small coding exercises''}.  Two instructors mentioned that the students they had observed using GenAI for writing code actually found the experience frustrating, noting that it would have been easier to write the code themselves.  A related, but less common theme, was around the use of GenAI tools as a conceptual learning aid.  A few instructors discussed students using AI to help them understand topics better from the class, and assisting with ideation for project work but not giving complete solutions (i.e. \emph{``They are using them to better understand topics from class (when they miss a meeting, get distracted, whatever)''}).

An interesting theme emerged around language enhancement and communication.  Several instructors observed students using generative AI tools to help improve their English language skills, both in their essays and in communicating with others online, such as writing emails or making posts on forums (i.e. \emph{``We have a variety of students using them to generate English text particularly among English language learners even for short textual interactions (like a brief regrade request)''}).  However, not all instructors viewed this use positively, with one commenting (i.e. \emph{``Students use it when they are not comfortable with their English skills, and the results of this is really frustrating/insulting to read''}).

\begin{custombox}{Potential for Academic Misconduct}
Where instructors have observed students using GenAI tools, there are concerning reports of academic misconduct, including generating complete solutions for assignments. On the other hand, some instructors observed students using GenAI productively for debugging, generating test data, understanding code, and improving English skills. 
\end{custombox}

Finally, it is worth noting that not all of the responses to this question appear to be derived from direct observation.  At least one response indicated that they had no proof but were ``pretty sure'' (relating to academic misconduct).  While instructor responses to this question do reveal the potential for GenAI tools to be used to aid student learning, they also highlight a concerning trend of academic misconduct and over-reliance.  This underscores the importance of providing clear guidelines to students in how to use such technologies productively and ethically in computing courses.

\subsubsection{Student use of GenAI}

Similarly, we asked students to describe the ways that they currently make use of GenAI tools in computing courses for both text generation and for code generation (Questions 18 and 19 in Appendix \ref{app:student-survey}). 

Many of the students in our survey had not used GenAI in their courses, with around 40\% of participants responding in this way for both text and code generation.  This proportion was similar to that of the instructors who had reported not using GenAI.  Of the students who reported not using GenAI, a small portion (fewer than 5\%) refused to do so for various reasons ranging from the risks around learning to it detracting from their joy of programming (e.g. \emph{``I do not use it at all. I love programming, I love to write programs, and I would not let anyone else do it for me''} and \emph{``I do not use GenAI tools in computing courses at all. Struggling and debugging is a valuable part of the learning process''}).  Several students were equally emphatic about not using such tools in the future (i.e. \emph{``i will never use GenAI for computing courses for code generation''}), and one student also refused to use GenAI tools on ethical grounds (i.e. \emph{``I do not feel that the output produced by a GenAI tool can safely be called `my own work', when GenAI tools use so many other people's work as input to produce their result''}).

\vspace*{2mm}
\begin{custombox}{Student Adoption of GenAI Tools}
Most students have explored GenAI tools for text or code generation. Those who do use GenAI tools most commonly apply them for paraphrasing or summarising text, for debugging errors in their own code, and slightly less often for code generation. However, some emphatically refuse to use these tools due to concerns around risks to learning and ethical issues about originality. This framing may help when providing course policies and explaining ethical considerations in syllabi (see Appendix \ref{appendix:student}).
\end{custombox}

With respect to text generation, the most common use (reported by 20\% of students) was paraphrasing or summarising existing text, for example to improve their own writing (e.g. \emph{``I use AI to summarise my own writing to see if the point I want to communicate is clear''} and \emph{``I will write something and then put it into ChatGPT to make it read better''}) or to produce a summary of a large quantity of text (e.g. \emph{``I use it to write summaries about books I've been reading''}).  A smaller proportion of students, fewer than 15\%, reported using GenAI tools for writing new text, with responses ranging from very short descriptions (i.e. \emph{``writing reports''}) to much more detailed processes including iterative development of written reports through multiple rounds of prompting (i.e. \emph{``I keep sending prompts for it to change this and that, add some topics, reword some sentences, explain to me what this sentence means so I understand''}).

The most common use of GenAI by students for coding-related tasks was debugging errors in code they had written, and this was reported by 25\% of respondents (i.e. \emph{``try to fix code when not working''}, \emph{``Helping search for bugs''} and \emph{``I copy and paste the codes that gives wierd error. I tell them what i am expecting but i am getting this then they tell me which part of codes are wrong''}). Code generation was the next most popular theme, with some students reporting using GenAI to generate solutions directly (i.e. \emph{``If I was solving a question that I don’t have the answers to, I would ask it to give me the solution''}), although others were more cautious about the outputs that were generated, with two students describing its use as a `last resort' (i.e. \emph{``I consider using it as a last resort. If I’m running short on ways that would solve a problem and have exhausted all the possible ideas I have then I ask for the explanation of the problem first and if that was unhelpful then I ask for a piece of code which I check for mistakes and incorporate in what I already have written.''}

\subsubsection{Student perceptions of the effects of GenAI tools on employment}

We asked students to describe the effects they think GenAI tools will have
on their prospects for future employment (Question 20 in Appendix \ref{app:student-survey}). There was a mix of positive and negative responses, which is consistent with the quantitative results of the corresponding Likert question (Question 17 in Appendix \ref{app:student-survey}). 

A considerable number of students (33) seemed concerned that GenAI tools will reduce job opportunities. Several were very pessimistic and went as far as to say that all jobs will be replaced by AI (e.g. \emph{``I think that if left unchecked as it is going right now, it will eventually take over all the jobs, no matter who you are really''}). Others were concerned that entry-level jobs will be affected more than senior-level jobs (e.g. \emph{``competition for entry level jobs is going to skyrocket''} and \emph{``entry-level opportunities will probably become quite rare''}). Along these lines, 19 students mentioned that GenAI tools will raise the expectations of employers and increase the difficulty of bootstrapping in the industry (e.g. \emph{``I do believe that the standards that companies require will be higher, as AI has proved to be above mediocre, perhaps affecting juniors and paid interns''}).

Some students argued that software engineering jobs are particularly at risk. For example, a student said: \emph{``programmers will be among the first group to see massive job losses. I believe this because the entirely text based nature of coding is well suited to LLMs. The tech industry is also faster to adapt than other industries.''}. Another student said: \emph{``I'm genuinely concerned about companies realising they only need 1/5 or 1/10 as many software engineers... especially since GenAI can read an entire codebase and easily put together working code from a prompt from a senior dev that also passes tests created by senior devs''}. Interestingly, a student argued that competition for software engineering jobs will increase because GenAI will make learning programming easier, which \emph{``will likely draw the attention of a lot of new people who would otherwise be uninterested in programming''}. 

A different concern raised by some students relates to the hiring process itself. Two students indicated that the use of AI tools to \emph{``judge r\'esum\'es''} might have a \emph{``massive impact''} on employment. According to one student, it feels \emph{``unethical and unfair''} and it is \emph{``incredibly draining to know that all that's between you and a job is a machine''}. Three students also mentioned that standing out to employers will become harder, given that many applicants might use GenAI tools to build portfolios that make them appear as solid candidates despite lacking the required skills. According to one student, such candidates \emph{``will flood the job market''}, and it will be harder for future employers to distinguish between them and those who genuinely have the required skills. These concerns were recently echoed by Armstrong et al. \cite{armstrong2023navigating} who explored the impacts of automated hiring systems as ``black boxes''.

While many students raised concerns, a good number of students (28) indicated that they are not concerned about AI taking over their jobs or about the job market being significantly impacted. Some of these students questioned the ability of GenAI to perform the tasks that humans are good at (e.g. \emph{``I do not think coding will become obsolete though, AI isn't even close to that good yet.''}). Other students questioned whether the job market will change at a fast enough pace to pose a threat to their employability in the near future (e.g. \emph{``I strongly believe that a proper programmer will most likely not be affected in terms of employment in the coming 7-12 years by GenAI''} and \emph{``at least for the next ten years employment will be fine, my skills are transferable and I am always happy to learn new things.''}).

\vspace*{2mm}
\begin{custombox}{Implications for Future Employment}
Students expressed mixed views on how GenAI tools might affect their future career prospects. While some believed job opportunities would decrease, others were optimistic that these tools would improve their productivity and give rise to new careers. 
\end{custombox}

Beyond this, many students thought GenAI would have a positive impact on their future employment. Eleven students mentioned that they expect more job opportunities to emerge because of advances in GenAI tools, and 32 students indicated that GenAI will have a positive effect on their productivity in the workplace. In fact, expecting an increase in work efficiency was the most commonly occurring comment amongst student responses. As one student put it: \emph{``it will make work so much easier; no more boilerplate code, or searching forever through StackOverflow''}. According to another student, GenAI tools will \emph{``allow for more creative flow, more interesting products because the hard work can be done easier, less research time on how to complete a job, and more time completing it''}. Additionally, several students indicated that GenAI tools help them learn better, and thus will improve the skills they need to get employed (e.g. \emph{``It can teach LeetCode pretty good :) so I'll have better chance to pass the technical interview''}). This attitude is orthogonal to that of some of the students whose responses showed negativity towards using GenAI tools while learning. According to those students, relying on GenAI tools may reduce their understanding of the material and thus affect their future chances of employment.

\subsubsection{Student and instructor perspectives on when GenAI tools should be allowed.} 

We asked instructors to elaborate on when they believe GenAI tools should be allowed or disallowed (Question 12 in Appendix \ref{app:instructor-survey}). The prevailing sentiment in the responses was that GenAI tools should not be used when students are learning the basics. Hence, many instructors indicated that GenAI tools should be disallowed in lower-level courses but allowed in upper-level courses. Some instructors argued that it is a function of complexity rather than course-level. For example, more complex assessments (regardless of the course level) are more appropriate for the use of GenAI tools than simpler ones that can be easily completed in their entirety by the tools. 

Another recurring and related theme was that allowing the use of GenAI tools depends on the course and assessment learning outcomes. For example (as described by an instructor), \emph{``if the assignment is to create a website with the goal of learning to apply HCI principles in the design, they should be able to use GenAI or other tools for the mechanical code generation''.} However, if the goal of the assignment is to see if they can write a piece of code, then they should not use GenAI tools to generate that piece of code. An instructor argued that \emph{``this is analogous to many other practices within the University; for example, a student would not normally have to build their own computer, but if they were on a hardware design course they might have to, and submitting a purchased machine would not satisfy the learning outcomes of the module''}. 

A minority of the instructor responses supported unconditionally allowing the use of GenAI tools. Some said that their use is fine as long as the student acknowledges that appropriately. Others argued that it is useless to attempt to disallow their use outside closed-exam conditions, as students will use them anyway. At the other end of the spectrum, a few responses supported always disallowing them, or disallowing them in all graded assessments (regardless of the level, type, topic, etc.).

\vspace*{2mm}
\begin{custombox}{Conditional Acceptance of GenAI Tools}
Both instructors and students suggested that whether the use of GenAI tools is permissible should depend on factors such as the course level, assessment type, purpose of the task, and how the tools are used. This indicates the need for nuanced guidelines and policies when dealing with GenAI in academic settings. 
\end{custombox}

We asked students the same question (Question 13 in Appendix \ref{app:student-survey}). Student answers were more diverse and more polarised than those of the instructors. Interestingly, a good number of students (n=29) argued for disallowing the use of GenAI tools in all coursework and exams. Some also argued for completely disallowing them even outside assessments (i.e. while learning). The arguments used by these students included a range of reasons, like ethical concerns regarding how the models were trained, concerns regarding the correctness of the tools, and concerns regarding the fairness of assessments if these tools are used. However, the majority of these students argued that the use of GenAI tools \emph{``harms learning''} and \emph{``defeats the purpose''} of assessments. Some of these students made strong statements indicating that GenAI tools \emph{``have no place in learning''}, are \emph{``completely counter-intuitive to going to University''}, and are \emph{``only used by people who aren't smart enough to solve problems on their own!''}.

Many of the students opposing the use of GenAI tools emphasised the importance of doing the assigned work and going through the full \emph{``discovery process''} without \emph{``taking shortcuts''}. A student said: \emph{``effort is the road to success and minimising effort can create a generation of couch warriors''}. This comment captures the gist of many of the responses that linked using GenAI tools with deficient learning. Another recurring argument was that using GenAI tools in coursework and exams defeats the purpose of assessments. A student likened it to continuously \emph{``looking to the back of a textbook for the answer''} and another likened it to having someone \emph{``sitting next to you and helping you''} complete the work on which you are being assessed.

On the other hand, fewer students argued for always allowing the use of GenAI tools. An argument made by several of these students was that GenAI tools are \emph{``the future of where the industry is going''} and thus learning how to use them is important for their success. One student said: \emph{``when we go into employment, we will need to use whatever resources we have available to us to be as productive and efficient as possible''}. 

The majority of students argued for a situational or a conditional use of GenAI tools. They provided a wide range of factors that affect (in their point of view) when the use of GenAI tools is acceptable. These factors include:
\begin{itemize}
    \item \emph{Course level}: GenAI tools should be allowed in upper-level courses, but not in lower-level courses when students are learning the basics.\vspace{.1cm}

    \item \emph{Assessment type}: GenAI tools should be allowed in coursework, but not in exams.\vspace{.1cm}
    
    \item \emph{Assessment weight}: GenAI tools should be allowed in minor assessments that carry a small weight, but not in major assessments.\vspace{.1cm}

    \item \emph{Task goal}: GenAI tools should be allowed if the goal is the application of already-learned concepts (e.g. to build an artifact). It should be disallowed if the goal is learning the concepts. \vspace{.1cm}
    
    \item \emph{Task size}: GenAI tools should be allowed if the task is large and complex, requiring stitching many pieces together. It should be disallowed if the task is small or trivial.
\end{itemize}

While these factors relate to the assigned task, some students felt the acceptability of the use of GenAI tools should be conditional on the way the tools are used, rather than on the task itself. For example:

\begin{itemize}
    \item \emph{How}: Using GenAI tools with understanding is fine. Blind copying and pasting of answers is wrong.\vspace{.1cm}

    \item \emph{How much}: Using bits and pieces or partial solutions generated by GenAI tools is fine. Using a complete solution generated fully by a GenAI tool is not fine. \vspace{.1cm}

    \item \emph{Why}: Using GenAI tools as a last resort, when stuck, or when there is no other way of getting help is fine. Relying on GenAI tools right from the beginning is not fine.
\end{itemize}

\noindent The last category above is interesting as it assumes that, in general, the use of GenAI tools is unethical unless it is out of necessity. The following are several quotes from the student responses that support this idea:

\begin{itemize}
    \item \emph{GenAI can be used as a last resort when the lecturer is rather difficult to explain a material and students use GenAI when they cannot understand what is being explained or assigned at all}.\vspace{.1cm}

    \item \emph{GenAI should be allowed if the courses force us to do work manually without any mentoring. Vice versa, if the mentor is giving course completely i think GenAI should be disallowed}.\vspace{.1cm}

    \item \emph{... when you run into a dead end and even after looking online and asking a friend and either don't know or you still don't understand to go and ask GenAI for an answer}. \vspace{.1cm}

    \item \emph{I believe GenAI should be allowed sometimes when you have no one else left to ask}.
\end{itemize}
\newboolean{hidesummaries}
\setboolean{hidesummaries}{true}
\ifthenelse{\boolean{hidesummaries}}
{\newcommand{\summary}[2]{}}
{\newcommand{\summary}[2]{
    \fbox{\bfseries\sffamily\scriptsize#1}
    {\sf\small$\blacktriangleright$ 
      {#2} $\blacktriangleleft$}}} 

\section{Curriculum and Assessment}
\label{sec:canda}

In the past 50+ years, a great body of research within the SIGCSE community addressed many trends, opportunities and challenges in Introductory Programming (CS1) courses~\cite{luxtonreilly2018introductory}. Among these are teaching and learning approaches, new forms of assessment, shifts in content, tools, and overall course design~\cite{becker2019fifty}. 
For example, at the turn of the millennium computing educators passionately debated whether to use an objects-first approach (or not)~\cite{cooper2003teaching}. Similarly, Alice, Scratch, Blockly, and other block-based programming languages have been the subject of much research~\cite{pausch1995alice,Resnick2009,seraj2019}. Although these developments were important for many reasons and groups of (present and future) students, they are not comparable, at least not in pace and ubiquity, to the rapid changes Large Language Models (LLMs) are currently triggering in higher education, the computing disciplines, and CS1 in particular. 
With LLMs available on nearly everyone's phone and laptop\footnote{Acknowledging that internet access is required to access LLMs, and that subscription-based services which could be superior to free ones present issues of access based on means, and potentially opening new divides.}, it is not only knowledge that is instantly retrievable but also problem explanations and solutions - in the form of programming code that is not necessarily correct. Given the pervasiveness of LLMs, this paradigm shift regarding the availability of knowledge, solutions, examples, and content (particularly in the form of code) is more comparable with the advent of the internet than other developments in the annals of how we teach and learn computing -- yet the speed of internet adoption was much slower as a whole.   

Given the myriad impacts of LLMs, it is important to acknowledge that educational systems are notoriously slow to change. Reasons for this are many, yet Lee Shulman~\cite{shulman2005signature} adds the pedagogical psychologist perspective, pointing out that the ``signature'' of a profession's teaching and learning is pervasive and perpetuated at three levels: surface, deep, and implicit (i.e., curricula, pedagogy, attitudes \& values). Now, however, with the seeming ubiquity of LLMs, it is inevitable that educators consider their impact on teaching, learning, assessment and delivery, leading to possible redesign of their courses at all of these levels.

Based on the concept of Constructive Alignment~\cite{biggs1996enhancing}, learning objectives need to be aligned with exercises, assignments, and assessment methods. Therefore, we discuss the relevance of LLMs for CS curricula and assessments with regard to course objectives, and course activities including formative and summative assessments. This discussion is centred on expert interviews we conducted with introductory programming educators, focusing on their changed educational views and practices to highlight how computing education is evolving (with what seems to be lightning speed) in light of LLMs. 


\subsection{Methodology for expert interviews}

To understand how computing curricula and assessments are currently being affected by the emergence of LLMs, we conducted an interview study with computing educators as experts in the field. The interviews were semi-structured, with an interview guide as a basis. 
Using the purposeful sampling method~\cite{patton_qual_2018} led to the selection of experts via the authors' networks, who were contacted via email. Moreover, an invitation was sent out to active contributors to a discussion thread from the SIGCSE mailing list concerned with LLMs. Another recruitment attempt was made via an open question in the instructor survey, where respondents willing to elaborate on their responses in an interview could enter their contact details (see Appendix \ref{app:instructor-survey}). 
The most important criterion for inclusion was that educators would have concrete plans or views toward changing their current course structure, assessment, or classroom practices in light of LLMs. This is one of the main ways the present work differs from that of Lau and Guo~\cite{lau2023from} discussed in Section \ref{sec:survey-prior}.

The interview guide included the following questions and follow-up questions: 
\begin{itemize}
    \item Which course(s) are you teaching in the next semester? [If they are teaching multiple courses, then try to talk about one particular course or at least make sure it is clear which course is under discussion at any point in the conversation.]
    \item Do you have an explicit set of written learning objectives/ competency goals for this course?
    \begin{itemize}
        \item If yes, will LLMs change these goals?
        \item If yes, what goal will change or be removed? What goal(s) will you add?
        \item If no but they have informal learning objectives, ask how they think these will change (or have changed).
    \end{itemize}
    \item Are you planning to change your pedagogy and/or learning activities because of LLMs?
    \begin{itemize}
        \item If yes, how (what did you use before, how do you change it, and why exactly)?
    \end{itemize}
    \item Are you planning to change the assessment mechanism?
    \begin{itemize}
        \item If yes, how (what did you use before, how do you change it, and why exactly)?
    \end{itemize}
    \item What is your vision for that course in the context of LLMs?
    \item Which opportunities for enhancing teaching, learning, and assessment can you think of?
    \item Which challenges come to your mind if you think about LLMs in the context of computing education?
\end{itemize}
These questions resemble some of the survey questions, but they allow for a more in-depth elaboration of instructors' practices. 
Interviews were scheduled to last between 20 and 60 minutes. In practice most were closer to 60. They were conducted via Zoom and automatically transcribed via speech recognition software, followed by a correction loop by a human (interviewers checked transcripts of other reviewers, not their own). After the transcripts were finalised, we deleted all audio and video recordings in accordance with the protocol submitted to the University of Toronto Research Ethics Board, who approved it prior to the study. Respondents were free to decide if they wish to remain anonymous or to be named. Those that are named in this report gave their consent for this. Affiliations are noted in the acknowledgements section and on the interviewees' first mention in this section. We also allowed participants to review a draft manuscript before final publication, to ensure that they are not misrepresented. 

The sample comprises 22 computing educators from nine countries and five continents. Table \ref{tab:countries} shows the locations of these 22 interviewees.

\begin{table}[!h]
\caption{Countries of Interviewees}
\begin{tabular}{rl}
\toprule
1 & Austria \\
1 & Brazil \\
1 & Denmark \\
2 & Germany \\
1 & India \\
1 & New Zealand \\
1 & The Netherlands \\
2 & UK \\
12& USA \\
\midrule
Total &  22 \\
\bottomrule
\label{tab:countries}
\end{tabular}
\end{table}


The interviews were fully transcribed verbatim and served as a basis for thematic analysis~\cite{braun2006using,ryan2003techniques} reflecting examples in the teaching community of practitioners teaching in the new context of LLMs. 
Initial themes were deductive, rooted in the literature review (see Section \ref{sec:literature_review}). As a next step, themes were reviewed and refined inductively based on the interview material, resulting in the structure presented in the following sections. Two of the authors reviewed both the verbatim interview transcripts and the identified themes for all interviews. Then themes were discussed and specified among four of the authors. Interview summaries were then compiled based on these themes and double-checked against the themes and transcripts by a second author.



\subsection{Learning objectives}

The rise of LLMs is causing many instructors to rethink their course learning objectives. For instance, as students increasingly use LLMs to write programs, instructors might need to put less emphasis on writing code and more emphasis on reading code. Some researchers have begun to investigate this new emphasis \cite{becker2023programming,vaithilingam2022expectation}. 
Although the ability to both read and trace code has always been important~\cite{lister2004a,CS2013curricula}, an examination of course syllabi showed that this is often an afterthought~\cite{becker2019what,kiesler2020towards}. Similarly, many topics in introductory programming and computing more widely have been heavily researched up to present day, yet their effects on the curriculum vary~\cite{becker2019fifty,kiesler2020on,clear2020:CC2020}. Large language models have been recently linked with many of these in various ways~\cite{wermelinger2023using, becker2023programming}. Such topics include metacognition~\cite{denny2019closer, loksa2022metacognition, prather2020what}, algorithmic/computational thinking~\cite{bao2023mind, saito-stehberger2022examples}, communication skills~\cite{willis2020developing, pollock2019collaborative}, and their respective dispositions.

In 2019, Becker and Fitzpatrick \cite{becker2019what} collected syllabi from 234 introductory programming courses from around the world and made their data available publicly for others to use (and contribute to). One of the types of data they categorised was explicit learning outcomes. Of the 234 syllabi, 154 contained explicit learning outcomes. The five most common learning outcomes were: ``testing and debugging'', ``writing programs'', ``selection statements (if/else, etc.)'', ``problem solving (including computational thinking terms)'', and ``arrays, lists, vectors, etc.''. These five objectives appeared on at least 40\% of the syllabi. Looking through the full list of learning objectives, the only one that was directly related to reading code was ``tracing program execution'', appearing on only 3\% of syllabi.  
Kiesler did a similar study in 2022 \cite{kiesler2020towards,kiesler_diss_2022} using syllabi from 35 German universities. She found that the most common objective was ``writing code'' and that the objective ``being able to read, explain and identify the output of (foreign) code'' appeared on less than 10\% of syllabi. 

In response to the rapid advance in LLM capabilities, educators are reconsidering their courses' objectives. In the following subsections, we present the respective themes identified in the interview transcripts and relate them to some recent studies in the field.


\subsubsection{How instructors are changing their learning objectives}
\label{sec:canda_objectives}
Several instructors discussed changes in their upcoming course learning objectives. However, given the rapid emergence of LLMs in computing education, in most cases, these changes are not yet reflected in official curricula or course syllabi. Instead, some educators are changing more fine-grained learning objectives in their courses. 

James Davenport (University of Bath, UK), for example, introduced two new sessions concerning the impact of LLMs on cyber-security in his class. Even though the official course objectives remain as they were due to their general nature,  
Davenport has started to teach students how defenders and attackers could take advantage of LLMs. 

Many educators acknowledge the dynamic nature of technology and anticipate potential adjustments to their learning objectives in the near future. Viraj Kumar (Indian Institute of Science, Bengaluru, India) expressed the need for flexibility, recognising that changes might be necessary as technology evolves even further: ``\textit{And even now I'm sort of holding my breath because now I'm saying, hey, let's put out these things, but you know, maybe things change.}'' Statements like this reflect the fast pace of advancing technologies in computing education and educators' openness to adapting their practices. Kumar recently updated their CS1 class of approximately 50 students to include the topic of LLMs' role in code generation.

Educators like Ewan Tempero (University of Auckland, New Zealand) emphasise the role of LLMs in automating routine tasks, enabling educators to shift their focus toward nurturing critical thinking skills: ``\textit{The more tools that [students] have to support doing the stuff that really isn't that interesting, the more [educators] can focus on the interesting stuff like critical thinking.}'' 
In this context, Briana Morrison (University of Virginia, USA) highlighted the importance of using citations with LLM-generated code and teaching students to evaluate LLM output in a critical manner. Even though educators are not teaching students how to write prompts at the University of Virginia, they are ``\textit{going to have a statement in the syllabus that using LLMs is allowed. However, we are going to require a reference, like a citation, that says this was generated by an LLM.}" 

Some educators, including Leo Porter (University of California San Diego, USA) note the importance of prompt engineering when using LLMs. Porter introduced new learning goals for his CS1 class, emphasising the non-deterministic aspect of LLMs, prompt engineering, and problem decomposition. Porter feels that crafting effective prompts is a competency students should develop as students should learn how to interact effectively with LLMs, ensuring that they obtain meaningful and accurate results. As for problem decomposition, Porter said: ``\textit{We didn't used to teach problem decomposition''} but felt that this was part of their hidden curriculum that he and his colleagues are pleased to see moving to the forefront.

In this context, Michael K{\"o}lling (King's College London, UK) adds that LLMs might force us to look at learning outcomes at the program level instead of the course level and that using LLMs should be explicitly taught. This aspect is also emphasised by Kristin Stephens-Martinez (Duke University, USA) who said ``\textit{We're going to have to help students understand how to use LLMs ... and you all [the students] need to understand that ChatGPT is fallible, and you need to be very critical of what it's doing.}'' Rodrigo Duran (Federal Institute of Mato Grosso do Sul Brazil, Brazil) explicitly encourages students to test and understand LLM-generated code, enabling them to evaluate if the LLM's answers are correct and to adapt their answers accordingly.

A recurring theme among the educators we interviewed was the elevation of code comprehension and critical thinking skills. Jean Mehta (Saint Xavier University, USA) highlighted the need to assess students' ability to read and understand code more thoroughly, a competency that has often been overlooked in the past. Mehta concludes that ``\textit{we should have more time to spend on these kinds of things.}'' 

It is possible that the impact of LLMs on introductory programming might go beyond changing course objectives and lead to the introduction of entirely new courses, syllabi, and learning objectives. An example of such a recent development is the ``Generative AI'' online course by DeepLearning.AI~\cite{deeplearning2023}. 



\subsubsection{Preserving core learning outcomes}
Computing education has always experienced change over time as new tools and technologies were introduced. 
Although Copilot may serve as a useful springboard for solving CS1 problems, students still need to dedicate time to learning algorithmic thinking, program comprehension, debugging, and communication skills~\cite{wermelinger2023using} in order to become not only proficient computing experts but also to use LLMs effectively. A number of interviewees shared this perspective and highlighted the need to preserve several core learning outcomes.

Educators who do not alter their learning objectives stress their focus on teaching fundamental programming concepts. The learning objectives of Peter Mawhorter's (Wellesley College, USA) introductory CS1 class remain stable and focus on fundamental concepts. Similarly, Frank Vahid (University of California, Riverside, USA) believes that it is still important to teach students how to define variables, use branches or loops to solve problems, how to use functions to keep code modular or use vectors to store data. Thus, students still need to learn to code and practice. In the era of LLMs, Vahid thinks ``\textit{... the most pressing thing is making it [assessment] harder... it [Generative AI] takes the work out of homework}.''. Vahid pointed out that students have for some time been getting help through other means such as Discord forums, Chegg, Stack Overflow, Piazza, etc., noting that ``\textit{We've had a leaky roof for a long time. And LLMs are the storm that finally causes us to be flooded.}''

Similarly, Leo Porter who recently published a textbook with Daniel Zingaro for teaching introductory programming with the help of LLMs from day one~\cite{porter2023learn} 
emphasised the importance of preserving core learning objectives that focus on teaching fundamental programming concepts. Porter noted that these objectives remain unchanged due to the foundational nature of the skills taught. They conclude that ``\textit{... things are staying the same for the Intro class because the skills I'm teaching there are so basic.}''

Some educators including Mark Liffiton (Illinois Wesleyan University, USA) views LLMs as valuable tools that can aid students in coding tasks. Liffiton intends to maintain hands-on coding practices while integrating LLMs into  Programming Languages and CS2 courses. They focus on building foundational knowledge that complements the capabilities of LLMs: ``\textit{I still want the students to be doing that work. I still want them to be practising the things that the tool could do to build the base of knowledge so they can later do the things that the tool can't do.}'' Another interviewee shares this concern: ``\textit{I'm very worried that if everybody forgets how to write code and think through this stuff, we will lose the ability to make new things.}'' At the same time, educators seem to be aware of the need to prepare students for industry expectations, which is likely to include LLM use. 

Dan Garcia (UC Berkeley, USA) pursues a similar approach. He recognises the potential of LLMs as educational tool, or aids, but emphasises the importance of teaching programming basics first. Garcia encourages the use of LLMs once students have mastered traditional programming concepts, concluding that ``\textit{... we can't stop teaching kids how to program.}'' This perspective is shared by Austin Cory Bart (University of Delaware, USA) who is not changing the learning goals in an introductory programming class for about 280 students. Nonetheless, they expect adjustments in subsequent courses as LLMs become more integrated into programming practices: ``\textit{But I look at almost every single course after mine as, Oh, yeah, that's probably going to need to change learning objectives.}''

To conclude, computing educators express the need to balance between leveraging LLMs for problem-solving while ensuring that students continue to develop competencies in algorithmic thinking, program comprehension, and debugging. Educators vary in their approaches, with some maintaining a focus on teaching fundamental programming concepts, while others see LLMs as complementary tools to enhance learning. The preservation of core learning objectives, particularly those related to basic programming, remains a consistent concern among educators.



\subsubsection{Towards conversational computing}
If computational thinking is the learning goal of a non-majors course, then using an LLM-based tool such as Github Copilot may be a useful approach as advocated by Denny et al.~\cite{denny2023conversing}. For students in non-computing majors who currently only take one (or just a few) programming course(s) to learn enough to write simple programs, it may be that using an LLM tool is all that is needed, making a programming-specific course unnecessary~\cite{piccolo2023many}. 
It may even be more effective than typical courses at introducing these students to programming and may also broaden participation in CS courses in the future. 

Michael Caspersen (It-vest \& Aarhus University, Denmark) believes that LLMs are forcing us to rethink what we actually teach our students, and encourages us to view them as an opportunity. According to Caspersen, LLMs do not add something qualitatively new, but quantitatively indeed! They emphasise issues that have always been present. Hence, LLMs may even contribute to increasing the quality of computing education, thereby making it more attractive to a broader range of students. 


\subsection{Course activities}
LLMs can be useful in generating several kinds of learning activities including novel variations of programming assignments~\cite{sarsa2022automatic}. However, this may introduce problems such as ensuring that students have been provided appropriate content in order to understand novel variations. 
LLMs can also generate good explanations of code~\cite{sarsa2022automatic, leinonen2023comparing, macneil2023experiences,hellas2023exploring}. This provides a mechanism for novices struggling to understand the run-time behaviour of novel problems to get auto-generated and hopefully helpful explanations. This use-case exemplifies the potential for LLMs to ease the burden often felt by teaching assistants, and could be a first line of help for students~\cite{sarsa2022automatic}. Another type of learning activity that can be provided by LLMs occurs in settings where students use them to help generate code solutions. Here, students can use LLMs in an iterative improvement loop. Students can continually alter prompts to refine the model output helping them to `build-up' a solution~\cite{yuan2022wordcraft}. In their interview Mark Liffiton took this concept one step further stating \textit{``So they're great educational tools because they can give something akin to one-on-one tutoring... and I think there's a ton of value in there.''}. This particular use-case of LLMs is less about what LLMs can produce, but what they can do for students (and educators) and aligns with what the Artificial Intelligence in Education (AIEd) community has been discussing for years~\cite{becker2017artificial}. 

Liffiton has also developed a tool called CodeHelp~\cite{liffiton2023codehelp} that uses LLMs that he is going to use with his students which can do some of what ChatGPT can do, but specifically does not solve the problem, and does not give students complete solutions - a kind of ``sanctioned'' access to the educational power of LLMs which could combat the concern of students becoming over reliant on them, and not actually learning from them. Mark sees this as adding to the course learning objectives to include working with this new tool and therefore LLMs. Similarly, Frank Vahid sees LLMs as tutors that will be available around the clock, and importantly, as tutors that will not judge students, stating \textit{``I'm very hopeful that it will become another TA for the class.''}

An important aspect of teaching is using carefully crafted examples to illustrate salient points.  If the goal is to use a \emph{real} or \emph{running} example, it can be tedious to have to deal with the many irrelevant (to the example at hand) aspects of a problem in order to ensure the example compiles and runs, in addition to keeping focus on the point desired. LLMs can help instructors with the tedium of such minutiae~\cite{sheard2017strategies, macneil2023experiences}.

Clearly, many of the student-initiated possibilities discussed above come with academic integrity concerns we discuss in Section~\ref{SEC:ACAD-INTEGRITY}. Several interviewees were aware of this, stressing the need for educators to emphasise the importance of students doing their own work. 

Perhaps the most extreme example of how Generative AI might be used in the introductory programming course comes by way of Dan Zingaro and Leo Porter's new textbook~\cite{porter2023learn}. The book begins by introducing students to the GitHub Copilot plugin within the Visual Studio Code IDE before students have learned to write a single line of Python code. Students create their first programs by typing English comments and letting Copilot generate the code. This corresponds to the \emph{sketch model} from Alves and Cipriano's Centaur Programmer~\cite{alves2023centaur} where the programmer generates the outline of the solution and the AI fills in the gaps. As Zingaro and Porter explain each line of LLM-generated code, they use the opportunity to teach the related Python syntax and programming concepts. But before they do this, they introduce functions and use this to motivate top-down design. Porter will be teaching 700 students using this approach in the upcoming semester (September~2023).

\subsubsection{In-class Activities}
LLM-based tools have motivated some changes to activities that specifically happen during classes. Denny et al.~\cite{denny2023conversing} found that Copilot's performance is substantially improved when it is prompted with individual problem-solving steps in natural language and encourage explicitly teaching prompt engineering to students. David H. Smith IV (University of Illinois, Urbana-Champaign, USA) will be using a tool and specific exercises for students to practice LLM prompts so they can use LLMs effectively. Leo Porter also stated that they will be adding a learning goal on prompt engineering as discussed earlier.

Although not planning on getting into prompt engineering, Briana Morrison is going to do more live coding with LLMs in the classroom including analysing why certain code is wrong. Viraj Kumar has also used Generative AI for live coding. J{\'e}r{\'e}mie Lumbroso (University of Pennsylvania, USA) created guided sessions using LLMs and shows examples of them giving incorrect answers. Similarly, Kristin Stephens-Martinez told us that they plan to work out examples in advance of using prompts that demonstrate a hallucinated answer or other technically-incorrect response to help students see for themselves that LLMs are not oracles. Austin Cory Bart is going to demonstrate the use of LLMs in class for three reasons: First, because LLMs are a great way to introduce AI topics, also providing a way to bring advanced material into the introductory course so students can look forward to what is coming later; Second, because Bart feels that if it is not discussed with students they'll just use it anyway, but in a more misguided manner. Finally, Bart stated: \textit{``At some point we have to start incorporating these tools... programmers are going to be putting these tools into the workflow. And I see that on my own... when I'm coding this summer having co-pilot auto-complete an entire function that I just started writing, that's too much of a game changer for it not to be addressed.''}

Interestingly, Leo Porter is the only interviewee who mentioned pair programming directly. Porter plans on continuing to use pair programming and peer instruction in class. Dan Garcia noted that LLMs should be used as nudgers, hint-givers or help-givers, but not oracles -- seeming to match the desired role of a human pair programmer, and Michael Caspersen noted that LLMs should be integrated in peer to peer learning. Others in the literature have speculated on what may come of pair programming in light of LLMs. For instance, Dakhel el al. noted that when used by experts, Copilot can be an asset as its suggestions could be `comparable to humans’ in terms of quality. However, it could become a liability when used by novices who may fail to filter its buggy or non-optimal solutions due to their lack of expertise~\cite{dakhel2023github}. Wermelinger stated that it is not surprising that GitHub dubs Copilot as `your AI pair programmer', even though the interaction is far more limited than with a human -- notably, Copilot does not provide a rationale for its suggestions~\cite{wermelinger2023using}. However, Frank Vahid did note a positive aspect of how many Generative AI tools present their output noting positively that LLMs generally do not judge the student. 

Do we stop teaching long division now that we have calculators? Do we let students use calculators when they are learning long division? No and no.

However, in the same way that they expect students to learn to use an IDE on their own time, some instructors have indicated that they will not be dedicating classroom time to explicitly teach students LLMs. Rodrigo Duran is not actively encouraging LLM use but at the same time is not actively discouraging their use. Dan Garcia does not plan on incorporating LLMs into their course, at least for the time being, choosing to keep their `ear to the ground' and see what others are doing, and stressing that the fundamentals are important: \textit{``Do we stop teaching long division now that we have calculators? Do we let students use calculators when they are learning long division? No and no.''} Peter Mawhorter does not plan to allow LLM use in their course out of fear of harming particular students, expecting that a small percentage of students -- likely those who are marginalised in other ways -- will have bad experiences due to biased, possibly sexist or racist output. It is also possible that these tools will give different output to different students, for instance based on the student's name if provided in starter code or a prompt.

\subsubsection{Unsupervised Activities}

Student use of LLMs in unsupervised learning activities, self-assessments, and other, self-directed scenarios have also been the subject of research. 
MacNeil et al.~\cite{macneil2023experiences}, for example, used GPT-3 and Codex to generate three types of code explanations (line-by-line explanations, lists of important concepts, and high-level summaries) from code snippets from an instructor-developed web software development e-book. They found that Codex generated less helpful and more verbose explanations than GPT-3. Moreover, Codex included code in the explanations, even though it was not desired. Students rated these explanations on a 5-point Likert scale, confirming that explanations matched the code and were useful for learning in general, whereas line-by-line explanations were rated as least helpful. 

An exploration of the potential of LLMs to generate formative programming feedback ~\cite{kiesler2023exploring} suggests that ChatGPT performs reasonably well for some introductory programming tasks and student errors indicating that students can potentially benefit in unsupervised learning scenarios. However, it is possible that it will fall on educators to provide guidance on how to use the generated feedback, as it can contain misleading information for novices. Several of the approaches discussed in the prior section are aimed at mitigating this. 

Interviewees mentioned unsupervised activities less than in-class activities in general. This could be at least partially due to the fact that this is the first academic year where Generative AI could be considered mainstream and unsupervised activities are more difficult to plan and hypothesise about. Most of the discussion around unsupervised activities centred around ensuring that students are doing their own work (even if using Generative AI as a tool for help) and that students understand the code that these tools produce. Frank Vahid believes that this could lead to increased instructor emphasis on tools that analyse student behaviour.  


In a high school introductory programming course, Christian Tomaschitz (TU Wien, Austria) had half of his students use ChatGPT and the other half use an e-book. ChatGPT was not introduced to students and students registered OpenAI accounts with their school e-mail addresses. All students had the same exam which was half open-book (including internet access), half closed-book. Tomaschitz noticed that the ChatGPT students had the possibility to interact (with ChatGPT), and these students were faster solving the problems, but it seemed as if they did not critically reflect the output, and understood less. Tomaschitz was concerned that the ChatGPT group was over-reliant on the tool and did not understand the output as well as the e-book group. 

Viraj Kumar used GitHub Copilot during class for live coding and told students they were free to use it, or any other Generative AI for coding. Students were polled after the course and asked if they were using ChatGPT. Most said that they were not. The consensus explanation seemed to be that they wanted to learn programming themselves and their thinking seemed to be that when they go for technical job interviews they will be expected to code without assistance. 

Several interviewees including Leo Porter and Michael Caspersen believe that LLMs will cause educators to focus more on code reading and less on writing from scratch. Caspersen noted that students read more than they write when learning to read natural language, but to-date this has not been the traditional approach when it comes to programming. Further, LLMs could support a `use, modify, create' approach for programming. Briana Morrison envisions more exercises about why certain code is wrong and fewer on writing code from scratch. 

Kristin Stephens-Martinez mentioned that they are concerned about LLM use by novices pushing more metacognitive practices earlier in the curriculum, perhaps before students are ready for this. For instance, students will have to ask themselves `am I going to take this shortcut or not?'. Getting students to recognise when it is a shortcut versus when it actually is not a shortcut is something that novices typically are not in a position to assess.



\subsection{Assessment}

Although we previously discussed activities that could have been assessed, in this section we focus specifically on formal assessment. Research on LLMs has demonstrated their ability to answer typical CS1 assessments that involve writing simple functions \cite{denny2023conversing,kiesler2023large} as well as more advanced material. Similarly, it has been shown that they perform in the upper quartile of real students on CS1 exams~\cite{finnieansley2022robots}. It has also been shown that they perform just as well on CS2 exams as CS1 exams~\cite{finnieansley2023my} suggesting that LLMs may soon be capable of effectively solving even more advanced problems. Indeed, recent results with GPT-4 suggest that it can solve most exercises in introductory programming courses~\cite{savelka2023thrilled}, which is also supported by our benchmarking work presented in Section~\ref{sec:benchmarks}.

However, evidence also shows that LLMs do not perform as effectively on computational thinking problems that do not involve code writing \cite{bellettini2023davinci}. Similarly while LLMs can often correctly answer more than half of coding-based multiple-choice problems, they answer double-digit percentages of multiple-choice problems incorrectly, leading to the hypothesis that either a combination of natural language and a code snippet, and/or chain-of-reasoning steps pose a challenge for LLMs~\cite{savelka2023can}. It might be tempting to think that even though LLMs can do well on relatively simple well-specified functions, longer and more complex problems are beyond the capabilities of LLMs -- although LLMs have been shown to perform well on more advanced (coding competition) problems~\cite{li2022alphacode}.

\subsubsection{Exams}

Some instructors are moving towards invigilated exams and assessments, and these may be worth more marks. For example, one interviewee changed the grade weighting of their programming assignments from 50\% of the course grade to 0\% of the course grade. They added an assessment category, ``coding interviews'', which ensures that students are not using LLM tools as part of the assessment.


Several educators mentioned oral exams. Jean Mehta plans to use 20-30 minute one-on-one oral exams at the end of every section. This is made possible by a combination of a flipped classroom with many videos and book/autograder technology such that there is less need for traditional lectures covering the material.  Michael K{\"o}lling said, ``\textit{I think there needs to be at least some assessment that includes an oral element because you know, there is at the moment, the problem is that submission of written work, whether it's text or programs, is taken as a proxy for intellectual achievement, right? And what we actually want to assess is intellectual achievement. We want to see some, you know, sort of intellectual work having happened there and we take the written work as evidence of that. And you know, these tools have removed that connection... The creation of written work is no longer evidence of intellectual work having happened.}''


Similar to personalising assignments, on open-ended writing assignments, Ewan Tempero requires more specific answers directly related to course materials: ``\textit{And it's not that we used strange terms or unusual terms, but we used specific terms in a particular way. And so, we expected answers to reflect that, to demonstrate that, yeah, they did actually understand what the course material was.}''

\subsubsection{Homework}

There may be less summative unsupervised work because instructors no longer trust that unsupervised work is the students' own. This may be accompanied by a tendency to not grade code assignments going forward. Along with increasing the weight of exams, many teachers are devaluing ``homework'' assignments altogether. Mark Liffiton said, ``\textit{I will sort of be operating in this assumption that some students may end up just getting a tool to do the work. And thus, I don't want to be putting too many points on that and giving an unfair advantage in those cases.}'' One instructor decreased the grade weighting of their programming assignments significantly and added this text to the assessment description, ``\textit{As the programming assignments are intended primarily for practice and learning, your program does not have to be fully correct to receive credit. The final evaluation of whether you have learned from your programming assignments is in your ability to solve problems on quizzes and the exam.}''

Some teachers no longer require ``writing assignments'' in the traditional sense. Jan Schneider (Goethe Universität, Germany) has a course that used to include writing a scientific paper, but it has been changed to allow students to produce it using LLMs. ``\textit{I mean, the main objective is to help people to start thinking in a more scientific way.  That's the overall objective.... I will not ask them to develop a mini paper by scratch where they need to write everything.}''

\subsubsection{Process over product}
Rather than assessing final solutions as products, some educators are increasingly focusing on assessing learning processes (which is common in other disciplines, e.g., teacher education) such as: submit-in-stages; solution reflection / commentary / critique; interviews; portfolios or learning journals; and presentations. 
Taking the approach of having students focus on the learning process through a diary or journal, Christian Tomaschitz replaced several previous exercises with reflection assignments in a diary for students to document their learning process.

In relation to more open-ended assignments, Leo Porter said, ``\textit{... gone are the days of us really just completely describing the exact behaviour of the functions... And then it's going to be a lot more work for us to grade because we're going to have to now look at the code or the PDFs of the code. Or look at the video of them showing how the code works.}'' This approach also shifts focus from memorising knowledge (or in a rote manner, the process that leads to the product without questioning the process) to application of skill and critical thinking. J{\'e}r{\'e}mie Lumbroso applies this shift to his Discrete Mathematics course, explaining that ``\textit{the idea is to focus less on the product and to focus more on process, on making sure that the students are able to explain what they are doing.}'' 
Some educators have already started using LLMs as part of the learning process. Briana Morrison explains a possible assignment thus: ``\textit{Here's the problem. Here's the prompt we gave ChatGPT, or Copilot, or whatever. Here's the output it gave us. It's wrong. Tell us why it's wrong.}''

Several educators worried that some classes do not lend themselves to approaches mentioned above.  For example, elementary theory courses are not really amenable to ``personalise'' a proof. Michael Caspersen, on the importance of basic competencies, even in the face of powerful LLMs stated \textit{``And that means that if you add a good programmer to large language models, you get two good programmers. Basically, that's the equation, right? If you add a mediocre programmer to large language models, you get just large language models. So, from that perspective... if you want to really be able to amplify the capabilities of humans, we need to make sure that the basic competencies remain.''}

Frank Vahid stated: \textit{``The biggest challenge is cheating ... you've gotta learn ... you've got to work to learn. You can't just let tools do your work for you... and this is true beyond computer science. In English, you've got to learn to write. Even though ChatGPT can do most of your writing for you... you've got to learn to write... that's how you think. That's why you're valuable as a human to a company... and so the same with computer science. So somehow, we've got to get that message out. It's just so tempting for students to save so much time.''}

\summary{belletinni2023davinchi}{
 \textbf{Michelle} \cite{bellettini2023davinci} DaVinci Goes to Bebras: A Study on the Problem Solving Ability of GPT-3
    GPT-3 did not perform well on Bebras contest questions based on computational thinking - answering correctly on about 1/3 of the questions. Even when the specific answers (particularly in multiple choice questions) are correct, the explanations reveal incorrect reasoning. This suggests that one avenue for assessment is to focus on Bebras-type questions for computational thinking rather than on writing code.
} 

\summary{li2022competition}{
 \textbf{Michelle: li2022competition \cite{li2022competition} Competition-level code generation with AlphaCode}
    In this paper, Li et al., demonstrate that their AlphaCode system pre-trained on GitHub data and then fine-tuned on competitive programming contest examples, earned a ranking in the top 54.3\% of human participants on problems from a simulated program competition. The problems in these competitions are significantly more complex than the types of problems on which LLMs have previously been demonstrated to work well. They are also designed to be novel for each competition. The authors claim that solving these competition problems ``requires understanding complex natural language descriptions, reasoning about previously unseen problems instead of simply memorising code snippets, mastering a wide range of algorithms and data structure, and precisely implementing submissions that can span hundreds of lines."  There are implications here for both pedagogy and assessment. We previously acknowledged that we could no longer use small self-contained well-specified functions for unsupervised student assessment and be confident that the submission represented the student's work, some instructors argued that LLM tools would be unable to provide correct solutions to longer more complex problems. This paper suggests otherwise.   
}

\summary{bull2023generative}{
\textbf{Natalie}  \cite{bull2023generative} 
An exploratory interview study with five professional developers \cite{bull2023generative} summarises opportunities and threats from an industry perspective, and discusses implications on software development education. Among them are changes of assessment formats, as LLMs perform very well in common tasks and for simple algorithms. They recommend more context-specific and applied problems, instead of generic ones. These suggested changes are perceived as advantageous for students, as applied problems still require "critical problem solving and creative thinking", and reduce the cheating potential by students using LLMs.  
}

\summary{idialu2023Whodunnit}{ - Assessment

copilot code is indeed difficult to detect.

"They suggest that the use of Codex is inevitable and that educators may have to accommodate its usage to ensure that students are learning, and this can be done better if educators can identify how much contribution such tools give in solving programming questions. In the study done by Puryear and Sprint, Copilot generated unique code that made it hard to detect any common solution using similarity detection tools."

"In light of our results, we think that there is a good possibility in training a machine learning model on a bigger dataset and being able to differentiate between code generated by AI versus code written by humans. Also, with a bigger dataset, the feature importance output might differ, which means that some features would be more important in bigger datasets. In addition, more features in bigger datasets could be eliminated, so the accuracy of the classifier might be as good in a big sample as it is in our small sample."

=> hmmm, I don't believe this :-)
}

\summary{savelka2023can}{ - Assessment

Can Generative Pre-trained Transformers (GPT) Pass Assessments in Higher Education Programming Courses?

For students, do stress foundational aspects.

"Yet, it is clear that a straightforward application of these easily accessible models could enable a learner to obtain a non-trivial portion of the overall available score (>55\%) in introductory and intermediate courses alike."

"These findings can be leveraged by instructors wishing to adapt their assessments so that GPT becomes a valuable assistant for a learner as opposed to an end-to-end solution."

"It rather appears that the level of depth in which the coding activities are exposing the topic to a learner matters when it comes to GPT’s ability to generate a correct solution. For example, the OOP activities focus on the use of foundational OOP constructs (e.g., defining a class, implementing a constructor). "
}

\summary{}{\textbf{Ray} Finnie-Ansley et. al. ~\cite{finnie-ansley2023my} use LLM's to solve existing coding assignments in CS2 courses. The models receive higher scores than the average student.}

\summary{}{\textbf{Ray} Savelka et. al.~\cite{savelka2023can} show that LLM's can struggle to answer multiple choice questions about code, especially questions that are more conceptual or questions that ask for the expected output of a code example.}

\summary{}{\textbf{Ray} Piccolo et. al. ~\cite{piccolo2023many} apply ChatGPT to all the programming assignments from an intro to bioinformatics course. They show that the LLM can successfully solve 97\% of the problems. The authors conclude that the models perform so well that bioinformatics students in the near future may no longer need to know how to write code.}

\subsection{Institutional initiatives, policy, and context}

Although some institutions (at least initially) have universally banned the use of LLM tools in student work~\cite{Nolan_2023}, others are starting to embrace them. There is little doubt that this will lead to an array of policies and initiatives that may be at the university, faculty, or class levels. Given that public awareness about LLMs occurred mid-academic year for many institutions (almost universally for North America and Europe in addition to many other parts of the world) this September marks the first academic year where LLMs are a nearly ubiquitous topic. Given that institutions are typically slow to react mid-year -- if they react at all, we have yet to arrive at a steady-state in terms of institutional initiatives. Nonetheless they are beginning to emerge as educators start thinking about the coming academic year. 

Peter Mawhorter's local policy is that LLMs are not allowed in class. In delivering this message to students Mawhorter aims to have a discussion with students on why that policy exists. On the other side of the coin Leo Porter is embracing LLMs and is aiming to use them, for instance, in quizzes. However Porter's institutional challenge is finding and organising enough computer-based testing facilities, and one example of how LLMs can have knock-on effects that could not only affect the course in question, but others via resource allocation and timetabling. Michael Caspersen set out a middle ground, starting with basic competencies and gradually building-in the use of LLMs letting higher levels of the SOLO taxonomy~\cite{biggs2014evaluating} come into focus.

Sven Strickroth (LMU Munich, Germany) noted another challenge that is caused by institutional policy -- significant changes to learning objectives are not easily possible due to the local accreditation cycle which dictates that this occurs only every five years.



\subsection{Other challenges and opportunities}

At the end of the interviews, we asked interviewees for their views on the challenges and opportunities they foresee in terms of the effects that LLMs will present in the introductory programming course. While some of these have already been discussed, several have not - some of these are presented here. 

\subsubsection{Challenges}
\begin{enumerate}
   \item Jan Schneider noted that everyone (including LLMs) have a lot of blind spots, and these can be found by writing -- either in code or in natural language -- like when other people or a LLM try to understand what we wrote, that is when blind spots appear, often in the form of ''\textit{whoa, there's a blind spot... something I didn't know that I was missing... and I have a big fear that if we start using these large language models, we will never acknowledge our blind spots. And we will miss a lot in learning and developing.''} 
    \item Mark Liffiton mentioned that it is a challenge now to make sure students are learning things that they could just have a tool do faster, and the concern of over-reliance -- not learning the things they would have if they did not use the tool to do it for them, noting that it is like cheating in a way -- not learning the things you would if you had done the work yourself. 
    \item Frank Vahid mentioned that ``you need to think''. A human is only useful because they are clever, and thoughtful, and intelligent. That is why we teach them programming -- because it is a way to help them learn to solve problems -- even though we have calculators we still should know how to do arithmetic, despite the existence of calculators. 
    \item Michael Caspersen fears that LLMs will enable disciplined students to become better, but for undisciplined students who are seeking the easy way out -- danger! Caspersen noted that it should not be considered the student's fault for taking the easy way out. It should be turned into a challenge for educators to come up with assessment systems that do not have an easy way out. Caspersen also noted that in many ways LLMs do not add something qualitatively new, but emphasise issues that we have been dealing with for a long time. This was corroborated by Michael K{\"o}lling who stated that it is the scale of these issues exploding that is novel, for instance the illusion of achievement (which is nothing new) and what that could do to learning at scale. K{\"o}lling is also concerned with intellectual laziness noting that learning is a struggle -- learning only happens when you intellectually struggle with something, and if LLMs offer an easy way out, then is learning happening? K{\"o}lling also mentioned cheating as a challenge, but that this is obvious, boring, and solvable. 
\end{enumerate}

\subsubsection{Opportunities}      
\begin{enumerate}
    \item Michael Caspersen sees a big opportunity in terms of rethinking what we are doing. \textit{``We should think deeply about what we actually teaching our students - and LLMs are doing that... [LLMs] radically challenge our reflections on what to teach, what to assess, how to assess. So that’s a great opportunity.''}
    \item Michael K{\"o}lling sees individualised learning in terms of progress, interests, feedback, and help as a big opportunity, noting that humans saw similar issues with books and the printing press. There was concern that people would not need to remember anything any more. However, we lived. ``\textit{In fact books made things better, right?''}
    \item Dan Garcia sees potential in scaling support which benefits educators, institutions and students, positing \textit{``Imagine an LLM that could examine every exam script and where mistakes were made give a whole concept map of what went wrong in the notional machine and where and how. I can try this for a few students but I have over 1,000. This could help scale support for everyone.''}
    
\end{enumerate}

\subsection{Discussion}

Above we have discussed issues raised by the expert instructors that we interviewed. In addition to those, we would like to discuss a certainly non-exhaustive set of issues we believe are important for instructors and CS program designers to consider presently. 

The potential biases reinforced by models trained on large datasets \cite{dignazio2020data} are concerning. 
This could be especially important for instructors who use LLMs to create course materials. 
Another issue is the presence of hidden or implicit learning objectives in existing computing programs -- something Leo Porter described regarding problem decomposition in Section~\ref{sec:canda_objectives}. There is little doubt that there are other such hidden learning objectives spanning not only knowledge, and skills, such as reading and tracing code but also dispositions, inter- and intrapersonal competencies, and other aspects relating to the whole person~\cite{raj2021professional} which the emergence of LLMs might bring to the fore. 
For example, many programs expect their graduates to be comfortable developing large programs using (new) debugging tools, work in a self-directed manner but also perform well in teams, and overcome challenges to pursue their goals long-term. But curricula often do not explicitly include these program-level outcomes in the learning outcomes for a particular course~\cite{becker2019fifty,kiesler_diss_2022}. The same is true for course activities and assessments. As individual instructors respond to the landscape changes induced by LLMs, it becomes more important than ever to consider and implement the constructive alignment~\cite{biggs1996enhancing} of individual course activities and desired program-level outcomes. This is not a new concern for educators, but one that has been amplified by the rapid change in teaching and assessment settings we now see happening (or about to happen). 

Related to this is the potential for LLMs to change the workplaces into which we are graduating students. While a number of researchers have looked at how current developers may use LLM-based tools~\cite{bull2023generative,barke2023grounded,dohmke2023sea} 
and how LLM-tools may be incorporated into professional software development tools~\cite{ernst2022ai-driven}, the participants in these studies have been programmers who learned to code initially without using LLMs. Although professional developers may benefit from AI's human-quality suggestions, novice developers lack the expertise to recognise and understand buggy or non-optimal solutions~\cite{dakhel2022github}. The use of AI tools could become a liability if inexperienced developers fail to remove or correct the tool's incorrect suggestions~\cite{dakhel2022github}. Potentially more dangerous is the fact that students can now execute code that they do not understand, yet designed through natural language prompts. 

It remains to be seen whether students who have LLMs available from the start will have the discipline and dispositions to develop programming competencies deeply or whether this will even matter. Reminiscent of the 1990s discussions of objects-first or objects-later, the opportunity to focus first on top-down design but have a working code for interesting problems completed by the AI is not universally recognised as a positive development. 

LLMs also impact peripheral and applied computing fields. For example, it was shown that LLMs can successfully solve 97\% of the programming problems in a bioinformatics course~\cite{piccolo2023many}. The authors conclude that the models perform so well that bioinformatics students in the near future may no longer need to know how to write (and likely not understand) code. As a consequence, several questions arise on the effects of LLMs not only on other applied areas of computing (e.g., data science; digital forensics; security; and games development) but also on programming as one of the core tiers of every computing degree. 
Related to that are questions about how the introduction of LLMs might affect participation in the computing field. Perhaps reducing the focus on syntax will make the field more attractive to traditionally under-represented audiences and increase retention rates. Additionally the influence of media coverage of computing topics is known to be a large factor in the decisions pre-university students make in terms of what courses to pursue at university. It remains to be seen what effects the intense media hype surrounding LLMs will have on future computing intakes. 

We anticipate more changes in learning objectives, course contents, learning activities, and assessments, which will, in turn, affect whom we teach and why in the (near) future. This might go hand in hand with long overdue changes in computing's signature pedagogy~\cite{shulman2005signature}, and its implementation on the surface, deep, and implicit dimensions. For decades, computing education researchers have presented excellent research on how to better teach our discipline, yet much of this has not made it into practice. We believe that the emergence of LLMs may finally force much-needed (and long ago intended but not yet fully implemented) change.

\vspace*{2mm}
\begin{custombox}{Advice for educators}
\begin{itemize}[leftmargin=1em]
\item Acknowledge the existence of LLMs with your classes regardless of if you embrace them or do not allow their use.
\item Make clear and discuss instutional and class policy, what it allows, what it does not allow, and why it is that way.
\item Assume that students are using LLMs even when not permitted. 
\item Do not underestimate the ability of LLMs to produce solutions to your activities (which may be indistinguishable from student-generated solutions).
\item Consider using an LLM tool to help in generating course materials. If you do this, be aware of possible bias in the output.
\item Reconsider your learning objectives in terms of their relevance to preparing those students who are aiming for careers in the software development industry (which is increasingly making use of LLMs in day-to-day work).
\item Reconsider your learning objectives (e.g., reading and understanding code), learning activities, and assessments to assure your courses remain constructively aligned.
\item Interrogate your learning objectives and ask what might be hidden or implicit and which LLMs might provide a vehicle for more focus. Correspondingly, interrogate your learning outcomes and ask which might be over-emphasised (e.g. code writing) that might need to be balanced with those that LLMs bring to the fore. 
\item Consider using LLMs in your course if only to provide a chance for students to receive more feedback, and practice independently, provided they are equipped to interpret LLM output in a way that facilitates learning. 

\end{itemize}
\end{custombox}

\subsection{Limitations and threats to validity}
We used three sources to build a list of educators to invite for interviews, the SIGCSE mailing list (via a reply to a message about LLMs), the opportunity for those responding to the instructor survey to volunteer to interview, and authors' own networks. Perhaps as a result of this our geographic representation is skewed. As expected the United States forms the bulk (55\%) of responses. Although our interview pool spanned five continents, we had no interviewees from Africa and only one interviewee each from Asia (India), Oceania (New Zealand), and South America (Brazil). Unfortunately, only 14\% (3) of the 22 identified as women.

Additionally, the interviews were semi-structured and although this is a common approach it can impose interviewer bias, although it also designed to result in a coherent set of interviews which focus on common topics while still allowing interviewers to express their own views and experiences. 




Finally, we did not attempt to verify claims that interviewees made about their classes, departments, or institutions, taking interviewee statements at face value.




\summary{alves2023centaur}{
\textbf{Brent} \cite{alves2023centaur}
We introduce the idea of Centaur Programmer, based on the premise that a collaborative approach between humans and AI will be more effective than AI alone, as demonstrated in centaur chess tournaments where mixed teams of humans and AI beat sole computers. The paper introduces several collaboration models for programming alongside an AI, including the guidance model, the sketch model, and the inverted control model, and suggests that universities should prepare future programmers for a more efficient and productive programming environment augmented with AI.

The programmer starts the process, indicating to the AI what the objectives and assumptions are, and the AI responds with solutions. With the help of the programmer, these solutions evolve until they reach the desired result. The final solution may or may not be better than a solution purely imagined by the programmer, but it will probably be achieved much more quickly. For now, we are mainly talking about efficiency gains rather than new solutions. But AI technologies are evolving rapidly, and, in the near future, we may actually have better-designed programs with fewer bugs and more efficient if created in a collaborative model.
In the "sketch model," the programmer outlines the program’s structure and the AI fills in the gaps. 
In the "inverted control model," the communication direction is reversed: the AI asks the programmer what they intend to do until it objectively understands the goal and constraints before implementation.

All of these models require a rethinking of the software development cycle.
}

\summary{becker2023programming}{\textbf{Michelle} Becker et al. \cite{becker2023programming} comes up again in generating materials so this summary is just about their ideas related pedagogy. They discuss the possibility of using LLMs to 1. provide explanations to algorithmic concepts, 2. generate some starter code which students can extend and 3. use to explain error messages. The last point is well covered in another paper by the same authors. The interesting point they hint is that we might need to change our courses to focus on teaching students to evaluate, rewrite, and extend code rather than writing it from scratch.\\ One final point they make related to pedagogy is the potential for students to be overly reliant on the tools and consequently learn LESS.}

\summary{ernst2022ai-driven}{
\textbf{Brent} \cite{ernst2022ai-driven}
AI-Driven Development Is Here: Should You Worry?

AIDE - "AI design environment" TabNine, Kite, SapFix, VSCode Copilot.

"A key developer task then is to carefully distinguish those tasks that are complex, and those that are obvious or complicated. AIDEs can remove the accidental complexity from what are obvious tasks, just like the code showed earlier."

"An important truism in software development is Fred Brook’s maxim “there is no silver bullet,” derived from his insight into essential (inherent) complexity versus accidental (self-imposed) complexity. Like any new approach to our challenging field, AIDE is unlikely to become a panacea for software development. But it does seem to portend an important shift in how we develop software, and just might remove some of the accidental complexity in our projects."
}
\summary{prather2023its}{
\textbf{Brent} \cite{prather2023its}
- time to more seriously study patterns, metacognitive, scripts?
- allow time to overcome initial hurdles
- shepherding, drifting
- do other tools help "exploration" and "acceleration"
- vs Hattie \& Timperley's feedback model: 1. where am I going? 2. how am I going? 3. where to next?

"We find that most students perceived that Copilot would help them write code faster, while also expressing concerns about not understanding the auto-generated code and becoming reliant on the tools – concerns also held by educators [22, 34]. We observed two new interaction patterns. The first was when students guided Copilot by utilising its auto-generated code prompts, shepherding it toward a solution instead of focusing on writing code from scratch and integrating Copilot’s suggestions. The second was when some students were moved along by some of Copilot’s incorrect suggestions, drifting from one to the next and therefore becoming lost. We also observed that students struggled with both cognitive and metacognitive difficulties when using the tool."

"Essentially, Copilot might enable the student to work on a higher level of abstraction where they can spend their mental effort on thinking about the semantics of the program instead of the syntax."

"Others faced the kind of metacognitive difficulties described by Prather et al. like feeling a false sense of accomplishment at having a lot of code, but still being far from a working solution."

"The first of these risks is over-reliance by developers on the generated outputs, which they suggest may particularly affect novice programmers who are learning to code. Indeed, this was the most common concern echoed by the participants in our study when asked to describe their fears and worries around this new technology. Our participants acknowledged that such over-reliance could hinder their own learning, a concern that has also been expressed by computing educators [32]. From the point of view of self-regulation, students will need better self-regulation skills to self-control their use of tools like Copilot to not develop an over-reliance on them – at least when they are freely available for use at the student’s discretion. In fact, we hypothesize that over-reliance on tools like Copilot could possibly worsen a novice’s metacognitive programming skills and behaviours."

"There is a need for systems like Copilot to help the user understand what it's doing and this could be especially effective for novice programmers. Explainable AI (XAI) is the idea that AI systems and the decisions those systems make should be understandable by people. Although XAI has been studied by researchers for nearly 40 years [23], it is increasing in importance as modern machine learning becomes more frequent in our daily lives."

"Therefore, we recommend that systems like Copilot should help users see a little bit into the black box, such as what it is using as input, a confidence value (or visualisation), and its own estimation of the skill level of the user. For example very recent models, notably OpenAI’s ChatGPT10, have begun to present user interfaces that support conversational dialogue and thus are ideally suited to explaining underlying decisions to users."
}

\summary{dakhel2022github}{
\textbf{Brent} \cite{dakhel2022github}
GitHub Copilot AI pair programmer: Asset or Liability?

Teaching requires instruction on "deciphering" and "adapting" suggestions.

"Based on our findings, if Copilot is used by expert developers in software projects, it can become an asset since its suggestions could be comparable to humans’ contributions in terms of quality. However, Copilot can become a liability if it is used by novice developers who may fail to filter its buggy or non-optimal solutions due to a lack of expertise."
}

\summary{leinonen2023using}{\textbf{Michelle} Leinonen et al.~\cite{leinonen2023using} has Codex generate an error message and suggested code fix for 3 programs producing each of 9 common Python error messages at 3 different temperatures (81 examples). They find that only about half of the explanations are correct and that less than half of the suggested code fixes are correct. They conclude that while there is opportunity to further explore variations in their approach (more prompt engineering or one-shot learning), the messages produced by current LLM tools (at that time code-divinci-002) would be too risky to provide automatically (without human intervention) to novice programmers.}


\summary{ahmed2022synshine}{Ahmed et al.~\cite{ahmed2022synshine} introduced SYNSHINE, an error correcting tool incorporated into a free and open-source version of VSCode that automatically repairs Java programs using output (error messages and warnings including error locations) from the javac compiler which can correct 75\% of programs with single errors in the Blackbox dataset including errors that IDEs (Eclipse, IntelliJ, BlueJ, and VSCode) cannot. Specifically, SYNSHINE utilizes a RoBERTa~\cite{liu2019roberta} (BERT-based~\cite{devlin2018bert}) masked-language model (MLM).}

\summary{macneil2023experiences}{\textbf{Brett} MacNeil et al.~\cite{macneil2023experiences} used GPT-3 and Codex to generate three types of code explanations (line-by-line explanations, lists of important concepts, and high-level summaries) from code snippets from an instructor-developed web software development e-book. five explanations were created for each of the three explanation types, for 13 code snippets – a total of 195 code explanations. These were integrated into the e-book. They found that:
\begin{enumerate}
\item students spent more time with longer explanations
\item explanations for the first code snippet in a chapter were the most viewed
\item more complex code snippets received more views than simpler ones
\item line-by-line explanations were viewed more than other explanation
types.
\end{enumerate}
They also found that in general, Codex generated less helpful and more verbose explanations than GPT-3, and that Codex also included code in the explanations which was not desired – the authors only wanted natural language explanations.  St

Students also rated explanations on a 5-point Likert scale and found that explanations matched the code and were useful for learning in general and also that line-by-line explanations were rated as least helpful. They performed a qualitative analysis of low-ranking explanations and found that they:
\begin{enumerate}
\item were overly detailed and focused on mundane aspects of the code
\item were possibly of the wrong type (for example, a concept explanation that read more like a line-by-line explanation)
\item mixed code and explanatory text
\end{enumerate}	 
Although about half of students who used the e-book interacted with the explanations, the authors envision several avenues to increase their utility: 
\begin{enumerate}
\item learnersourcing and continuous improvement of explanations
\item live (on-demand) code explanations
\item integrating LLM-generated code explanations outside the confines of e-books, for example as a browser extension. 
\end{enumerate}	
}








\section{Ethics}
\label{sec:ethics}

Ethics of algorithms (including AI systems such as LLMs) is concerned with the societal context around algorithmic systems and how these systems affect both individuals and society.  Our focus must therefore rest on aspects such as the provenance and quality of training data, the usage of AI systems, and associated costs (in the widest sense), rather than on how to ``integrate ethics into the system'' \cite{bender2021on,johnson2023ethical}.  Moreover, Mittelstadt et al.~\cite{mittelstadt2016the} point out that algorithmic systems tend to be large, complex and highly modularised, which makes it difficult in general to assign responsibilities.  Hence, with an increasingly opaque training procedure of LLMs and a lack of clear responsibility in terms of authorship~\cite{samuelson2020AI_BB}, the use of large language models naturally raises a number of concerns pertaining to academic integrity.  Furthermore, using a large language model has been shown to affect the user's own opinions~\cite{jakesch2023co}.

While a full discussion of the ethics of large language models in computing education is beyond the scope of this paper, we would like to highlight three crucial aspects: the role and stance of us as professionals in computing education, the policies brought forward by academic institutions, and the question of academic integrity in the context of large language models (which we will discuss in Section~\ref{SEC:ACAD-INTEGRITY}).

\subsection{Ethics in LLM literature}


In a study on the values encoded in the ML research literature, Birhane et al.\ identified a number of ethical values~\cite{birhane2022the}, from which the papers would draw motivation.  In a total of 100~highly cited papers, the study found that \emph{performance} was clearly the most frequent value and pointed out that performance is not a neutral term but comes with ethical implications.  For instance, performance is typically measured with respect to specific benchmarks and data sets, thereby introducing bias---particularly when the dataset is thought to represent the ``real world''.

A full study of all values and their ethical implications found in the computing education literature is beyond the scope of this paper.  However, we extracted explicitly stated motivations from the papers in our literature review as a first approximation to a full extraction and coding of values.

In line with Birhane et al.\ we observed that \emph{performance}, \emph{generalisation}, \emph{efficiency}, \emph{novelty}, and \emph{scalability} were often mentioned.  Many papers cited the ``impressive'' or ``human-level'' performance of current large language models.  In contrast to what Birhane et al.\ report, the focus on performance in our dataset seems to be secondary as a means to highlight the timeliness of the research. This is particularly the case since the papers in our study did not propose performance improvements, but rather built on available performance.  Furthermore, while performance in the ML research community typically relates to specific (and often well-known) benchmarks, we found that performance in the papers we reviewed usually referred to either the LLMs' ability to solving exercises, assignments or passing exams, or the LLMs' production of teaching materials, say, such as exercises.

With this focus on application of large language models rather than the design of a new AI system, we also found that several papers pointed out the potential to ``save time'' or help instructors cope with growing class sizes.  While these motivators may be understood as scalability issues, we believe that there is a difference in how this term would be understood in the paper surveyed by Birhane et al.  Despite ostensible similarities concerning the underlying values encoded  in the research literature, there might be some notable differences.

The free availability of large language models (i.e.\ that students can access LLMs at no cost) was a recurring theme in our surveyed literature.  We would like to highlight this notion as an example of an ethically problematic assumption for two reasons.  On the one hand, availability at ``no cost'' is often inaccurate because the costs might be hidden and paid, e.g., through provision of private data.  On the other hand, ChatGPT offers a range of models with different performance characteristics, not all of which are free.  Some students might therefore have access to more powerful tools than others.  Investigating the assumptions, beliefs, and values held by the computing education community itself might therefore be well warranted.  We call on the community to do so in future work.

\subsection{Code of ethics implications}
\label{sec:code-of-ethics}
The IEEE Code of Ethics~\cite{ieeecodeofethics} comprises three sections.  The first focuses on ethical standards, behaviour and conduct.  The second section focuses on ethical treatment of other people, and the third focuses on compliance.   The AAAI Code of Professional Ethics and Conduct~\cite{aaaiethics} was adapted from the ACM Code of Ethics and has the same structure. The ACM Code of Ethics~\cite{acmcodeofethics} comprises four sections. The first section outlines the fundamental ethical principles that all computing professions should use to guide thinking. Section 2 describes the ethical responsibilities of computing professionals, section 3 covers ethical leadership, and section 4 focuses on compliance.  In the following sections, we use the ACM general ethical principles to frame the discussion of ethical issues raised by the use of large language models in computing education. 







\subsubsection*{ACM General Ethical Principles}
In this section we review policies from major universities around the world on the use of large language models in education. As of this writing, many universities do not currently have an official policy publicly available online. Instead, many universities are still presently working through the policy implications of large language models through task forces and other initiatives, such as at the University of Virginia \cite{UVATaskForce}, which is a top-ranked school for computer science in the USA. See Figure \ref{fig:student_responses} for responses to the question ``The policies at my university are clear regarding what is allowed and what is not allowed in terms of using GenAI tools'', which illustrates this point. 

To structure our review, we consider the first part of the ACM General Ethical Principles. Parts two through four of these principles were too specific for most university policies and therefore not as relevant to the present work. Most universities did not explicitly mention coding in their policies, with the exception of Yale University ~\cite{YalePolicy} and University of Adelaide ~\cite{AdelaidePolicy}.

To select universities for consideration of their LLM policies, we first found popular rankings of universities worldwide for computer science programs. Next, we examined policies at these universities, specifically those from Canada, USA, UK, and Australia: University of Toronto~\cite{UTorontoPolicy}, Duke University~\cite{DukePolicy}, Yale University~\cite{YalePolicy}, Massachusetts Institute of Technology (MIT) ~\cite{MITPolicy}, University of California Los Angeles (UCLA) ~\cite{UCLAPolicy}, University of Adelaide~\cite{AdelaidePolicy}, Monash University~\cite{MonashPolicy}, and Oxford Brookes University~\cite{OxfordBrookesPolicy}. We do not consider this a systematic attempt, nor would that be presently possible, given the conditions described above. Instead, this represents a purposeful sampling of top-ranked universities around the world to get an overall picture of how universities are responding to the appearance of LLMs.

The ACM General Ethical Principles are divided into the following sections:

\paragraph{Contribute to society and to human well-being, acknowledging that all people are stakeholders in computing.}
Only MIT suggested that the arrival of LLMs into education is an opportunity to think about student academic well-being ~\cite{MITPolicy}. No universities sampled contained anything about general human well-being or the idea that we are all stakeholders in computing, or by extension, education or society in general. It seems the ideas in these policies are limited to the specific implications of using LLMs, rather than the general. We find this to be an unfortunate oversight because institutions can help guide students' concerns beyond themselves and their immediate circumstances to the broader collective human project of the pursuit of knowledge.

\paragraph{Avoid harm.}
Use of LLMs can lead to poor outcomes due to over-reliance, ease of breaching academic integrity, and use of incorrect information leading to poor products. The universities we sampled all drew attention to the potential to create harm to others, institutions, and even the students themselves. Harm to others could come in the form of using the work of others without citation \cite{DukePolicy}. Harm to institutions can come in the form of students using LLM tools to generate work that is incorrect, offensive, or otherwise inappropriate and therefore dragging the university into potential altercation. It can also call into question the legitimacy of the learning outcomes and verification of them, which can jeopardise accreditation and institutional reputation or the community's trust in that institution. 

Universities seemed most interested in helping students to understand the potential harm that students could incur upon themselves through inappropriate use of AI resources. \citeauthor{isserman2003plagiarism} writes that ``plagiarism isn’t a bad thing simply because it’s an act of intellectual theft --- although it is that. It’s a bad thing because it takes the place of and prevents learning.''~\cite{isserman2003plagiarism}.  One of the most significant concerns is the ease with which LLMs can produce the output that we ask students to produce.  As educators, we are not interested in the output \textit{per se} -- rather we want students to engage in activities and processes that result in learning.  Using a tool to produce the required output circumvents that learning, and deprives students from the opportunity to learn~\cite{isserman2003plagiarism}. Many of the policies we sampled mentioned that students can prevent the growth of critical thinking skills and competencies in crucial areas by taking shortcuts through inappropriate use of these models ~\cite{UCLAPolicy,YalePolicy,MonashPolicy,OxfordBrookesPolicy}. This is particularly important when writing code ~\cite{YalePolicy}. Not only does this harm their short-term ability to pass the course, but it robs them of their long-term ability to master the subject matter and preparedness for future work. Finally, harm could come from inadequate understanding or preparation to self, others, and institutions in courses where safety concerns are paramount, such as a chemistry lab. Skipping crucial learning could harm everyone involved ~\cite{AdelaidePolicy}. While most computing courses do not share similar safety concerns, it is possible in industry that inadequate learning or over-reliance on these tools could expose self, others, and institutions to harm, such as writing malformed code for self-driving cars.

\paragraph{Be honest and trustworthy.}
Most universities that we sampled wanted to make it clear that LLM tools are not reliable and can produce incorrect or fake results. Duke warned faculty and students that LLM ``output is only as good as its input''~\cite{DukePolicy}. Others warned of the now well-known phenomenon of LLM ``hallucination'' where they will reply with incorrect data when not enough is available ~\cite{UCLAPolicy} or create sources and facts even when enough data is available ~\cite{YalePolicy, MonashPolicy, UTorontoPolicy}. It could also be that the data is incorrect simply because it is out of date ~\cite{AdelaidePolicy}, since many LLMs do not have access to data created after they were trained. It is clear that universities are thinking of honesty and trustworthiness in terms of the output of the tool and when evaluated this way LLMs lack credibility.

\paragraph{Be fair and take action not to discriminate.}
Almost all university policies and guidelines made it a priority to discuss biases that exist in AI in general and LLMs in particular. Duke, UCLA, and Oxford Brookes pointed out that the models themselves often discriminate based on their training data, which means it receives all the stereotypes and misinformation from whatever data it is based on ~\cite{DukePolicy,UCLAPolicy,OxfordBrookesPolicy}. Monash University warned faculty and students that LLMs may take data out of context, cannot predict future events with any amount of accuracy, and could present certain sensitive data inappropriately ~\cite{MonashPolicy}. Finally, MIT noted that fairness could be at risk based on who has access to the models ~\cite{MITPolicy}. While access to several popular models is currently free, this may not always be the case, and therefore requiring students to use them could become discriminatory in the future.

\paragraph{Respect the work required to produce new ideas, inventions, creative works, and computing artefacts.}
University policies seemed to mostly skip this criterion. However, there were a few statements related enough to discuss here. For instance, Duke emphasised that creative writing or coding meant doing the hard work of developing each person's specific voice. Using AI tools could undercut that endeavour and cheat the student of their ability to develop innovative ideas, computing or otherwise ~\cite{DukePolicy}. This could be because, as Adelaide noted, AI tools lack originality and common sense ~\cite{AdelaidePolicy}. Related to new ideas and innovation, UCLA noted in their policy that AI tools will often reflect outdated information that may fail to represent the progress of social movements since the training of the model was completed ~\cite{UCLAPolicy}. If certain sets of rights were not legal when the model was trained, one should not expect the model to suggest that they should be.

\paragraph{Respect privacy.}
The lack of control over these tools, as well as what data they keep about their users, was mentioned in most policies. AI tools could invade users' privacy ~\cite{DukePolicy, MITPolicy}, violate FERPA (a student privacy protection law in the United States) if student records are handed to it \cite{UCLAPolicy}, do not assume that users are at least 18 years old ~\cite{MonashPolicy}, and are not bound by university ethics rules and policies. Furthermore, AI tools will often take user data and use it to train their models, whether users want that or not. Monash encouraged students and faculty to consider that their data could be stored by the model and used in other contexts ~\cite{MonashPolicy}.

\paragraph{Honor confidentiality.}
Ethical issues identified by universities included threats to confidentiality, though each one took a slightly different approach. Creating an account and using AI tools could bother students who are worried about the models stealing their intellectual property and may therefore be unwilling to use them ~\cite{UCLAPolicy}. The tools also do not respect confidentiality of the people from whom they have taken the data to train the models, which could result in unintentional plagiarism for both students and faculty \cite{MonashPolicy}. Finally, these models do not respect confidentiality with regard to legal issues and data sent to them may be turned over to law enforcement agencies or other third party vendors and affiliates without user consent \cite{MITPolicy}.

\section{Academic Integrity Implications}
\label{SEC:ACAD-INTEGRITY}

Recent work on student use of generative AI tools has raised alarms about academic integrity violations~\cite{prather2023its}. \citeauthor{jones2008cyber} summarises several common practices that are deemed to be cheating, distinguishing between plagiarism, collusion, and falsification~\cite{jones2008cyber}.   We add contract cheating (as described by Deakin University~\cite{deakinpolicy}), and use of unauthorised resources (as described by University of Auckland~\cite{aucklandpolicy}) to this list of practices that breach academic integrity. These terms are described by the respective documents as:
\begin{description}
\item [Plagiarism:] A student incorporates another person’s or body’s
work by unacknowledged quotation, paraphrase, imitation or
other device in any work submitted for assessment in a way that
suggests that it is the student’s original work~\cite{jones2008cyber}.
\item [Collusion:] The collaboration without official approval between
two or more students (or between student(s) and another
person(s)) in the presentation of work which is submitted as the
work of a single student; or where a student(s) allows or permits
their work to be incorporated in, or represented as, the work of
another student~\cite{jones2008cyber}.
\item[Contract cheating:] A student requests another person or service (including, according to Deakin University, artificial intelligence content production tools) to produce or complete all or part of an assessment task to submit as their own work~\cite{deakinpolicy}.
\item [Falsification:] Where the content of any assessed work has been
invented or falsely presented by the student~\cite{jones2008cyber}.
\item[Unauthorised resources:] Using software, websites, materials or devices not explicitly permitted~\cite{aucklandpolicy}.
\end{description}

We discuss each of these academic integrity concerns with respect to generative AI.  

\subsection{Plagiarism}
Plagiarism is the use of the work of others without appropriate attribution.  This raises the issue of who is the author of work created by generative AI.  There are several possibilities:
\begin{enumerate}
\item The community that produced the source content used as input to the generative AI model is the author of the work.
\item The generative AI software is the author of the work.
\item The user of the generative AI software is the author of the work.
\end{enumerate}
Although some in the literature treat the use of AI tools as plagiarism ~\cite{orenstrakh2023detecting}, we argue that although the community providing source material has influenced the generated content, this is similar to the natural process of writing in which authors read source material and use the information to generate new content, based on existing literature.  Generative AI is almost always creating content \emph{based on} the training data.  In this case, the community has not authored the work generated by the model, so using AI generated content would not be considered plagiarism of the original authors of work that was used as input to the generative AI model.  In some rare cases, GenAI tools can generate the work of someone else exactly, which would in fact be plagiarism.  Since this is extremely rare, we do not consider it here other than to acknowledge it.


Although it may be tempting to consider generative AI to be the author of the work in all cases, academic publishers take an opposing view.  Examples of statements include:
\begin{quote}
    ``AI tools cannot meet the requirements for authorship as they cannot take responsibility for the submitted work. \ldots ~Authors are fully responsible for the content of their manuscript, even those parts produced by an AI tool, and are thus liable for any breach of publication ethics.'' (Committee on Publication Ethics)~\cite{copeposition}
\end{quote}
\begin{quote}
    ``AI does not meet the Cambridge requirements for authorship, given the need for accountability. AI and LLM tools may not be listed as an author on any scholarly work published by Cambridge.'' (Cambridge University Press)~\cite{cambridgepolicy}
\end{quote}
\begin{quote}
    ``Authors should not list AI and AI-assisted technologies as an author or co-author, nor cite AI as an author. Authorship implies responsibilities and tasks that can only be attributed to and performed by humans.'' (Elsevier)~\cite{elsevierpolicy}
\end{quote}
\begin{quote}
    ``Artificial Intelligence Generated Content (AIGC) tools --- such as ChatGPT and others based on large language models (LLMs) --- cannot be considered capable of initiating an original piece of research without direction by human authors. \ldots --- these tools cannot fulfil the role of, nor be listed as, an author of an article.'' (Wiley)~\cite{wileypolicy}
\end{quote}
\begin{quote}
    ``Generative AI tools and technologies, such as ChatGPT, may not be listed as authors of an ACM published Work.'' (ACM)~\cite{acmpolicy}.
\end{quote}

Our position is aligned with those of publishers that the \textit{user} of generative AI tools should be considered the author of the work.  This is consistent with the view of Pamela Samuelson, who states ``The pragmatic answer to the AI authorship puzzle, \ldots, the [author is the] user who is responsible for generating the outputs. If anyone needs to be designated as owner of rights in the outputs, it should be the user.'' \cite{samuelson2020AI_BB}
As such, we do not believe that use of AI-generated content by students should be considered \textit{plagiarism}, and should not be referenced or cited as an independently authored piece of work.  The student using the generative AI tool should be treated as the author of the work.

It should be noted that, in the apparent effort to combat plagiarism, there have been numerous tools that attempt to detect AI-generated material, such as CopyLeaks, GPTKit, GLTR, and GPTZero. However, given the probabilistic nature of generative AI that is used to both generate the assignment in question and check the assignment, these tools are unreliable and produce many false positives~\cite{orenstrakh2023detecting}. Orenstrakh et al. found that these detectors are even worse when evaluating code~\cite{orenstrakh2023detecting}.

\subsection{Collusion}
Collusion occurs when a student works together with another person to create work that they subsequently claim as their own.  This requires both parties to willingly agree to work together and therefore assumes that both parties have agency.  As generative AI has no agency, we do not consider a student who submits generated content to be engaged in collusion.

\subsection{Contract cheating}
Contract cheating is traditionally described as a student requesting another person to produce work that they submit as their own.  The definition by Deakin University extends this view of contract cheating to explicitly include the use of generative AI tools~\cite{deakinpolicy} as do some other universities ~\cite{MonashPolicy}.  However, we disagree with this position, as it contradicts the view of the user of generative AI as the author --- the position taken by academic publishers.  

We recognise that generative AI models are capable of generating content that is more extensive than other software tools, but as a matter of principle, we consider the user to be the author (as discussed previously).  generative AI is a tool which may be used by a student to produce work, much as calculators and other software tools are used.  We therefore do not believe that use of generative AI software should be treated as contract cheating.

\subsection{Falsification}
Falsification occurs when a student invents or misrepresents data or results.  The possibility that generative AI invents ``facts'' is well-known, and typically described by the term \textit{hallucination}~\cite{ji2023survey}. Many university policies recognise that AI tools like LLMs can produce incorrect data and facts, which we discuss in Section \ref{sec:code-of-ethics} under the ACM requirement to "be honest and trustworthy."  The view of publishers is that an author is responsible for the accuracy of the content in their work.  Examples of policy from publishers include:
\begin{quote}
    ``Authors are fully responsible for the content of their manuscript, even those parts produced by an AI tool, and are thus liable for any breach of publication ethics.'' (COPE)~\cite{copeposition}
\end{quote}
\begin{quote}
    ``Where authors use generative AI and AI-assisted technologies in the writing process, these technologies should only be used to improve readability and language of the work. Applying the technology should be done with human oversight and control and authors should carefully review and edit the result, because AI can generate authoritative-sounding output that can be incorrect, incomplete or biased. The authors are ultimately responsible and accountable for the contents of the work.'' (Elsevier)~\cite{elsevierpolicy}
\end{quote}
\begin{quote}
    ``The author is fully responsible for the accuracy of any information provided by the [generative AI] tool and for correctly referencing any supporting work on which that information depends.'' (Wiley)~\cite{wileypolicy}
\end{quote}
\begin{quote}
    ``Authors are accountable for the accuracy, integrity and originality of their research papers, including for any use of AI.'' (Cambridge)~\cite{cambridgepolicy}
\end{quote}

We take the position that students who use generative AI are responsible for the content they include in their work.  Inaccuracies, citations for non-existent papers, and other hallucinations that may arise from use of generative AI are the responsibility of the student author.  Therefore, the category of \textit{Falsification} is a relevant academic integrity issue for students using generative AI. As discussed above, this could cause harm to students, anyone publishing their work, and to the university as a whole.

\subsection{Use of unauthorised resources}
In an educational context, students are required to engage in tasks that may have specific constraints.  For example, the use of calculators in general is acceptable, and we are comfortable with the notion that using a calculator for data analysis does not impact authorship, or breach any notion of academic integrity.  However a student in a calculus class may be asked to solve a differential equation without the use of a calculator, and access to calculators (or types of calculations) may be restricted in secure assessments such as exams.  Such constraints are typically imposed to ensure that learning outcomes are met (e.g., that the student can solve a differential equation without outside assistance).  A student who used a calculator for a given assessment when it was not permitted  would be an issue for academic integrity because the student uses unauthorised resources.

Our position is that this is the appropriate category for use of generative AI in computing education.  Figure \ref{fig:student_responses} shows results to various ethical questions where it seems that many instructors and students agree that using GenAI tools to create an entire answer is wrong.  In some courses, such as introductory programming, the use of generative AI can be undesirable as it can solve problems with minimal intellectual input from students.  Not only does this violate academic integrity, but as discussed above, students who utilise these resources inappropriately in lower division courses may cause harm to themselves by preventing their own preparation for upper division coursework.  However, in upper division courses it may be an appropriate productivity tool that students would be permitted to use, which the survey data from Figure \ref{fig:student_responses} seems to corroborate given responses to questions about generating pieces of an assignment, help with style, or fixing bugs.  

However, this requires teaching staff to be explicit in their course syllabus, or assessment description, about which resources students are permitted to use.



 


\vspace*{2mm}
\begin{custombox}{Advice for educators}
\begin{itemize}[leftmargin=1em]
\item We encourage educators to teach students about appropriate ethical use of generative AI throughout the curriculum, and to allow the use of such tools where it is pedagogically appropriate. See Appendix \ref{appendix:student} for an example.
\item When educators assign assessed work for students, any restrictions on the use of tools such as generative AI should be explicitly stated. 
\item Students who use generative AI to complete assessed work should be required to include a statement about how it was used, consistent with academic publication requirements.
\item Students who use generative AI when they are not permitted, or in ways that are restricted, are engaged in misconduct by using unauthorised resources. Academic consequences as a result of this behaviour should be made clear in the course syllabus.
\end{itemize}
\end{custombox}


\subsection{Advice for students}
\citet{simon2018informing} highlights the importance of educating students about academic integrity, and revealed a wide variety of ways that academic integrity is communicated to students.  
Given the disruptive nature of generative AI, we recommend that a guide for students is developed by teachers and distributed to students to provide explicit advice about appropriate use of generative AI tools.  

We recommend that any guidelines developed for students should:
\begin{itemize}
    \item adopt professional practices and standards where possible; and
    \item adapt professional practices where needed to ensure good pedagogical practices are maintained.
\end{itemize}

Publishers typically require acknowledgement where generative AI has been used in development of a manuscript.  We believe this would be useful for teachers, and for students, to reflect on how generative AI was used, so we recommend that assessments require students to include a statement about how generative AI was used in assessment tasks (where permitted).

After considering the relevant findings from this report, we have developed a resource that provides guidance about the use of generative AI for students.   It is by no means comprehensive or complete and reflects our perspectives on what students should know before using these tools. 
Our recommendations are informed by the risks identified in the literature, the ACM code of ethics, survey results, and the academic integrity documents that we analysed.  We offer this to teachers as a resource that may be adapted and/or distributed to students. 

\vspace*{2mm}
\begin{custombox}{Guide for students}
See Appendix~\ref{appendix:student} for a sample handout or text that could be adapted and included in a course syllabus.
\end{custombox}
\section{Benchmarking Large Language Models for Computing Education}
\label{sec:benchmarks}

In this section we focus on the performance of LLMs in the context of computing education. Teachers are interested in how good LLMs actually are in conducting tasks such as solving programming problems, explaining code, generating test questions, and providing feedback to students. 
However, the speed at which new models arrive and old models are deprecated is staggering. As we demonstrate, the currently published literature may underestimate what the newest models, such as GPT-4, can do. Some recent work has found that GPT-4 outperforms earlier models in tasks such as visual programming~\cite{singla2023evaluating}, Socratic questioning of novices~\cite{al2023socratic}, solving multiple-choice questions and programming exercises~\cite{savelka2023thrilled}, and that performance can be close to human tutors for some tasks~\cite{phung2023generative} while not for others~\cite{al2023socratic}.

In addition to being interested in how much the capabilities of LLMs in computing education related tasks have increased, we are interested in analysing the suitability of existing benchmarks for computing education as they might not translate to computing education settings. For instance, we expect that many students will be able to fix small mistakes made by LLMs themselves. Similarly, tasks in existing benchmarks might not match those typically found in computing education courses.

Another issue with current papers lies in our ability to validate results. A wide variety of parameters, prompts, and evaluation approaches have been used, and they are not always reported in detail. Furthermore, a slight variation in a prompt might generate quite different results. In this section we explore how we assess LLMs in the computing education context. We choose to focus on the task of generating a solution to a programming problem, because this is a major task for students and is a focus of existing literature.

First, we review datasets that are available for evaluating LLMs and tag problems in several of those datasets to assess where they fit in the context of computing education. Second, we take one of the first papers on LLMs that appeared in the computing education context~\cite{finnieansley2022robots} and replicate it using multiple different more recent models. In the replication, we use the state-of-the-art GPT-4 model to give insight into how rapidly the performance of models has increased, the GPT-3.5-turbo model that powers the free version of ChatGPT which many students are likely to use, and GitHub Copilot which is free for students and educators and can be used as a plugin for popular IDEs. In addition, we openly release the problem descriptions and test cases for the dataset used in the original study~\cite{finnieansley2022robots} and our replication to facilitate future replication.\footnote{The data can be found here: \url{https://osf.io/bu9h3/?view_only=a16f3e474be94188b884aa0dca02041f}} Finally, we report on our experiences running two analyses using openly available datasets, revealing the difficulties we encountered and their possible effects on results.




\subsection{Review of empirical datasets}

\begin{table*}
 \begin{center}
  	\caption{Prominent datasets available to evaluate LLM systems and tools.}
	\label{tab:datasets}
\rotatebox{-90}{
  \begin{tabular}{|p{2.5cm}|p{4.5cm}|l|l|l|l|p{7cm}|}
\hline
Name & Description & Language & Size & Testcases & Solutions & Link\\\hline
Mostly Basic Python Programs (MBPP)~\cite{austin2021program} & Natural language descriptions of introductory problems & Python & \~1000 & Yes & Yes & \url{https://paperswithcode.com/dataset/mbpp}\\\hline
Search-Based Pseudocode-to-Code (SPOC)~\cite{kulal2019spoc} & Pseudocode descriptions of coding problems & C++ & 18356 & Yes & No & \url{https://paperswithcode.com/dataset/spoc}\\\hline
Blackbox~\cite{brown2014blackbox,brown2018blackbox} & Traces of editing and IDE interactions & Java & - & No & No & \url{https://bluej.org/blackbox/}\\\hline
Deepfix~\cite{gupta2017deepfix} & Student-generated code with syntax errors  & C & 6922 & N/A & N/A & \url{https://paperswithcode.com/dataset/deepfix}\\\hline
Automated Programming Progress Standard (APPS)~\cite{hendrycksapps2021} & Natural language descriptions of problems of various difficulties & Various & 10000 & Yes & Yes & \url{https://paperswithcode.com/dataset/apps}\\\hline
HumanEval~\cite{chen2021evaluating} & Docstring descriptions of (mostly introductory) programming problems & Python & 164 & Yes & No & \url{https://paperswithcode.com/dataset/humaneval}\\\hline
Grounded CoPilot~\cite{barke2023grounded} & Programming tasks provided in an observation study & Python or Rust & 4 & No & No & \url{https://github.com/michaelbjames/copilot-study}\\\hline
PyFiXV CodeForce~\cite{phung2023generating} & Buggy submissions to programming contest problems for program repair & Python & 240 & No$^\dag$ & No & \url{https://github.com/machine-teaching-group/ edm2023_PyFiXV/tree/master}\\\hline
ChatGPT\_ Bioinformatics~\cite{piccolo2023many} & Natural language descriptions of introductory bioinformatics problems & Python 
& 184 & No & Partial & \url{https://github.com/srp33/ChatGPT_Bioinformatics}\\\hline
EvalPlus~\cite{liu2023your} & Re-release of the HumanEval code-generation dataset with additional testcases & Python & 164 & Yes & No & \url{https://github.com/evalplus/evalplus}\\\hline
LeetCode Assistant & LeetCode problem prompts, responses generated by LLMs, and attempts to repair buggy prompts & Python & 1209 & No$^\dag$ & No & \url{https://zenodo.org/record/7792965\#.ZCvyv-xBwUE}\\\hline
StudentEval~\cite{babe2023studenteval} & Student-generated prompts for introductory programming problems & Python & 1749 
& Yes & Yes & \url{https://huggingface.co/datasets/wellesley-easel/StudentEval}\\\hline
FalconCode~\cite{defreitas2023falconcode} & Natural language descriptions of introductory programming problems & Python & 661 & Yes & Yes & \url{https://falconcode.dfcs-cloud.net/index.php}\\\hline
CS1QA~\cite{lee2022cs1qa} & Naturally-occurring questions asked by students to LLMs with the LLM response & Python & 17698 
& N/A & N/A & \url{https://github.com/cyoon47/CS1QA}\\\hline
AD2022~\cite{petersen2022dataset} & Problems with a data structures course with student solutions & Python & 16 & Yes & Yes & \url{https://www.inf.uni-hamburg.de/en/inst/ab/lt/resources/data/ad-lrec}\\\hline
Digital TA~\cite{demidova2023dataset} & Small exercises generated by a digital TA and associated solutions & Python & 11 
& No & Yes & \url{https://doi.org/10.5281/zenodo.7799971}\\\hline
Socratic Questioning~\cite{al2023socratic} & Socratic dialogues with a human instructor & Python & 86 & N/A & N/A & \url{https://aclanthology.org/2023.bea-1.57/}\\\hline
\multicolumn{7}{l}{$\dag$ While testcases are not provided, the dataset consists of publicly available contest problems which have automated testing.}\\
        \end{tabular}
} 
    \end{center}
\end{table*}

Table~\ref{tab:datasets} presents a set of openly accessible datasets that have been or could be used to investigate questions about programming exercises or tasks in computing education contexts. To obtain the datasets in this table,  we reviewed all of the papers in our literature review (Tables~\ref{tab:ref-papers} and~\ref{tab:lit}) to identify any data that they used. We do not claim that this list is exhaustive, but it reflects the datasets in use when we conducted our literature review. In addition to these datasets, we are aware of one other attempt to review existing benchmarks for a particular task that might be performed by LLMs: natural language to code generation~\cite{zan2023large}. We believe the datasets listed in Table~\ref{tab:datasets} represents a more broad set of applications and illustrates the kinds of questions being investigated and the breadth of educational contexts being examined.

A review of Table~\ref{tab:datasets} suggests a number of limitations in the data available for pursuing LLM research in an educational context. Most of the datasets contain exercises (described in more or less structured natural language). This reflects a focus on the question of whether LLMs can solve typical programming problems. In contrast, relatively few datasets contain chat logs where students interact with an LLM or student-submitted code with syntax errors (for code repair tasks), but additional publicly available data for questions beyond code-generation would be beneficial, as that would allow researchers at small educational contexts, where collecting sufficient data may be difficult, to engage in LLM work~\cite{kiesler2022delfi,kiesler2023opendataiticse}. 

Even for code generation tasks, more data would be beneficial. Almost all of the datasets focus on Python (e.g., providing docstrings as input, Python starter or solution code, or student chats featuring Python code), with a smaller number featuring C/C++. Relatively little public data appears to be available for other languages. Also, as described in the next section, many of the available datasets focus on small exercises used in introductory programming, with relatively little data available to examine larger programming problems or content for more advanced courses. 

Finally, as previously identified by \citet{liu2023your}, many of the published datasets do not provide robust evaluation of the exercises they include. \citet{liu2023your} provide the EvalPlus dataset, which enhances previously published datasets with additional test cases; they found that the limited tests available meant that incorrect ``solutions'' were accepted as passing. We also found evaluation to be limited, with some datasets requiring manual intervention to complete evaluation. We provide more detail on issues we encountered when using these datasets in Section~\ref{sec:novel_analysis}.


\subsection{Problem context}

\begin{table}
 \begin{center}
  	\caption{Fraction of instructor-assigned tags (Introductory (Intro), ObjectOriented (OO), or DataStructures (DS)) assigned to exercises in various datasets.}
	\label{tab:dataset-tagging}
  \begin{tabular}{l|ccc}
  \toprule
  & Intro & OO & DS\\\midrule
  HumanEval~\cite{chen2021evaluating} & 98.8\% & 0 & 1.2\%\\
  FalconCode~\cite{defreitas2023falconcode} & 100\% & 0 & 0\\
  My AI CS2~\cite{finnieansley2023my} & 83.3\% & 9.3\% & 7.4\%\\
  \bottomrule
        \end{tabular}
    \end{center}
\end{table}

To categorize the types of exercises present in the datasets, we manually tagged two prominent datasets, HumanEval~\cite{chen2021evaluating} and FalconCode~\cite{defreitas2023falconcode}. The datasets were tagged by a single author, an experienced instructor who has taught introductory programming, data structures, and advanced systems programming. The author tagged the exercises as being suitable for (a) an introductory programming course (Intro), as they use builtin data structures like strings and do not introduce complex algorithmic logic; (b) an introductory course using classes (OO), as they introduce classes or methods; or (c) a data structures or algorithms course, as they use some abstract data types (e.g., trees, queues, graphs) or complex algorithmic logic (e.g., subsequence matching, linear programming). The results are presented in Table~\ref{tab:dataset-tagging}. We found that in these two datasets, the vast majority of problems only cover material suitable in an introductory programming setting. 

Many of the other datasets in Table~\ref{tab:datasets} are similar to the two datasets we analysed, in that they appear to focus on introductory material. We requested access to the data used in~\citet{finnieansley2023my}'s paper, as the topic was a more advanced (CS2) course. This time, the dataset was tagged by two authors (one who had tagged the previously discussed datasets and a second experienced instructor); we calculated Cohen's kappa and found near perfect agreement (0.94). Again, we found that the majority of the content was primarily suitable for an introductory course, with relatively few questions asking about object-oriented code or data structures. In addition, while reviewing the problems, we found that almost all were examples of small exercises, with relatively few requiring multiple functions to solve. The FalconCode~\cite{defreitas2023falconcode} dataset includes a few exceptions to this general trend.

Finally, we examined the Automated Programming Progress Standard (APPS) dataset~\cite{hendrycksapps2021}, as it explicitly advertises that it includes exercises of various difficulties. This is a large set (10000 problems), so we manually tagged a sample of 200 exercises and then used keyword searching to identify the usage of classes and common data structures. This means that our estimates will underestimate the complexity of the exercises. Many of the problems in this set \textit{are} more complex, as expected as they were largely drawn from programming contests. However, they may not be problems typically seen in educational contexts. The instructor reviewing the problems would not use many of these problems in any course, as they introduce issues like floating point error, exceeding the maximum representable integer, or linear-time pattern searching. At the same time, relatively few explicitly reference OO topics (4.1\%) or common data structures (6.6\%). 

Taken together, this analysis suggests that many of the available datasets -- and as a result, the published results -- reflect an early introductory programming context (CS1) with relatively few examples of common CS2 material (\citet{cipriano2023gpt} is a recent exception demonstrating research on OO topics) and even fewer covering any more advanced topics. Some datasets do include more challenging tasks, but these may not reflect the kinds of problems student programmers solve in upper year courses, as they appear to be inspired by (or were drawn directly from) programming contest sites.

\subsection{Replication: The robots are coming}
\label{sec:robots-replication}

\citet{finnieansley2022robots} published the first paper that examined the performance of large language models in solving introductory programming exercises. They found that Codex\footnote{More specifically, the first version of the model `code-davinci-001'.} had better performance than the median student on the same exam exercises. In order to understand how the performance of large language models in this task has improved in the past two years, we partially replicated their study.

\textbf{Method:} We contacted the authors of the original study and received the problem descriptions and test cases used in the study. The original study had a total of 30 exercises; 23 used in two exams and seven variants of the Rainfall-problem~\cite{fisler2014recurring}. In the original study, the method used for the Rainfall-problems differed from the method used for the exam questions. In our replication, we follow the method originally used for the exam problems for both the exam and the Rainfall problems. We generate up to ten solutions for each problem, stopping if the LLM creates a solution that passes the tests, or includes a ``trivial formatting error''. As in the original article, we manually fixed the trivial errors and considered this step as an additional trial. All models were prompted with the problem description surrounded by triple-quotes as was done in the original~study.

We evaluate three models in our replication: GPT-4, GPT-3.5-turbo, and GitHub Copilot. At the time of writing, GPT-4 is the state-of-the-art LLM, so it is interesting to see how it performs on solving the problems. GPT-4 also powers the paid version of ChatGPT. GPT-3.5-turbo is the model that powers the free version of ChatGPT, which is likely what many students will be using, and thus it is also interesting to include. Lastly, GitHub Copilot is free for students and educators, and is directly embedded into the IDE, so it is possible that students will be using it too, which is why we decided to include it in the replication. 

Copilot works slightly differently to the other two models, as it cannot be directly `prompted'. For evaluating Copilot, we used the Visual Studio Code Copilot plugin, and provided the prompt (problem description surrounded by triple-quotes) in an empty file. We would then wait for Copilot to provide a suggested completion, and would accept the first suggestion that Copilot provided. In the rare cases where no suggestion was provided based on just the problem description, we would write `def' to start the function, after which Copilot would suggest code if it had not before.

\begin{figure*}[t!]
    \centering
    \begin{subfigure}[t]{0.5\textwidth}
        \centering
        \includegraphics[height=2.0in]{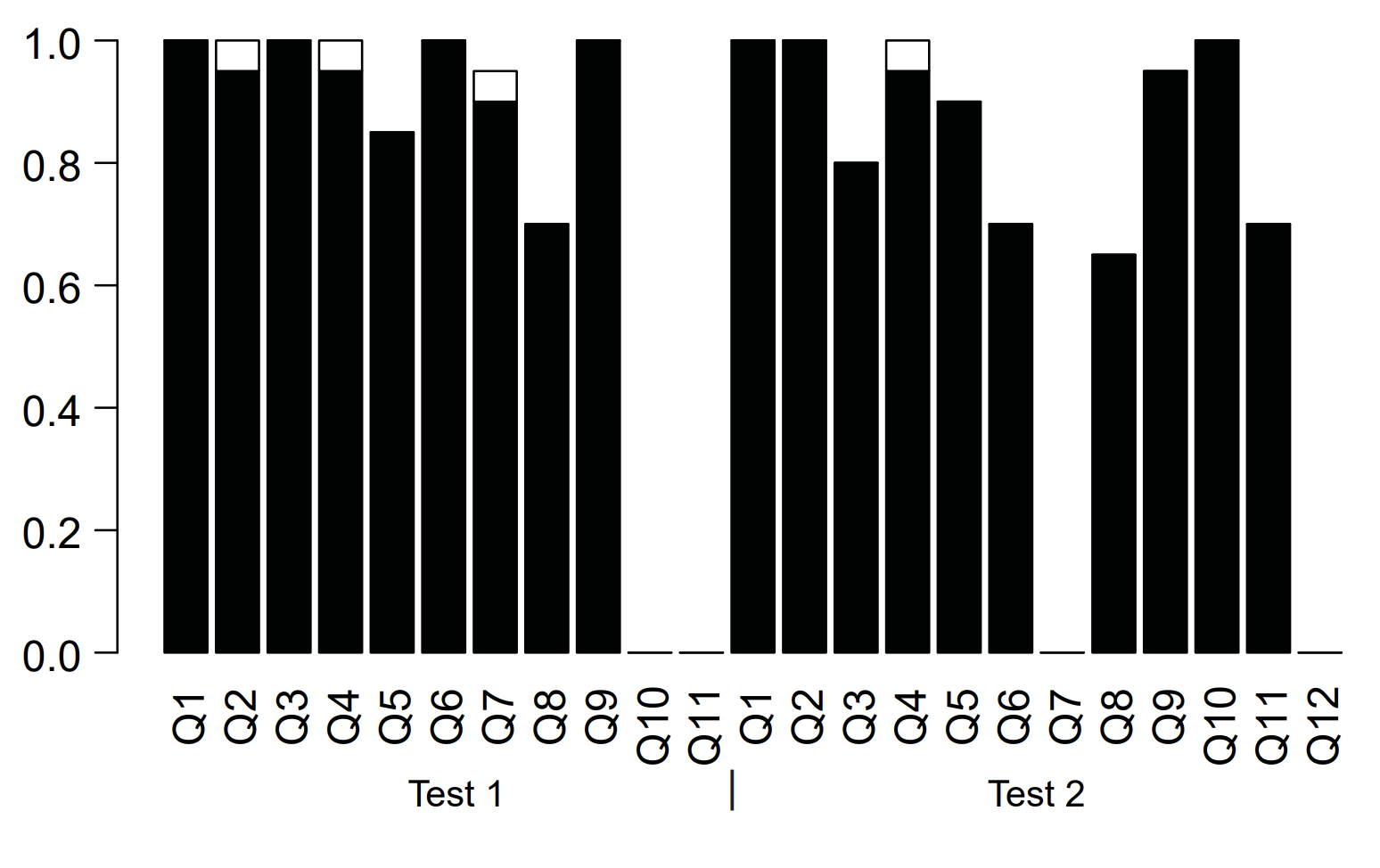}
        \caption{Results of the original ``Robots Are Coming'' paper that used Codex~\cite{finnieansley2022robots}.}
    \end{subfigure}%
    ~ 
    \begin{subfigure}[t]{0.5\textwidth}
        \centering
        \includegraphics[height=2.0in]{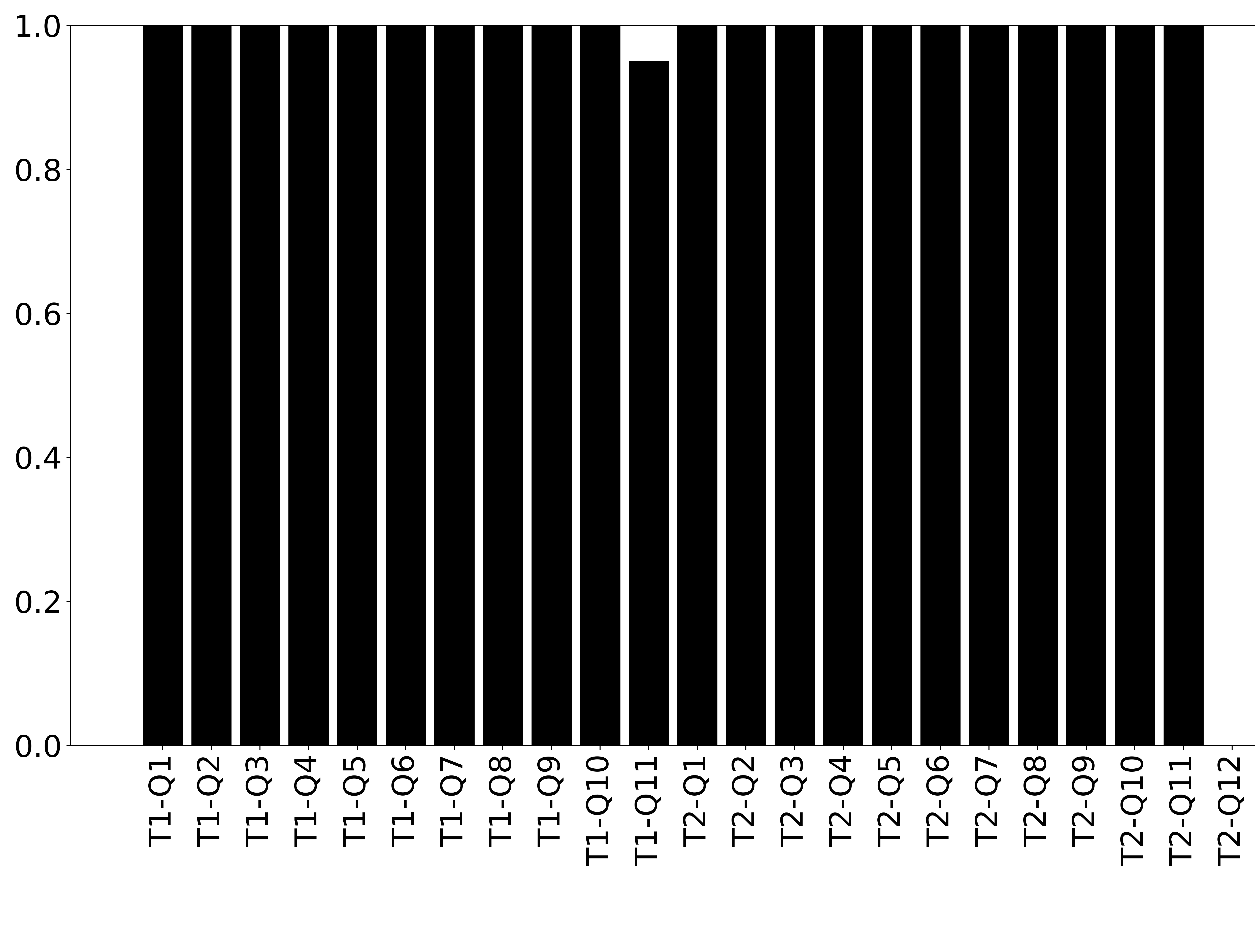}
        \caption{Results of our replication of~\cite{finnieansley2022robots} with GPT-4.}
    \end{subfigure}
    \caption{A comparison of the original results and the score achieved by GPT-4 on the two CS1 tests and Rainfall-problem variants presented in~\cite{finnieansley2022robots}.
\label{fig:replication_results}}
\end{figure*}

\textbf{Results (GPT-4):} For GPT-4, we used the temperature value of 0.9 similar to the original study. The results of the replication are presented in Figure~\ref{fig:replication_results} (only GPT-4) and Table~\ref{tab:replication_results} (all three LLMs). It is clear that GPT-4 outperforms Codex, which is the model used in the original study. All problems were solved in under ten attempts, except for the last question in the second test (T2-Q12). As noted in the original study, some of the Rainfall variants had somewhat vague problem descriptions, leading to GPT-4 having more trivial formatting issues with the outputs. In the original study, the results of Codex were similar to students in the top quartile; in our replication, GPT-4 would have been one of the top students in the class. The only problem that GPT-4 was not able to solve was the last problem of the second test (T2-Q12), which involved drawing bar graphs using text where the shape of the bar graph was determined by values in a dict passed to the function.

\textbf{Results (GPT-3.5):} Similar to GPT-4, we used the temperature value of 0.9 for GPT-3.5. GPT-3.5 performed only slightly worse compared to GPT-4. It was able to solve most problems on the first try, although many attempts included trivial formatting issues. For example, inadvertently leaving out a period at the end of a printout such as having ``print("The sum is", sum)'' when the tests expected a period at the end of the string. Compared to GPT-4, GPT-3.5 was unable to solve one of the Rainfall variants, specifically the one from Simon. Looking into why GPT-3.5 struggled, the biggest issue for the model was that the problem description stated that ``A day with negative rainfall is still counted as a day, but with a rainfall of zero.'' This was not taken into account in the code that GPT-3.5 generated as the code for all ten completions would simply ignore any days with negative rainfall values. Similar to GPT-4, GPT-3.5 was not able to solve the last question of the second test (T2-Q12).

\textbf{Results (Copilot):} Copilot performed the worst out of the three evaluated models, successfully solving 20 out of the 23 exam problems and four out of the seven rainfall variants. We looked into the issues in code for the problems that Copilot was unable to solve. T1-Q10 involved printing all words in a given sentence that start with a given character. The problem description explicitly stated that ``The sentence will end with a full-stop.'' The code generated by Copilot would not remove the full-stop from the end of the sentence, resulting in failure in the edge case when the last word of the sentence needed to be printed where the tests assumed the full-stop is not included in the word, but Copilot did not take this into account. T1-Q11 involved sorting four numbers given as a parameter using only the ``min()'' and ``max()'' functions (using of lists, if/elif/else, and loops was forbidden). While Copilot took the constraints into account, all ten completions had logical flaws and did not pass all the tests. Finally, similar to GPT-4 and GPT-3.5, Copilot was unable to solve T2-Q12 which involved printing bar graphs made of text. For T2-Q12, many completions from Copilot would not compile (the completion would be incomplete), and when it did, the code was nowhere near correct. For example, many of the completions would always print the same bar graph regardless of input. 

For the rainfall variants, Copilot was unable to correctly solve the one from Soloway, the one from Simon, and the one from Guzdial. For the variant from Simon, the issue was the same as for GPT-3.5 in that negative values were always ignored entirely, even though the problem description asked for these to be counted as days with a rainfall of 0. For both the Soloway variant and the variant from Guzdial, none of the completions generated by Copilot handled the edge case where the list is empty, leading to division by zero.

One interesting finding was that Copilot would sometimes write just the function signature with a comment such as ``\# write your code here'' followed by ``pass'', or the completion would not have any code at all but only suggestions/hints for how to start work on the problem as comments.

\textbf{Discussion:} Overall, all models performed quite well, having performance that would have allowed them to pass the exams. Unsurprisingly, GPT-4 outperformed GPT-3.5 and Copilot, which corroborates previous work where GPT-4 has outperformed other LLMs for various tasks~\cite{singla2023evaluating,al2023socratic,savelka2023thrilled,phung2023generative}. The problems where any model struggled were either complex (e.g., involving printing bar graphs as text) or had vague problem descriptions, leading to some edge cases being ignored (e.g., rainfall variants that did not specify what should happen when the list is empty). As noted in the original study, LLMs somewhat struggle with trivial formatting issues, such as missing some formatting or having extra printouts not asked for in the problem description. We noticed that there was less variation in the completions generated by Copilot compared to the ones from GPT-4 and GPT-3.5. This might be due to the use of a relatively high temperature value for these models -- previous work has found similar results and speculated that Copilot likely uses a lower temperature value~\cite{wermelinger2023using}.





\begin{table}[ht]
\centering
\caption{The replication results for~\cite{finnieansley2022robots}. An asterisk indicates that the solution required trivial modifications, which was counted as an additional attempt (i.e., ``2*'' means that the problem was solved on the first try, but had trivial mistakes e.g. in formatting of strings such as a missing period or extra unnecessary prints). A `-' means that the problem was not solved within 10 attempts.}
    \label{tab:replication_results}
\begin{tabular}{lccc}
\toprule
     Problem &  \multicolumn{3}{c}{Solved on attempt} \\
     & GPT-4 & GPT-3.5 & Copilot \\
\midrule
      T1-Q1 &                  2* & 1 & 2* \\
      T1-Q2 &                  1 & 1 & 1 \\
      T1-Q3 &                  1 & 1 & 1 \\
      T1-Q4 &                  2* & 2* & 2* \\
      T1-Q5 &                  1 & 2* & 1 \\
      T1-Q6 &                  1 & 2* & 1 \\
      T1-Q7 &                  2* & 1 & 1 \\
      T1-Q8 &                  1 & 1 & 1 \\
      T1-Q9 &                  1 & 1 & 1 \\
     T1-Q10 &                  1 & 1 & - \\
     T1-Q11 &                  3 & 1 & - \\
      T2-Q1 &                  1 & 2* & 1 \\
      T2-Q2 &                  1 & 2* & 1 \\
      T2-Q3 &                  1 & 1 & 1 \\
      T2-Q4 &                  1 & 2 & 1 \\
      T2-Q5 &                  1 & 2 & 4* \\
      T2-Q6 &                  1 & 1 & 1 \\
      T2-Q7 &                  1 & 1 & 1 \\
      T2-Q8 &                  2* & 2 & 4 \\
      T2-Q9 &                  1 & 1 & 1 \\
     T2-Q10 &                  1 & 1 & 1 \\
     T2-Q11 &                  2 & 1 & 1 \\
     T2-Q12 &                  - & - & - \\
 RF-Soloway &                  2* & 2* & - \\
   RF-Simon &                  2* & - & - \\
  RF-Fisler &                  2* & 2* & 10 \\
RF-Ebrahimi &                  2* & 2* & 2* \\
 RF-Guzdial &                  2* & 3 & - \\
 RF-Lakanen &                  2* & 2 & 3* \\
  RF-Apples &                  1 & 1 & 1 \\
\bottomrule
\end{tabular}

\end{table}

\subsection{Novel analysis}
\label{sec:novel_analysis}

\subsubsection{APPS}
\label{sec:apps}
Hendrycks et al. created the Automated Programming Progress Standard (APPS) dataset~\cite{hendrycksapps2021} as a benchmark for program generation. The dataset consists of 10.000 programming problems of varying difficulty, manually extracted from the online coding websites Codewars, AtCoder, Kattis, and Codeforces. The average number of lines for the solution is 18. 
The dataset has been divided into a training set and a test set, both containing 5000 problems. The following elements are provided for each problem:
\begin{itemize}
    \item Problem description, including a description of the expected input and output of the problem and some concrete examples.
    \item A JSON file with inputs and corresponding output. On average, each problem has 21.2 test cases.
    \item A JSON file with metadata, with a difficulty level (introductory, intermediate, competition) and the url to the website where the problem is hosted.
    \item A list of solutions from humans.
\end{itemize}

Hendryck et al. have tested the dataset in 2021 with GPT-2, GPT-3, and GPT-NEO. They found that GPT-NEO performed the best by passing almost 15\% of test cases on introductory problems.

\paragraph{Method} We aim to use the dataset from this paper and assess how newer models are able to handle these problems. We use a simplified method, employing the following steps:

We run the models (GPT-3.5-turbo16k, GPT-4) with the following parameters: temperature=0.0, max\_tokens=4000, and default values for Top P (1), Frequency penalty (0), and Presence penalty (0).
The system prompt we use is as follows:

\begin{verbatim}
You are a highly intelligent coding bot that can easily
handle any Python programming task. Given a natural
language instructions you can always implement the
correct Python solution. Your focus is the solution
code only. You are not allowed to provide explanations.

Make sure to use input statements for input, and do not
give a method definition

Example (toy) instructions:
Implement a Python program to print "Hello, World!" in
the hello.py.

Example bot solution:
=== hello.py ===
x=input()
print(x)
===    
\end{verbatim}

As the user input, we provide the full problem description as described above.

We process the generated code by running the code and providing the inputs from the first 25 test cases to each solution. We then compare the output to the expected output with an exact comparison, after stripping the results. After noticing that differences in characters for line ends caused problems (`\symbol{92}r\symbol{92}n' versus `\symbol{92}n'), we fixed this in the output comparison. We consider a test case to fail after detecting a timeout of 5 seconds. For each problem, we store a list of test results and calculate a success rate as the percentage of passed test cases.

For simplicity, we skipped the problems that had starter code. We also skipped problem descriptions that could not be read. 
We ran a sample of 100 interview-level problems for GPT-4, running 25 test cases for each problem. We manually assessed the failing test cases, to check if the problems were actual coding problems, or were caused by formatting mistakes. Minor issues were corrected.

The final runs were conducted in September 2023.

\begin{figure}[h!]
    \centering
    \includegraphics[width=0.5\textwidth]{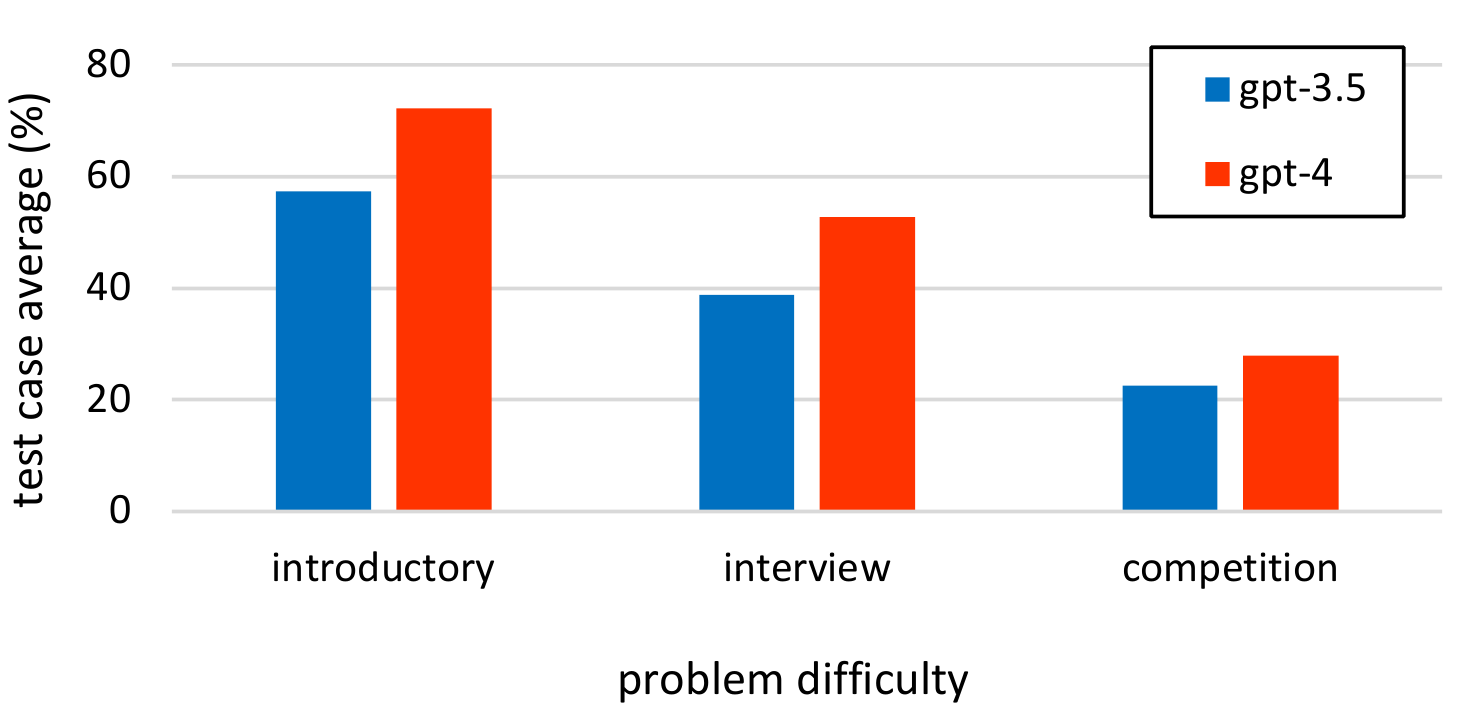}
    \caption{GPT success rate for different exercise types.}
    \label{fig:apps-results}
\end{figure}

\paragraph{Results} Table~\ref{tab:apps} and Figure~\ref{fig:apps-results} show the results. GPT-4 performs quite well overall, with an average score of 51.5\% on test cases. Comparing to GPT-3.5, which scored 39.2\%, the performance has clearly improved. Keeping a strict pass/fail criterion (all tests should pass), only 36.1\% of the GPT-4 solutions pass all tests, and 21.0\% for GPT-3.5. We also observe a large difference between problem types, with GPT-4 solutions to introductory problems scoring as high as 72.2\% on test case average, but the most difficult, competition level problems score only 28.7\%. 

\begin{table*}[ht]
\caption{Average test case score for APPS problems, ran with max. 25 test cases.}
\label{tab:apps}
\begin{tabular}{l r r | r r | r r  }

 &&& \multicolumn{2}{c|}{\textbf{Test case average}}  & \multicolumn{2}{c}{\textbf{Strict accuracy}}  \\

\textbf{Type} & \textbf{Count} & \textbf{Avg nr of tests} & \textbf{GPT-3.5} & \textbf{GPT-4} & \textbf{GPT-3.5}  & \textbf{GPT-4} \\
\midrule
Introductory & 974 & 8.9 & 57.3\% & 72.2\% & 44.5\% & 62.9\% \\
Interview & 2972 & 15.3 & 38.8\% & 52.9\% & 18.7\% & 35.8\%\\
Competition & 1000 & 9.3 & 22.6\% & 28.7\% & 5.2\% & 10.8\% \\[2px]
\textit{Overall} & 4946 & 12.9 & 39.2\% & 51.5\% & 21.0\% & 36.1\%\\

\\

\end{tabular}
\end{table*}

\paragraph{Discussion}

A major downside of the APPS dataset is that it is a public dataset with problems from popular online coding websites. There is a large chance that solutions for it have been included in training recent models.

Overall, APPS is a high-quality dataset with extensive test suites for many of the problems. 
However, in the context of computing education, several problems in this dataset might not be suitable for novice programmers. Even the introductory set contains some complex problems, and are targeted more at people participating in programming competitions.

We included an explicit instruction in the prompt to `use input statements for input, and do not give a method definition'. Omitting this in a first test run showed many tests failing, because the model returned a method definition. This shows we need to be sure the model creates solutions in the exact same format as expected.

There are different aspects to be considered when running an analysis of a model on a certain task. The number of attempts to get a solution from the model should be specified. The `Robots are coming'-replication (Section~\ref{sec:robots-replication}) performed multiple attempts, while we only use one attempt for this dataset.

\subsubsection{FalconCode}
\label{sec:falcon}
FalconCode is a novel collection of over 1.5 million Python programs from over 2,000 undergraduate students at the United States Air Force Academy \cite{defreitas2023falconcode}. The dataset is not available online but is provided to anyone who applies following the instructions at the dataset website.\footnote{\url{https://falconcode.dfcs-cloud.net/index.php}} The dataset contains 661 introductory Python programming problems used in 4 courses. To understand how well selected LLMs (GPT-3.5/4) perform on these problems we (i) extracted the problem statements and unit tests; (ii) performed de-duplication of problem statements; (iii) utilised the LLMs to generate solutions; (iv) ran the unit tests provided in the dataset against the outputs of the LLMs; and (v) analysed the test results.

\paragraph{Data extraction} The dataset is provided in the convenient tabular format where the table we work with has the 661 problems. Among other fields, there are the \emph{prompt} and \emph{testcase} columns that contain the problem statement and unit tests to evaluate the correctness of the solution. The problem statements are in HTML format which we preserved. The unit tests are runnable Python code. Hence, we extracted those into separate Python files.

\paragraph{De-duplication} The original dataset lists 661 problems. Apparently, some problems are re-used across courses. Since the id of the problem appears to be preserved it is possible to de-duplicate the problems based on the id which reduces the number of problems to 422. However, it turns out that the problem set is further reduced to 310 if one de-duplicates based on the problem description (prompt field). Interestingly, de-duplicating only on the prompt field, disregarding the id, yields 385 unique problem descriptions which suggests that there are several problems with the identical id that have different descriptions. Extracting plain text descriptions from the original HTML format and removing superficial white space results in 344 unique problem descriptions. For this study, we opted for de-duplicating on the original (HTML) prompt (just in case the white space difference could be meaningful). Hence, there are 385 problems we use in our experiments.

\paragraph{Generating solutions} To generate a solution to the given problem, we include the problem statement (HTML) into the same prompt as used with the APPS dataset experiment (Subsection \ref{sec:apps}). We decided against extracting plain text of the problem statements because the HTML may encode important information through formatting that a state-of-the-art LLM such as GPT-3.5/4 may be capable of leveraging. Note that many of the problem statements were referring to an external resource (e.g., starter code or a file with data) which has not been made available as part of the dataset. Hence, we cannot include the resource into the prompts submitted to the LLMs. We extract the completion of the submitted prompt and save it into a Python file named with the id of the problem (needed for the unit tests to discover the solution). In a non-negligible number of cases, the completion was wrapped in \verb|```python ...```| tokens. While this issue could likely be fixed via further prompt engineering, we simply removed these tokens using a regular expression. Leaving the tokens in would make the Python file not runnable (syntax error), which would automatically fail all the unit tests despite the solution being correct.

\paragraph{Testing} To test correctness of the automatically generated solutions, we executed the provided unit tests. The unit tests rely on the solutions being present in the same directory in a file named with the id of the problem. Additionally, there is an external \emph{cs110} grading (testing) library that needs to be installed through \emph{pip}.\footnote{\url{https://pypi.org/project/cs110/}}

We saved the output of the unit testing for each individual problem into a separate file. The last line of these had a predictable format: \emph{Unit Test Returned: \#}, where \# is replaced with number from 0 to 100. We utilised simple regular expressions to extract this final result of the evaluation from each of the output files. Note that assignments with identical problem statements could have been associated with different unit tests. Therefore, for each problem we ran all available unit tests, and took the average result of the tests as the score. 

\paragraph{Results analysis} Table \ref{tab:falcon} shows the performance of GPT-3.5/4 on the 385 FalconCode problems across three different types of assignments:

\begin{itemize}
    \item \emph{Skill} -- Small (1–-3 lines) programs focused on specific programming skill. 
    \item \emph{Lab} -- Medium (10–-50 lines) programs focused on utilisation of one or more skills.
    \item \emph{Project} -- Larger (50--300 lines) programs solving an open-ended problem.
\end{itemize}

\noindent GPT-4 performed markedly better than GPT-3.5. In the subsequent analysis, we focus on the better performing GPT-4 model. The overall performance of 45.4\% suggests a rather weak performance of the LLM in handling these introductory programming problems. In order to understand the causes of the low performance we analysed each case where GPT-4 did not achieve the full score. Specifically, we performed a thematic analysis in which causes of each failed assignment were extracted as codes and then collated into higher-level themes \cite{braun2006using}. The results of the analysis are shown in Table~\ref{tab:falcon_fail}.

\begin{table}[t]
\setlength{\tabcolsep}{3pt}
\caption{Results for FalconCode. The middle column describes the raw success rates for each category of problem, and the rightmost column describes the success rates after problems where insufficient information is provided were removed.}
\label{tab:falcon}
\begin{tabular}{l|rrr|rrr}
               & \multicolumn{3}{c|}{\textbf{Full}}& \multicolumn{3}{c}{\textbf{Clean}}\\
\textbf{Type} & \textbf{Count} & \textbf{GPT-3.5} & \textbf{GPT-4} & \textbf{Count} & \textbf{GPT-3.5} & \textbf{GPT-4}\\
\midrule
Skill    & 162 &30.3\%& 40.0\% & 81& 57.6\% & 80.0\%\\
Lab      & 214 &29.8\%& 51.4\%   &155& 47.5\% & 71.0\%\\
Project  &   9 & 0.0\%& 0.0\%  & 0 & -      & -\\
\midrule
\bf Overall  &\bf 385 &\bf 29.3\%&\bf 45.4\% &\bf 236&\bf 51.4\% &\bf 74.1\%\\
\end{tabular}
\end{table}

We first analysed the performance on Projects since it amounted to 0.0\%. It turns out that the problem statement included in the dataset only points to an external pdf file that contains the actual instructions, e.g.:

\begin{quote}
Objective: Create a drone simulation that can scan a battlefield for targets and engage them.

Instructions: Read writeup (airstrike.pdf) and use the template file to begin work.
\end{quote}

\noindent Since the pdf file has not been released with the dataset we could not provide the LLM with adequate instructions to generate a solution. This is the case for all 9 projects.

\begin{table}[t]
\caption{Reasons for LLM (GPT-4) failures on FalconCode problems. We used this analysis to filter the dataset down to a clean version that is appropriate to use for the evaluation. The clean column signifies which reasons were considered as the failure on the LLM part. These data points were included in the clean dataset.}
\label{tab:falcon_fail}
\begin{tabular}{lrrrr}
\textbf{Failure Cause} & \textbf{Clean} & \textbf{Skill} & \textbf{Lab} & \textbf{Project} \\
\midrule
Missing Instructions   & &    &      &  9     \\
Missing Starter Code   & & 58 & 7    &        \\
Missing Data File      & & 20 & 46   &        \\
No Unit Tests          & & 2  & 2    &        \\
Incorrect Unit Tests   & & 1  & 4    &        \\
Unexpected Library     &\checkmark& 1  & 2    &        \\
Incorrect Structure    &\checkmark&    & 25   &        \\
Incorrect Solution     &\checkmark& 22 & 28  &        \\
\midrule
Overall Failed         & &104 & 114  & 9      \\
Overall Failed (clean) & &23  & 55   &       \\
\end{tabular}
\end{table}

As Skill assignments are supposed to require only small solutions, not exceeding several lines of code, the performance of 40.0\% is rather unexpected. Our analysis revealed that the main causes of the poor performance are related to the same cause detected with the Project tasks. There are a substantial number of situations where the Skill assignment required an external data file, and even more commonly starter code was needed to complete the assignment, e.g.:

\begin{quote}
    You have been provided with a list called list\_of\_animals. Write a program that prints out each of the items in this list (one item per line).
\end{quote}

\noindent As the starter code has not been included in the dataset, the LLM does not have the complete information to produce the desired output. We also detected instances where the LLM used an unexpected library (not part of the Python standard library) and hence the program would crash, i.e., the unit tests would fail. In a few cases, the unit tests would be missing or incorrect. We also identified several instances where GPT-4 generated a genuinely incorrect solution to the problem statement that provided sufficient information.

The most common cause of failed unit tests for Lab assignments was also a missing data file and/or starter code (not released with the dataset). Another common cause was an incorrect structure of the (possibly correct) solution. A typical example would be a solution containing a function that returns a value whereas it was supposed to be a script asking a user to provide an input and print the output to the terminal. In the remaining cases GPT-4 generated a genuinely incorrect solution.

Based on the above analysis, we report another set of results (Clean) on the subset of 236 FalconCode problems that provide sufficient information for the LLM and are associated with valid test cases. The success rate increases from 45.4\% to 74.1\%. It is worth emphasising that if we would also disregard the cases where GPT-4 produced the correct solution using an unexpected structure (e.g., a function returning a value instead of a program asking user for an input and printing to a terminal) or utilised an unexpected library, the success rate would increase to 86.8\%. Finally, a large portion of the genuinely incorrect solutions are rather superficial problems, such as not rounding to a single decimal as demonstrated in the example provided in the instructions. A simple change to the prompt adding the specific instruction would certainly fix such issues. Hence, one can conclude that we only observed minimal amount of cases where GPT-4 would produce a truly incorrect solution to the problem (most certainly in fewer than 5\% of cases).


\subsection{Discussion}

In this section, we provide an overview of the issues we encountered while performing our replication and analyses. Our experiences could provide valuable insights to researchers who want to study LLM performance as well as to teachers who are interested in their performance in an educational context.

\paragraph{Higher LLM performance than identified in the literature} 
Our replication of the \citet{finnieansley2022robots} paper suggests that new LLM models are significantly more capable than is currently reported in the literature. Our experiment with the FalconCode dataset further supports this conclusion, and while success rates are lower on the APPS dataset, those problems are significantly more challenging than those typically provided in the CS1 and CS2 courses typically discussed in the literature.

\paragraph{Challenges using publicly available datasets} 
We encountered a number of challenges applying LLMs to publicly available datasets, even though the datasets themselves are of high quality. In particular, we anticipate that future researchers will find that very few datasets will have been produced specifically to support LLM code generation research, so they are likely to not include critical information like starter code, data files, or formatting instructions.

Researchers will also encounter challenges even with datasets produced for code generation tasks. As suggested by Liu et al.~\cite{liu2023your}, existing datasets may need to be augmented to evaluate LLM-generated code accurately. The number and quality of the test cases provided might not completely cover all exercise requirements and edge cases, therefore giving a false positive result. Alternately, test cases can even be incorrect, or too strict, exceeding what is required in the instructions, lowering the potential performance of models.

\paragraph{Homogeneity in available datasets}
Finally, future researchers may struggle to find appropriate datasets. Most datasets we found could support code generation, using Python, of CS1 problems. To assess model performance on multiple types of tasks, for different programming languages, or at different levels, will require new datasets. The community will need to reward the effort of curating and maintaining such datasets, as providing a complete and well-evaluated dataset is challenging (as noted above) and is important for enabling research by a diverse set of groups.

\subsubsection{Limitations}

We focused our efforts on replicating code generation tasks, but there are many other research questions that are potentially even more challenging to replicate. Testing generated code can be easily automated by running test cases, although this might not capture all aspects relevant to computing educators, such as code quality and suitability of the solution with regards to which concepts the student has learned so far. These latter aspects could also be assessed automatically, but we have not attempted to do so in our study.

Outside of code generation, assessing solutions for other types of exercises common in CS (e.g. regular expressions, UML-diagrams, automata), LLM-generated exercises, feedback, and explanations require additional datasets and may require a qualitative framework that could be difficult to provide or to transfer to another research team.

\vspace*{2mm}
\begin{custombox}{Advice for users and creators of LLM CSEd Datasets}
\begin{itemize}[leftmargin=1em]
\item Creators: Include full and precise problem descriptions, so that the LLM can be given sufficient information for solving the problem.
\item Creators: Include full test cases, ideally in a format where they are easy to run for others, so that LLM performance can be easily evaluated.
\item Creators: Include any resources needed to complete the assignments, e.g., the starter code or data files.
\item Creators: Make it easy to update or extend the data set, and report issues.
\item Users: Make LLM parameters clear for replication.
\item Users: Clearly describe the prompts used with the models, ideally providing example prompts.
\end{itemize}
\end{custombox}

\section{Conclusions}

This report is the output of an ITiCSE Working Group that explored how the emerging generative AI revolution will impact the future of computing education.  The first time that the group met was in April 2023 -- three years after the release of the ground-breaking GPT-3 large language model; less than two years after the release of the Codex model (a variant of GPT-3 specifically fine-tuned for coding tasks); less than one year since the Copilot plug-in (for generating code directly within an IDE) was made available for free to students worldwide; five months since the release of ChatGPT (providing a convenient chatbot interface); and just one month after the release of GPT-4, a powerful multi-modal large language model.  Against this backdrop of rapid advancements, our working group came together at a time when the computing education community was just beginning to grapple with the widespread use of generative AI tools by students as well as the general public.  Many urgent questions were being asked about how to adapt to the challenges and opportunities presented by these new models and tools.  In particular, if students are able to generate solutions to all of their programming coursework, how will this impact what is taught, how it is taught, and how students will remain motivated to learn? 

Our overarching goal is for this report to serve as a focal point for researchers and practitioners who are exploring, adapting, using, and evaluating LLMs and LLM-based tools in computing classrooms. We now return to the list outlined in Section \ref{sec:contributions} to summarise our main contributions:

\begin{enumerate}[leftmargin=*,align=left]

\item \textbf{A review of the literature:}
We provide a detailed review of the literature on LLMs in computing education, current as of August 2023.  Using a keyword search of relevant databases and two rounds of forward and backward snowballing, we synthesise findings from 71 primary articles.  Due to rapid changes in the field and the slow pace of publishing in traditional venues, much of this work was available only as pre-prints on platforms such as arXiv.  We included all such literature in our review and assessed every article with respect to a set of quality metrics.  The most common type of paper to date involves evaluating the performance of LLMs when applied to tasks such as programming exercises.  A key finding, which justifies some of the widely voiced concerns around academic misuse of LLMs, is that current models tend to perform at least as well as most students on typical introductory-level programming tasks.  We also reviewed papers that discussed possible opportunities and challenges of LLMs, that studied how end-users (including students) interacted with LLMs, and that used LLMs to generate high-quality learning resources.  Among the risks that were identified, the most common concern expressed by authors was that students would become overly-reliant on using LLMs to generate and debug code. 

\item \textbf{Prevailing attitudes:}
To understand how LLMs are currently being perceived and used, we conducted a survey involving 171 students and 57 instructors from computing courses spanning 20 distinct countries.  We found that, in general, students and instructors had similar perceptions about LLMs with respect to questions around experiences, expectations and beliefs. However, they differed in their perceptions of how clear course policies were about the allowed use of LLMs, with instructors -- somewhat surprisingly -- finding these policies to be less clear than students.  Many of the respondents to our survey had very little experience using generative AI tools at the current time, although we expect familiarity to grow rapidly in the coming years.  Some instructors were concerned that their students were using such tools inappropriately, and a small fraction of students refused to use generative AI tools for ethical reasons and due to concerns about harming their learning.   In many cases perceptions were well-aligned -- both students and instructors felt strongly that there should be some restrictions on the allowed usage of generative AI tools for coursework. 

\item \textbf{New instructional approaches:}
Although many instructors are only just beginning to think about the impacts on their teaching, some have already made concrete changes to their curricula and assessments.  In order to document these recent adaptations, we conducted 22 in-depth interviews with instructors on five continents who already had concrete plans in place to change some aspect of their teaching.  We found that some instructors were beginning to place a greater emphasis on `process over product'.  That is, instead of just grading a final artefact, there is an evaluation of the processes used by students when working on a product.  In addition, there was also a trend towards placing a greater emphasis on invigilated assessments such as exams, with a reduction to the grade weighting placed on unsupervised homework assignments.  We anticipate further changes to learning objectives, course content and assessment practices in the near future, and we see an important need to swiftly disseminate best practices that emerge. 

\item \textbf{Academic integrity: Policies \& recommendations}
We reviewed academic integrity policies that mentioned generative AI from major universities around the world and found that these explicitly addressed many of the principles stated in the ACM code of ethics.  However, it is unclear how students are being educated about the ethical use of generative AI in the classroom.  This appears to be an important area for future work given the findings from our survey which revealed that students and instructors have quite different views regarding the clarity of current policies. Further work in this area is needed to understand how to effectively embed these principles in computing classrooms.  

We follow our review of policies with concrete recommendations for both students and instructors.  We agree with the position articulated by many publishers that the user of the LLM should be considered the author of the generated text. This has implications for academic integrity, in that \textit{plagiarism} is not usually a concern but instead users would be responsible for any \textit{falsification} produced from uncritical use of LLM-generated artefacts.   We encourage instructors to teach students about the ethical use of generative AI throughout their courses, clearly stipulating any restrictions on use for assessed work.  Should students use such tools for graded tasks, we recommend they include a statement detailing its usage, and any violations should be regarded as academic misconduct with the penalties explained clearly.  Institutions and faculty will need to communicate these expectations explicitly, and thus it is imperative that we provide students with resources to understand how to use LLMs appropriately.  To this end, we have prepared a sample handout that can be adapted and included in a course syllabus on the ethics of using generative AI tools for assignment work (see Appendix~\ref{appendix:student}). 

\item \textbf{Encouraging replication:}
Given that instructors are naturally interested in how well LLMs can solve typical tasks that are set for students, a common thread of work to date has been to evaluate the performance of various models.  However, replicating prior work using newer models is difficult, given that a wide variety of parameters, prompts, and evaluation approaches have been used, and not all methods are reported with sufficient detail. Producing a dataset that contains everything necessary for high-quality LLM research (in particular, accurate evaluation of the artefacts generated by LLMs) is challenging and needs to be encouraged by the community.  We therefore identify a seminal paper on LLM evaluation for programming tasks, and prepare and release the problem descriptions and test cases in order to facilitate future replication work.  Our own replication of this prior work, using a state-of-the-art model, shows an extraordinary performance improvement over the span of two years since the original work was carried out.

\end{enumerate}

As we collectively face the changes being ushered in by the AI revolution, it is clear that LLMs present significant challenges but also new opportunities for computing educators.  We present this report not only as a snapshot of the current state at this relatively early stage, but also as a call to action: to encourage broad exploration of the use and impacts of LLMs and LLM-based tools in computing classrooms, to adapt teaching methods and update academic integrity policies, and to develop best practice and to share them widely with the computing education community.  The future of computing education is rapidly evolving, and shaping it must be a collective effort.

\begin{acks}
Thank you to the following:
\begin{itemize}
    \item Michael Caspersen for joining this effort in the early days and graciously bowing out when more prestigious matters intervened.
    \item All of the students and educators who responded our surveys.
    \item All of the educators who participated in interviews (in alphabetical order by surname): 
    \begin{itemize}
    \item Austin Cory Bart (University of Delaware, USA)
    \item Michael Caspersen (It-vest \& Aarhus University, Denmark)
    \item James Davenport (University of Bath, UK)
    \item Rodrigo Duran	(Federal Institute of Mato Grosso do Sul Brazil)
    \item Dan Garcia (UC Berkeley, USA)
    \item Michael K{\"o}lling (King's College London, UK)
    \item Viraj Kumar (Indian Institute of Science, Bengaluru, India)
    \item Mark Liffiton (Illinois Wesleyan University, USA)
    \item J{\'e}r{\'e}mie Lumbroso (University of Pennsylvania, USA)
    \item Peter Mawhorter (Wellesley College, USA)
    \item Jean Mehta (Saint Xavier University, USA)
    \item Briana Morrison (University of Virginia, USA)
    \item Leo Porter (University of California San Diego, USA)
    \item Jan Schneider (Goethe Universit{\"a}t, Germany)
    \item David H. Smith IV (University of Illinois, Urbana-Champaign, USA)
    \item Kristin Stephens-Martinez (Duke University, USA)
    \item Sven Strickroth (LMU Munich, Germany)
    \item Ewan Tempero (University of Auckland, New Zealand)
    \item Christian Tomaschitz (TU Wien, Austria)
    \item Frank Vahid (University of California, Riverside, USA)
    \item those who wished to be de-identified in this report.

    \end{itemize}
\end{itemize}
\end{acks}

\balance
\bibliographystyle{ACM-Reference-Format}
\bibliography{main,extraction,CandAextra}

\appendix 

\clearpage
\newpage
\section{Paper Extraction Form} \label{app:paper-extraction}

The following questions were placed on a form that the group used when evaluating the papers that were included in the literature review.

\begin{enumerate}
    \item \textbf{Paper title} [text entry]
    \item \textbf{Bibtex entry} [text entry]
    \item \textbf{Article type} [multi-select]
    \begin{itemize}
        \item Position / discussion paper
        \item Supervised study (a study that is conducted in a highly controlled environment like a research lab)
        \item Unsupervised study (a study that is conducted in a less restricted environment, such as online)
        \item New tool paper (presenting a new tool / LLM)
        \item Evaluation paper (evaluating an existing tool / LLM)
        \item Other [text entry]
    \end{itemize}
    \item \textbf{Author affiliation} [multi-select]
    \begin{itemize}
        \item Academic
        \item Industry
    \end{itemize}
    \item \textbf{Country of human participants}: If data is collected from human participants, provide a comma separated list of countries of where participants were located (if not explicitly mentioned, put ``unclear"). [text entry]
    \item \textbf{Level of human participants} [multi-select]
    \begin{itemize}
        \item Uncontextualized
        \item Primary school (e.g. elementary, intermediate, middle school)
        \item Secondary school (e.g. high school)
        \item Tertiary education (e.g. college, university)
        \item Informal education (e.g. MOOCs)
        \item Professional developers
        \item Not applicable
        \item Other [text entry]        
    \end{itemize}
    \item \textbf{Number of human participants from whom data was collected} (if available, type into other) [multi-select]
    \begin{itemize}
        \item Not applicable
        \item Unclear
        \item Other [text entry]
    \end{itemize}
    \item \textbf{Description of participants}: A copy-paste (or paraphrased) description of participants from whom data is collected which may be useful to a more detailed thematic analysis.  This information can often be found at the beginning of a Methods section. [text entry] 
    \item \textbf{How do the authors motivate the work}: A copy-paste (or paraphrased) description of the motivation for the work as expressed by the authors. [text entry]
    \item What LLM / tool is used? [multi-select]
    \begin{itemize}
        \item GPT-3
        \item GPT-4
        \item Codex
        \item Copilot
        \item Unclear
        \item N/A
        \item Other [text entry]
    \end{itemize}
    \item \textbf{What are the explicit research questions / research goals / hypotheses in the article?} A copy-paste (or paraphrased) description of the RQs, goals, hypotheses. [text entry]
    \item \textbf{What programming languages are involved in the study?} [multi-select]
    \begin{itemize}
        \item Java
        \item Python
        \item C
        \item C++
        \item Not programming language focused
        \item Other [text entry]
    \end{itemize}
    \item \textbf{How does the article evaluate the data collected?} [multi-select]
    \begin{itemize}
        \item Qualitatively
        \item Quantitatively
        \item N/A
    \end{itemize}
    \item \textbf{Quality assessment}: An assessment of the research ``quality".  [Yes / No / Vague radio grid]
    \begin{itemize}
        \item \textbf{Is there a clearly defined research question/hypothesis?}
        \item \textbf{Is the research process clearly described?}
        \item \textbf{Are the results presented with sufficient detail?}
        \item \textbf{Are threats to validity / limitations addressed in an explicit (sub)section?} (code as ``vague" if discussed, but not in a separate subsection)
    \end{itemize}
    \item \textbf{What is the contribution / what are the key results of the article?} Provide a short summary of the main findings. [text entry]
    \item \textbf{Curriculum changes: Provide any potential effects on computing curriculum that could result from this work} (when answering this question, please feel free to provide some of your own commentary - it is fine to mention curriculum changes which the paper prompted you to think about, even if they aren't explicitly mentioned by the paper authors). [text entry]
    \item \textbf{Opportunities: Provide any potential opportunities for computing education that could result from this work} (when answering this question, please feel free to provide some of your own commentary - it is fine to mention opportunities which the paper prompted you to think about, even if they aren't explicitly mentioned by the paper authors). [text entry]
    \item \textbf{Threats: Provide any potential threats for computing education that could result from this work} (when answering this question, please feel free to provide some of your own commentary - it is fine to mention threats which the paper prompted you to think about, even if they aren't explicitly mentioned by the paper authors). [text entry]
    \item \textbf{Additional notes}: Can be used to note any interesting aspects of paper or anything else relevant that isn't captured in the extraction fields above. [text entry]
\end{enumerate}
\clearpage
\newpage
\section{Student Survey Questions} \label{app:student-survey}
The following questions were used in the survey filled out by students.

\begin{enumerate}

\item{
\textbf{Gender}
\begin{itemize}
    \item[\Circle] Man
    \item[\Circle] Woman
    \item[\Circle] Non-binary
    \item[\Circle] Other:
\end{itemize}
}

\item{\textbf{Country} (drop-down list)}

\item{
\textbf{Level of Study}
\begin{itemize}
    \item[\Circle] First Year
    \item[\Circle] Second Year
    \item[\Circle] Third Year
    \item[\Circle] Fourth Year
    \item[\Circle] Fifth Year
    \item[\Circle] Other: 
\end{itemize}
}

\item{
\textbf{Degree Major / Specialization}
\begin{itemize}
    \item[\Circle] Computer Science
    \item[\Circle] Software Engineering
    \item[\Circle] Information Technology
    \item[\Circle] Computer Engineering
    \item[\Circle] Bioinformatics
    \item[\Circle] Other:
\end{itemize}
}

\item{\textbf{Number of courses with a programming component which you have completed} (text entry)}

\item{
\textbf{Rate your agreement with the following statements:} (Likert)
\begin{itemize}
    \item I regularly use GenAI tools when working with text (e.g.: writing emails, reports, summaries)
    \item I regularly use GenAI tools when working with code (e.g.: generating code or explanations, writing programs, debugging, ...)
    \item I regularly use GenAI tools when working with images (e.g.: generating new pictures, ...)
\end{itemize}
}

\item{
\textbf{After generating code using GenAI tools, I mostly:}

\begin{itemize}
    \item[\Circle] Not applicable (I have not used GenAI tools to generate code)
    \item[\Circle] Use the code immediately.
    \item[\Circle] Skim through the code briefly to make sure that it looks correct.
    \item[\Circle] Read it carefully (with skepticism) to ensure that it is correct.
    \item[\Circle] Read it carefully (with skepticism) and also write code to test it.
\end{itemize}
}

\item{
\textbf{Rate your agreement with the following statements:} (Likert)
\begin{itemize}
    \item I expect to use GenAI tools increasingly in my learning practices in the future
    \item Using GenAI tools frequently to generate code is harmful for my learning of programming
    \item GenAI tools can provide guidance for coursework as effectively as human teachers
    \item GenAI tools will replace human teachers in the future
\end{itemize}
}

\item{
\textbf{Students must be taught how to use GenAI tools well for their future careers} (Likert)
}

\item{
\textbf{When you have a question regarding the material you are studying or are stuck on a problem, in what order do you do the following?} (ranking question)
\begin{itemize} 
    \item Ask using GenAI tools
    \item Ask on the course discussion forum
    \item Search online (e.g. Google)
    \item Ask a friend
    \item Ask on online forums like StackOverflow
    \item Ask the course instructor/TA
\end{itemize}
}

\item{
\textbf{Rate your agreement with the following statements} (Likert)
\begin{itemize}
    \item The policies at my university are clear regarding what is allowed and what is not allowed in terms of using GenAI tools
    \item The policies in the courses I took last semester were clear regarding what is allowed and what is not allowed in terms of using GenAI tools
    \item There should be no restrictions on the use of GenAI tools in coursework
\end{itemize}
}

\item{
\textbf{For programming assignments, I believe GenAI should be:}

\begin{itemize}
    \item[\Circle] Always allowed
    \item[\Circle] Allowed in some assignments, disallowed in others (based on the assignment type, course level, etc.)
    \item[\Circle] Always disallowed
\end{itemize}
}

\item{
\textbf{Can you elaborate on when you believe GenAI should be allowed or disallowed?} (text entry)
}

\item{
\textbf{To what extent do you think students at your school are using GenAI tools in ways that your instructors would not approve of?}
\begin{itemize}
    \item[\Circle] Almost everyone
    \item[\Circle] Many
    \item[\Circle] Some
    \item[\Circle] A few
    \item[\Circle] Almost none
\end{itemize}
}

\item{
\textbf{In the absence of an explicit course policy on the use of GenAI tools, which of the following do you consider as NOT ethical? (Mark all that apply):}

\begin{itemize}
    \item[\Square] It is unethical to auto-generate a solution for the whole assignment (or a large portion of it) and submitting it without understanding it.

    \item[\Square] It is unethical to auto-generate a solution for the whole assignment (or a large portion of it) and submitting it after reading it and completely understanding it.
    
    \item[\Square] It is unethical to auto-generate a solution even for small parts of the assignment.
    
    \item[\Square] It is unethical to use GenAI tools to "explain" to you (step-by-step) how to solve the problem.
    
    \item[\Square] It is unethical to provide your code to GenAI tools and ask them to help you fix a bug.
    
    \item[\Square] It is unethical to ask GenAI tools to comment, tidy and improve the style of your code.
    
    \item[\Square] It is unethical to write the solution in a programming language (other than the one used in the course) and asking GenAI tools to translate it for you to the language of the course (and then submitting the translated code).
    
\end{itemize}
}

\item{
\textbf{If everyone in class is using GenAI tools, but it is against the rules to use them, then I would still use them.} (Likert)
}

\item{
\textbf{Rate your agreement with the following statements} (Likert)
\begin{itemize}
    \item GenAI tools will negatively impact my future job prospects
    \item GenAI tools will harm the development of generic or transferable skills such as teamwork, problem-solving, and leadership
    \item I am concerned that I will become over reliant on GenAI tools
    \item I trust the code written by GenAI tools more than the code I write
    \item My instructors can detect code that was written by GenAI tools
    \item My instructors actively check for unauthorized use of GenAI tools
    \item Even if my instructors disallow GenAI tools in my programming assignments, it is fine for me to use them to generate code as long as I understand the code very well. It is unethical only if I copy without understanding.
\end{itemize}
}

\item{
\textbf{Describe the ways you currently use GenAI tools in computing courses for text generation (e.g.: writing reports, summaries, etc)} (text entry)
}

\item{
\textbf{Describe the ways you currently use GenAI tools in computing courses for code generation (e.g.: debugging, writing, etc)} (text entry)
}

\item{
\textbf{Describe the effects you think GenAI tools will have on your prospects for future employment:} (text entry)
}

\item{
\textbf{What are your views on the allowed usage of GenAI tools in coursework/exams?} (text entry)
}

\end{enumerate}

\clearpage
\newpage
\section{Instructor Survey Questions} \label{app:instructor-survey}
The following questions were used in the survey filled out by instructors.

\begin{enumerate}

\item{
\textbf{Gender}
\begin{itemize}
    \item[\Circle] Man
    \item[\Circle] Woman
    \item[\Circle] Non-binary
    \item[\Circle] Other:
\end{itemize}
}

\item{\textbf{Country} (drop-down list)}

\item{
\textbf{Teaching Experience (years)} (text entry)
}

\item{
\textbf{Select the sizes of classes you taught in the most recent semester}
\begin{itemize}
    \item[\Square] 1-10 students
    \item[\Square] 11-30 students
    \item[\Square] 31-50 students
    \item[\Square] 51-100 students
    \item[\Square] 101-250 students
    \item[\Square] 251-500 students
    \item[\Square] 500+ students
\end{itemize}
}

\item{
\textbf{Select the type of department/school/faculty}
\begin{itemize}
    \item[\Circle] Computer Science
    \item[\Circle] Software Engineering
    \item[\Circle] Information Technology
    \item[\Circle] Computer Engineering
    \item[\Circle] Bioinformatics
    \item[\Circle] Other
\end{itemize}
}

\item{
\textbf{Rate your agreement with the following statements:} (Likert)
\begin{itemize}
    \item I regularly use GenAI tools when working with text (e.g.: writing emails, reports, summaries)
    \item I regularly use GenAI tools when working with code (e.g.: generating code or explanations, writing programs, debugging, ...)
    \item I regularly use GenAI tools when working with images (e.g.: generating new pictures, ...)
\end{itemize}
}

\item{
\textbf{After generating code using GenAI tools, I mostly:}

\begin{itemize}
    \item[\Circle] Not applicable (I have not used GenAI tools to generate code)
    \item[\Circle] Use the code immediately.
    \item[\Circle] Skim through the code briefly to make sure that it looks correct.
    \item[\Circle] Read it carefully (with scepticism) to ensure that it is correct.
    \item[\Circle] Read it carefully (with scepticism) and also write code to test it.
\end{itemize}
}

\item{
\textbf{Rate your agreement with the following statements:} (Likert)
\begin{itemize}
    \item I expect to use GenAI tools increasingly in my teaching practices in the future	

    \item Using GenAI tools frequently to generate code is harmful for my students' learning of programming	

    \item GenAI tools can provide guidance for coursework as effectively as human teachers	

    \item GenAI tools will replace human teachers in the future	

\end{itemize}
}

\item{
\textbf{Rate your agreement with the following statements} 
\begin{itemize} 
    \item Students must be taught how to use GenAI tools well for their future careers	

    \item I plan to change my assessment practices now that GenAI tools are commonly available	

    \item I plan to change my curriculum now that GenAI tools are commonly available	
\end{itemize}
}

\item{
\textbf{Rate your agreement with the following statements} (Likert)
\begin{itemize}
    \item The policies at my university are clear regarding what is allowed and what is not allowed in terms of using GenAI tools
    
    \item The policies in the courses I taught last semester were clear regarding what is allowed and what is not allowed in terms of using GenAI tools	

    \item There should be no restrictions on the use of GenAI tools in coursework
\end{itemize}
}

\item{
\textbf{For programming assignments, I believe GenAI should be:}

\begin{itemize}
    \item[\Circle] Always allowed
    \item[\Circle] Allowed in some assignments, disallowed in others (based on the assignment type, course level, etc.)
    \item[\Circle] Always disallowed
\end{itemize}
}

\item{
\textbf{Can you elaborate on when you believe GenAI should be allowed or disallowed?} (text entry)
}

\item{
\textbf{To what extent do you think students at your school are using GenAI tools in ways that you would not approve of?}
\begin{itemize}
    \item[\Circle] Almost everyone
    \item[\Circle] Many
    \item[\Circle] Some
    \item[\Circle] A few
    \item[\Circle] Almost none
\end{itemize}
}

\item{
\textbf{In the absence of an explicit course policy on the use of GenAI tools, which of the following do you consider as NOT ethical for students to do? (Mark all that apply):
}

\begin{itemize}
    \item[\Square] It is unethical to auto-generate a solution for the whole assignment (or a large portion of it) and submitting it without understanding it.
    
    \item[\Square] It is unethical to auto-generate a solution for the whole assignment (or a large portion of it) and submitting it after reading it and completely understanding it.
    
    \item[\Square] It is unethical to auto-generate a solution even for small parts of the assignment.
    
    \item[\Square] It is unethical to use GenAI tools to "explain" how to solve the problem step-by-step.
    
    \item[\Square] It is unethical to provide code to GenAI tools and ask them to help fix a bug.
    
    \item[\Square] It is unethical to ask GenAI tools to comment, tidy and improve the style of the code.
    
    \item[\Square] It is unethical to write the solution in a programming language (other than the one used in the course) and asking GenAI tools to translate it to the language of the course (and then submitting the translated code).
\end{itemize}
}


\item{
\textbf{Rate your agreement with the following statements} (Likert)
\begin{itemize}
    \item GenAI tools will negatively impact my students' future job prospects	

    \item GenAI tools will harm the development of generic or transferable skills such as teamwork, problem-solving, and leadership	

    \item I am concerned that my students will become over reliant on GenAI tools	

    \item I trust the code written by GenAI tools more than the code I write
    
    \item I can detect code that was written by GenAI tools	
    
    \item I actively check for unauthorized use of GenAI tools	
\end{itemize}
}

\item{
\textbf{Describe the ways you currently use GenAI tools in computing courses for text generation (e.g.: writing reports, summaries, etc)} (text entry)
}

\item{
\textbf{Describe the ways you currently use GenAI tools in computing courses for code generation (e.g.: debugging, writing, etc)} (text entry)
}

\item{
\textbf{Please describe any changes you have made, or plan to make, to your teaching approaches in courses you are teaching:} (text entry)
}

\item{
\textbf{Please describe any changes you have made, or plan to make, to your assessment approaches in courses you are teaching:} (text entry)
}

\item{
\textbf{If you have already implemented changes,  describe how successful you think they were.} (text entry)
}

\item{
\textbf{What new content/courses do you think should be taught/added to the curriculum?}  (text entry)
}

\item{
\textbf{Does your institutional academic integrity policy explicitly mention generative AI?}
\begin{itemize}
    \item[\Circle] Yes
    \item[\Circle] No
\end{itemize}
}

\item{
\textbf{Can you provide a link to your institution's policy on GenAI tools?} (text entry)
}

\item{
\textbf{Do you have a policy explicitly stated in your syllabus about when GenAI tools should or could be used in your course? }
\begin{itemize}
    \item[\Circle] Yes
    \item[\Circle] No
\end{itemize}
}

\item{
\textbf{If you have a statement about acceptable use of generative AI in your course syllabus documents, and you are willing to share that statement, please paste it below:} (text entry)
}

\item{
\textbf{Have you observed any students using these models? If so, how are they using them?} (text entry)
}

\item{
\textbf{Are you willing to be interviewed regarding the use of GenAI tools in computing classes?  If so, please provide a preferred contact email address.} (text entry)
}

\end{enumerate}

\clearpage
\newpage
\section{Student Guide}\label{appendix:student}



Generative AI refers to a kind of artificial intelligence software that is capable of generating information in response to prompts.  The software is trained on source data, and uses that training data as input to a sophisticated model that predicts the appropriate response to the prompt.  It does not understand the prompts, but it produces a convincing simulation of understanding.  Examples of generative AI systems that use text include ChatGPT and Bard, and generative AI models capable of generating images include Midjourney and DALL-E.

Generative AI tools can be used in ways that increase productivity and help you to learn.  However, they may also be used in unproductive ways that provide answers without helping you to learn.


\vspace*{2mm}
\begin{custombox}{Policy on generative AI}
\begin{itemize}
\item You may use AI tools to help you learn during lab exercises and assignments.
\item You will NOT be permitted to use AI tools in secure assessments (i.e., the Test and Exam).
\end{itemize}
\end{custombox}

\subsection*{Examples of productive use}
Generative AI tools are used in industry so you will be likely to use them regularly in your future work after graduation.  Therefore, you should learn to use them appropriately so you receive the most long-term benefit.  As a student, effective uses of generative AI  tools are centered on helping you understand course material, and may include asking generative AI to:
\begin{itemize}
    \item Explain a given topic, or to provide an example of how programming constructs are used.
    \item Explain your program one line at a time.
    \item Produce an example that is similar to assignment questions.
    \item Explain the meaning of error messages.
    \item Generate code to complete tasks that you have already mastered from previous coursework.  
\end{itemize}

\subsection*{Examples of inappropriate use}
Some uses of generative AI do not typically help you learn, and such uses are likely to result in worse long-term outcomes (e.g., you will not be able to complete Test and Exam questions, or to continue to following courses that expect a mastery of early programming content).  Examples of these uses are: 
\begin{itemize}
    \item Using AI tools on official assessments where it has been forbidden.
    \item Asking generative AI to complete laboratory questions or assignments for you.
    \item Asking generative AI to debug code that has errors.
    \item Writing a code solution in a language you know and then asking an AI tool to translate that code into the language required for the assignment.
\end{itemize}


\subsection*{Risks of generative AI}
There are many risks associated with the use of generative AI.
\begin{description}
    \item[Accuracy]  If you are using generative AI tools for learning then you should always double-check the content.  For example, if you are assigned to write a program that uses a specific algorithm, AI tools may generate a solution that arrives at the correct answer but does not use the required algorithm.  If you use generative AI to assist in the creation of assessed content then you are responsible for the accuracy and correctness of the work that you submit.
    \item[Quality]  Content generated may be of poor quality, and generic in nature.  Code may have security flaws and may contain bugs.  It is important that you understand how any generated code works and you evaluate the quality of the content.
    \item[Learning]  Generative AI can be a powerful productivity tool for users who are already familiar with the topic of the generated content because they can evaluate and revise the content as appropriate.  Tasks assigned by your teachers are designed to help you learn, and relying on AI tools to complete tasks denies you the opportunity to learn, and to receive accurate feedback on your learning.
    \item[Over-reliance]  Using AI tools to do your work for you may achieve the short-term goal of assignment completion, but consistent over-reliance on AI tools may prevent you from being prepared for later examinations, subsequent coursework, or future job opportunities.
    \item[Motivation] You may experience lack of motivation for tasks that generative AI can complete.  It is important to understand that you need to master simple tasks (which generative AI can complete) before you can solve more complex problems (which generative AI cannot complete).  Stay motivated!
\end{description}

\subsection*{Impact on others}
There are many consequences to inappropriate usage of AI tools. Some of these consequences may be unintended, and could potentially harm others.  For example:
\begin{description}
    \item [Other students] You could expose other students to harm by preventing their learning or including content in a group assignment that violates academic integrity.
    \item [Faculty] Violating academic integrity standards through the use of AI tools requires time and energy, and is emotionally draining to teachers and administrators, to enforce these standards.
    \item [Institutional] Including code from AI tools that you do not understand could expose the university to loss of reputation or even financial harm through lawsuits.
\end{description}

\subsection*{Academic misconduct}
Using generative AI in ways that are not permitted will be treated as academic misconduct.  This will have serious consequences.

\end{document}